\newif\ifsubmode 
\submodefalse

\ifsubmode
\documentclass[10pt,preprint]{aastex}
  \received{}
  \revised{}
  \accepted{}
\else
  \documentclass[iop,numberedappendix,revtex4]{emulateapj}   \usepackage{apjfonts}
\fi

\usepackage{natbib}
\usepackage{epsf}
\usepackage{color}
\usepackage{amsmath}
\usepackage{enumitem}
\usepackage{lscape} 
\usepackage{graphicx}
\usepackage[colorlinks=true,citecolor=blue,urlcolor=blue,linkcolor=blue]{hyperref}
\newcommand{\herschel}{\textit{Herschel}}
\newcommand{\spitzer}{\textit{Spitzer}}

\newcommand{\chandra}{\textit{Chandra}}

\newcommand{\lsim}{\lesssim}
\newcommand{\gsim}{\gtrsim}

\newcommand{\etal}{et al.}
\newcommand{\eg}{e.g.}

\newcommand{\msol}{\hbox{$M_\odot$}}

\newcommand{\snr}{\hbox{$\mathrm{SNR}$}}

\newcommand{\uJy}{\hbox{$\mu$Jy}}
\newcommand{\ujy}{\hbox{$\mu$Jy}}

\newcommand{\mone}{\hbox{$[3.6]$}}
\newcommand{\mtwo}{\hbox{$[4.5]$}}

\newcommand{\sdssu}{\hbox{$u$}}
\newcommand{\sdssg}{\hbox{$g$}}
\newcommand{\sdssr}{\hbox{$r$}}
\newcommand{\sdssi}{\hbox{$i$}}
\newcommand{\sdssz}{\hbox{$z$}}

\newcommand{\shela}{SHELA}
\newcommand{\texp}{\hbox{$t_\mathrm{exp}$}}
\newcommand{\myhref}[1]{\href{#1}{#1}}

\ifsubmode
\else

\fi
\newcommand{\ed}[1]{{{#1}}}
\newcommand{\edtoo}[1]{{#1}}

\shorttitle{Spitzer HETDEX Exploratory Large Area Survey}
\shortauthors{PAPOVICH ET AL.}

\begin{document}

\slugcomment{Accepted for Publication in the Astrophysical Journal Supplements}
\ifsubmode
\else
\title{The Spitzer-HETDEX Exploratory Large-Area Survey}
\fi

\author{\sc C.~Papovich\thanks{papovich@tamu.edu}\altaffilmark{1,2}, 
H.~V.~Shipley\altaffilmark{1,2,3}
N.~Mehrtens\altaffilmark{1,2}, 
C.~Lanham\altaffilmark{1}, 
M.~Lacy\altaffilmark{4}, 
R.~Ciardullo\altaffilmark{5,6},
S.~L.~Finkelstein\altaffilmark{7}, 
R.~Bassett\altaffilmark{8}
P.~Behroozi\altaffilmark{9},
G.~A.~Blanc\altaffilmark{10,11,12}, 
R.~S.~de Jong\altaffilmark{13},  
D.~L.~DePoy\altaffilmark{1,2},
N.~Drory\altaffilmark{7}, 
E.~Gawiser\altaffilmark{14},
K.~Gebhardt\altaffilmark{7}, 
C.~Gronwall\altaffilmark{5,6},
G.~J.~Hill\altaffilmark{7}, 
U.~Hopp\altaffilmark{15,16},  
S.~Jogee\altaffilmark{7},
L.~Kawinwanichakij\altaffilmark{1,2},
J.~L.~Marshall\altaffilmark{1,2},
E.~McLinden\altaffilmark{7}, 
E.~Mentuch~Cooper\altaffilmark{7}, 
R.~S.~Somerville\altaffilmark{14},
M.~Steinmetz\altaffilmark{13},  
K.-V.~Tran\altaffilmark{1,2},  
S.~Tuttle\altaffilmark{7}, 
M.~Viero\altaffilmark{17}, 
R.~Wechsler\altaffilmark{17},
G.~Zeimann\altaffilmark{5,6,7}
}
%
\affil{$^{1}$Department of Physics and Astronomy, Texas A\&M University, College
Station, TX, 77843-4242 USA; papovich@physics.tamu.edu}
\affil{$^{2}$George P.\ and Cynthia Woods Mitchell Institute for
  Fundamental Physics and Astronomy, Texas A\&M University, College
  Station, TX, 77843-4242 USA}
\affil{$^{3}$Department of Physics \& Astronomy, Tufts University, 574 Boston Avenue Suites 304, Medford, MA 02155, USA}
\affil{$^{4}$North American ALMA Science Center, NRAO Headquarters, Charlottesville, VA 22903, USA}
\affil{$^{5}$Department of Astronomy and Astrophysics, The
  Pennsylvania State University, University Park, PA 16802, USA}
\affil{$^{6}$Institute for Gravitation and the Cosmos, The
  Pennsylvania State University, University Park, PA 16802, USA}
\affil{$^{7}$Department of Astronomy, The University of Texas at Austin, Austin, TX 78712, USA}
\affil{$^{8}$International Centre for Radio Astronomy Research, University of Western Australia, 7 Fairway, Crawley, WA 6009, Australia}
\affil{$^{9}$Space Telescope Science Institute, 3700 San Martin Dr.,  Baltimore, MD 21218}
\affil{$^{10}$Departamento de Astronom\'ia, Universidad de Chile, Camino
  del Observatorio 1515, Las Condes, Santiago, Chile}
\affil{$^{11}$Centro de Astrof\'isica y Tecnolog\'ias Afines (CATA), Camino del Observatorio 1515, Las Condes, Santiago, Chile}
\affil{$^{12}$Visiting Astronomer, Observatories of the Carnegie
  Institution for Science, 813 Santa Barbara St, Pasadena, CA 91101,
  USA}
\affil{$^{13}$Leibniz-Institut f\"ur Astrophysik Potsdam (AIP), An der Sternwarte 16, D-14482 Potsdam, Germany}
\affil{$^{14}$Department of Physics and Astronomy, Rutgers, The State University of New Jersey, 136 Frelinghuysen Road, Piscataway, NJ 08854, USA}
\affil{$^{15}$Max-Planck-Institut f\"ur Extraterrestrische Physik,
  85741, Garching, Germany}
\affil{$^{16}$Univerist\"atssternwarte M\"unchen, LMU Munich, Scheiner Str. 1, D 81679 Munich, Germany}
\affil{$^{17}$Kavli Institute for Particle Astrophysics and Cosmology, Stanford University, 382 Via Pueblo Mall, Stanford, CA 94305, USA}



\begin{abstract}  

\noindent 
We present post--cryogenic \spitzer\ imaging at 3.6 and 4.5~\micron\
with the Infrared Array Camera (IRAC)  of the \spitzer/HETDEX
Exploratory Large-Area (SHELA) survey.  SHELA covers $\approx$24
deg$^2$ of the Sloan Digital Sky Survey ``Stripe 82'' region, and
falls within the footprints of the Hobby-Eberly Telescope Dark Energy
Experiment (HETDEX) and the Dark Energy Survey.  The HETDEX blind $R
\sim$~800 spectroscopy will produce $\sim$~200,000 redshifts from the
Lyman-$\alpha$ emission for galaxies in the range $1.9<z<3.5$, and an
additional $\sim$~200,000 redshifts from the [\ion{O}{2}] emission for
galaxies at $z<0.5$. When combined with deep $ugriz$ images from the
Dark Energy Camera, $K$-band images from NEWFIRM, and other ancillary
data,  the IRAC photometry from \spitzer\ will enable a broad range of
scientific studies of the relationship between structure formation,
galaxy stellar mass, halo mass, AGN, and environment over a  co-moving
volume of $\sim$0.5~Gpc$^3$ at $1.9<z<3.5$.  Here, we discuss the
properties of the SHELA IRAC dataset, including the data acquisition,
reduction, validation, and source catalogs.  Our tests show the images
and catalogs are 80\% (50\%) complete to limiting magnitudes of 22.0
(22.6) AB  mag in the detection image, which is constructed from the
weighted sum of the IRAC 3.6 and 4.5~\micron\ images.  The catalogs
reach limiting sensitivities of 1.1~\ujy\ at both 3.6 and
4.5~\micron\ (1$\sigma$, for $R=2\arcsec$ circular apertures).  As a
demonstration of science, we present IRAC number counts, examples of
highly temporally variable sources, and galaxy surface density
profiles of rich galaxy clusters.  In the spirit of \spitzer\
Exploratory programs we provide all images and catalogs as part of the
publication. 
\end{abstract}
 
\keywords{ catalogs --- galaxies: clusters: general --- infrared:
  galaxies --- surveys}


\section{INTRODUCTION}

\setcounter{footnote}{1}

The launch of the \spitzer\ Space Telescope \citep{werner04} allowed
for large surveys of galaxies at near-IR wavelengths, which are free
from foreground terrestrial thermal emission and are sensitive to the
rest-frame peak of the stellar emission in galaxies
($\lambda_\mathrm{rest} \sim 1.6$~\micron) over redshifts $z\sim 1-2$
\citep[\eg,][]{eise08,papo08,muzz13a}.    During the cryogenic mission,
\spitzer\ executed a variety of initial, wide-area surveys
\citep[\eg,][]{lons03,ashby09,wilson09}, and the post-cryogenic
(``warm'') mission enabled much larger surveys with increasingly 
larger combinations of depth and area \ed{
\citep{maud12,ashby13a,ashby13b,ashby15,labbe13,labbe15,timlin15,baro16}.}

\ifsubmode
\begin{figure}[t]
\epsscale{1.1}
\else
\begin{figure*}  
\epsscale{1.25} 
\fi 
\plotone{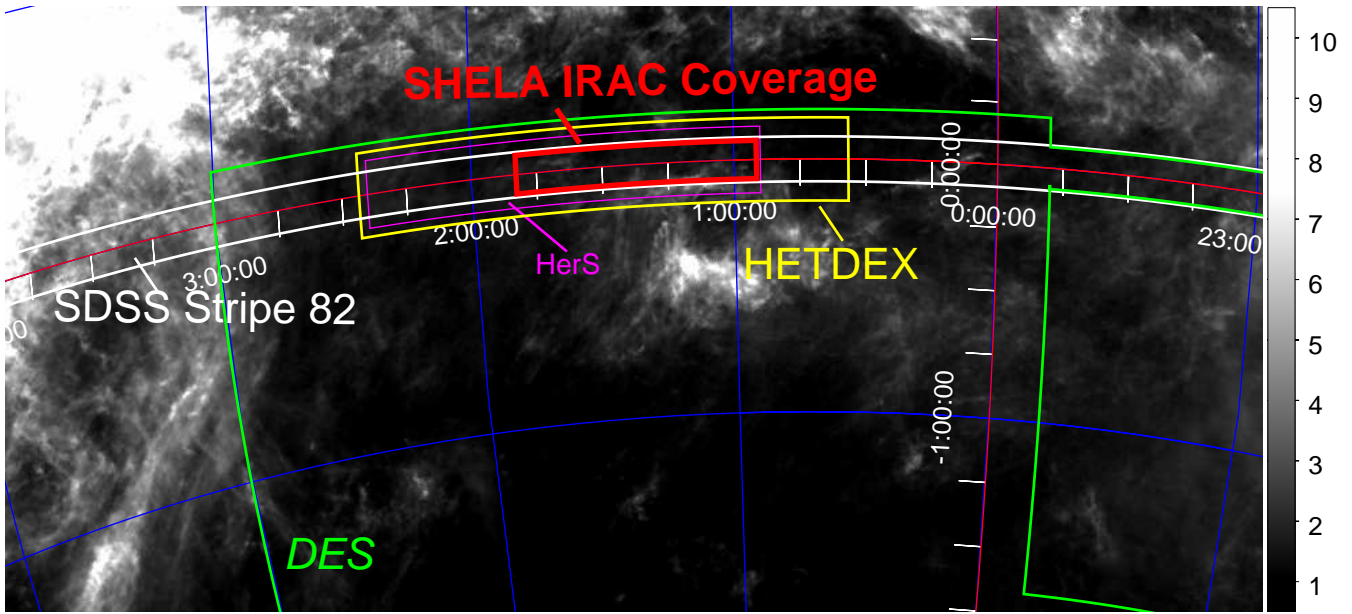}
\caption{Field layout for SHELA. The
lines show the coverage of our SHELA IRAC data (thick red-lined region),
the SDSS Stripe 82 coverage (white-lined region), the planned coverage
of the fall-field of the HETDEX survey (yellow-lined region), the HerS
\herschel\ coverage (magenta-lined region) and the
planned coverage of the DES (green-lined  region).   The lines of
constant R.A.\ and Decl.\ are labeled.   The background image
shows the IRAS 100~\micron\ map of a portion of the south Galactic
Pole \citep{schl98}.  The image intensity scales with surface
brightness as indicated in the color bar in units of MJy/sr.  }\label{fig:field}
\epsscale{1}
 \ifsubmode
\end{figure}
\else
\end{figure*}
\fi

  The size and depth of the near-IR imaging surveys carried out by
\spitzer\ have expanded our knowledge of how dark matter halos
accumulate baryons and convert them into stars.  The physics governing
this formation involves a range of complex processes \citep[see the
recent review by][and references therein]{some15}.  The processes for
the growth of galaxies include baryon and dark-matter accretion
histories, gas cooling, star formation,  and galaxy mergers, while the
processes that inhibit this growth include energetic winds from
massive stars and supernovae, radiation and kinematic feedback from
AGN, and shock heating of gas in large dark matter halos.
Distinguishing the importance of these different processes, and their
dependence on halo and stellar mass, redshift, and environment is one
of the  main goals of galaxy formation theory.

\spitzer\ has allowed us to test some of the theoretical physical
processes by comparing measurements of the galaxy stellar mass
distribution with model predictions.   These processes should
  manifest themselves as a function of galaxy
stellar mass, halo mass,  redshift, and environment.   By connecting
galaxies to their halos, we can identify and constrain the relative
importance of the physical processes responsible for galaxy growth at
different stages of their evolution.  For example, recent studies have
attempted to measure the ratio of galaxy stellar mass to halo mass
(SM--HM) as a function of halo mass
\ed{\citep[\eg,][]{moster10,leau12,moster13,behr13b,krav14}}.  The SM--HM relation
provides a powerful mechanism for connecting the predictions for the
halo mass function \citep[which is well understood,
e.g.,][]{spri05a,tinker08,behr10,behr13b} to the observed stellar mass
functions and mass--dependent spatial clustering of galaxies
\ed{\citep[\eg,][]{wein04,redd13,hear14,skibba15}. }

Expanding tests of galaxy formation derived from the SM--HM relation
requires observational measurements of galaxies over large areas to
measure both the bulk statistics and scatter in the halo- and
stellar-mass distributions.   This drives the need for larger extragalactic
surveys that cover areas containing the full range of environments 
in which galaxies form, and during the epochs when the physical
processes manifest.  


The post-cryogenic \spitzer\ mission has enabled such surveys.  Here,
we describe one such program, the \spitzer/HETDEX Exploratory Large
Area (SHELA) survey, which is designed to measure the evolution of the
nature of the SM--HM relation for galaxies over a large baseline in
redshift, $1.9 < z< 3.5$.  SHELA targets a $\approx$24 deg$^2$ field
in the Sloan Digital Sky Survey (SDSS) Stripe 82 field
\citep{annis14}, and covers a portion of the footprint of the
Hobby-Eberly Telescope (HET) Dark Energy eXperiment
\citep[HETDEX,][]{hill08}.   The SHELA field contains a large amount
of ground-based \ed{imaging,  including $griz$ data from the Dark
Energy Survey, additional $ugriz$ data from our own DECam imaging
program (Wold \etal\ 2016, in preparation), $K$-band data from the
NEWFIRM instrument (Stevans \etal\ 2016, in preparation), and data} in
the far-IR, sub-mm, and X-ray wavelengths
\ed{\citep{lama13a,lama13b,viero14}.}   The large SHELA field covers
nearly 0.5 Gpc$^3$ in cosmological volume at both moderate redshifts,
$0.5 < z < 2.0$, and at high redshifts, $2.0 < z < 3.5$, and opens the
distant Universe in the way that large-area, shallow surveys, such as
the Sloan Digital Sky Survey \citep[SDSS, ][]{sdssDR12} have expanded
our knowledge of the local Universe.  As an equatorial field, the
SHELA field is accessible to terrestrial telescopes in both
hemispheres, which gives it a high and lasting legacy value for
studies of galaxy evolution, AGN, and large-scale structure.  

\subsection{Overview of Paper}

Here we present the overview of the \spitzer/IRAC imaging dataset and
catalogs for SHELA.   The outline for this \textit{Paper} is as follows.  In \S~2,
we describe the SHELA survey field and the survey strategy with the
\spitzer\ space telescope.  In \S~3, we describe the data reduction
and mosaicking of the \spitzer\ dataset, and we describe astrometric
and photometric quality checks on the imaging data.  In \S~4, we
discuss the construction of the source catalog, and the catalog
properties, including source completeness.  We also discuss estimates
of photometric errors.  In \S~5, we discuss basic scientific results,
including source number counts, temporally varying objects, and the
galaxy surface density of rich clusters.   In \S~6, we summarize the
work.  

Throughout, we denote photometric magnitudes measured in the
IRAC channel 1 and channel 2 as \mone\ and \mtwo, respectively.
Unless stated otherwise, all magnitudes here are relative to the AB
system \citep{oke83}.  For convenience, we provide  conversions
between the AB system and the system relative to Vega,
$\mone_\mathrm{AB} - \mone_\mathrm{Vega} = 2.79$~mag and
$\mtwo_\mathrm{AB} - \mtwo_\mathrm{Vega} = 3.26$~mag, derived from a
comparison to the spectrum a A0V spectral type star.    Users of the
catalog may apply these to the flux densities in the catalog to
convert them to the magnitude system relative to Vega.
For any derived, physical quantity, we assume a cosmology with $\Omega_m=0.3$,
$\Omega_{\Lambda}=0.7$, and $H_0=70$ km s$^{-1}$ Mpc$^{-1}$,
consistent with the \textit{WMAP} seven-year data \citep{koma11} and
Planck 2013 data \citep{planck14}.   

\subsection{Overview of Data Products}

\ed{Included with this paper we release science--quality versions of
the reduced IRAC imaging and catalogs.  All data products are
available through the NASA/IPAC Infrared Science Archive
(IRSA\footnote{\myhref{http://irsa.ipac.caltech.edu/data/SPITZER/SHELA}\label{footnote:IRSA}}).  
A full description of the imaging is given in \S~\ref{section:mosaics}, and the
contents of our catalogs are detailed in \S~\ref{section:catalogs}.  Users may wish to skip
to those sections.  Here, we provide a high-level overview and
some recommendations for the use of these data.  }

\ed{The catalogs contain source flux densities and their associated
errors in units of $\ujy$, where the absolute bolometric magnitude is
then given by   $m_\mathrm{AB} = 23.9 - 2.5 \log(f_\nu/\mathrm{\uJy})$
(see \S~1.1).    The catalog contains four different flux density
measurements: one defined using a circular $4\arcsec$ diameter
aperture (extension \texttt{4ARCS}), one derived using a circular
$6\arcsec$ diameter aperture (extension \texttt{6ARCS}), one formed
using each objects' isophotal apertures (extension \texttt{ISO}), and
one created using an elliptical aperture defined from the objects'
light profiles using the \citet{kron80} definition (extension
\texttt{AUTO}).  Each of these flux estimates has an associated
uncertainty.  Note that we have used the IRAC point-response functions
(PRFs; see \S~\ref{section:prf} and \S~\ref{section:apcor}) to
correct our finite aperture (i.e., the \texttt{4ARCS} and
\texttt{6ARCS}) measurements  for light falling outside the defined
aperture; thus, these data should represent the total fluxes for point
sources.   In contrast, the isophotal  aperture (i.e., \texttt{ISO})
measurements have \textit{not} been corrected for missing light.  (By
definition, the flux densities measured in the Kron [\texttt{AUTO}]
aperture are ``total'' and require no correction.)}

\ed{The choice of aperture will depend on the exact user requirements
  of the application.  For faint point sources, we recommend using
  the flux densities measured in 4\arcsec-diameter apertures
  (\texttt{4ARCS}),  as    these data contain the fewest low
  signal-to-noise ratio (\snr) pixels, and hence 
  have the highest \snr\ overall.  For brighter objects, or sources more extended 
  than $\sim 2\arcsec$, the larger aperture measurements are more
  appropriate.   When the choice is uncertain, we recommend that users
  compare the $4\arcsec$-diameter aperture flux measurements with those derived using
  the other apertures and check for evidence of light loss. }

\section{Field and Survey Characterization}

\subsection{HETDEX}\label{section:hetdex}

\ed{HETDEX is a survey which will measure the redshifts of
$8 \times \sim 10^5$ Ly$\alpha$ emitting galaxies (LAEs) between 
$1.9 < z < 3.5$ using a suite of 78 wide-field integral field units (IFU)
spectrographs covering the wavelength region $350 - 550$~nm
\citep{hill08}.  The goal of these observations will be to provide
sub-percent level measurements of the Hubble expansion parameter
and the angular diameter distance at $z \sim 2$ via the large scale
distribution of galaxies in the redshift range of HETDEX.  The
result will be a significant constraint on the evolution of dark energy
that is competitive with (and independent of) values based on surveys
of the Ly$\alpha$ forest} \edtoo{\citep[e.g.,][]{slosar13,delu15}.}

\ed{The entire HETDEX survey will cover 420~deg$^2$ with a 1/4.5
  filling factor over two fields:  a $\sim 300$~deg$^2$ northern
  field, and a $\sim 140$~deg$^2$ equatorial region.  The $\approx
  24$~deg$^2$ SHELA field falls within the equatorial region, and, within
  its borders, HETDEX will increase its fill factor to unity (i.e.,
  every portion of the  SHELA field will be targeted for
  spectroscopy).  The $10 \sigma$ detection limit for these spectra
  will be $3.4 \times 10^{-17}$~ergs~cm$^{-2}$~s$^{-1}$ at 500~nm, or
  equivalently for continuum objects, $g_{AB} = 21.9$~mag.}

\ifsubmode
\begin{figure}[t]
\epsscale{1.1}
\else
\begin{figure}  
\epsscale{1.2} 
\fi 
\plotone{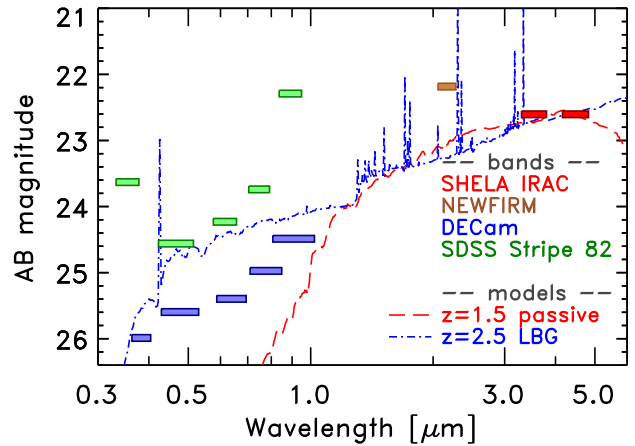}
\caption{\ed{Comparison of optical and IR magnitude limits for the SHELA
  field compared with fiducial model stellar populations.  The colored bars
  show the $3\sigma$ limits for the SHELA IRAC imaging derived in this
Paper, with preliminary values from our DECam and NEWFIRM imaging, as
well as the 50\% completeness limits for the SDSS Stripe 82 survey
data.   The curves show model stellar populations of a $z=2.5$
star-forming ``Lyman-break galaxy'' (LBG) with stellar mass $\log(M/M_\ast) = 10$ and moderate extinction, $E(B-V)=0.25$,  and a $z=1.5$
passively evolving galaxy with age $\log t/\mathrm{yr}$ = 9.3 and
stellar mass $\log(M/M_\ast) = 10.3$.} }\label{fig:fluxlimits}
\epsscale{1}
 \ifsubmode
\end{figure}
\else
\end{figure}
\fi

\ed{Figure~\ref{fig:field} shows the location of the equatorial HETDEX
field and the SHELA field.  The SHELA/IRAC imaging will detect
$>$200,000 galaxies at $1.9<z<3.5$ down to a limiting stellar mass of
$\simeq 2-3\times 10^{10}$~\msol. Figure~\ref{fig:fluxlimits} shows
how the SHELA IRAC $3\sigma$ flux limits compare to the spectrum of a
star-forming galaxy at $z=2.5$ (expected in both the LAE and non--LAE
populations), including the effects of nebular emission lines
\citep{salmon15},  and a spectrum of a passive galaxy  at $z=1.5$.
The figure also shows the preliminary 3$\sigma$ magnitude limits from
our DECam data (Wold et al.\ 2016, in preparation) and NEWFIRM data
(Stevans et al.\ 2016, in preparation) and the 50\% completeness
limits for the optical imaging in the SDSS Stripe 82 field
\citep{annis14}.   The combined depths of the optical ($ugriz$),
near-IR ($K_s$), and mid-IR (3.6--4.5~\micron) data will enable one to
measure stellar masses of galaxies to down to these flux limits.  This
enables the goal  the combined HETDEX--SHELA dataset, which is
measure the stellar masses $z \sim 2.5$  galaxies past the
characteristic mass, $M^\ast$, at these redshifts
\citep[\eg][]{muzz13c,tomc14}.  Moreover, the combined HETDEX-SHELA dataset
will enable the measurement of the relationship between halo mass
(constrained by the HETDEX density field) and stellar mass (derived
from the optical/IR photometry) over a survey volume large enough to
minimize statistical uncertainties.}
 
Results from the HETDEX pilot survey using a prototype of the HETDEX
IFU on the McDonald 2.7m illustrate the impact of joint spectroscopy
and stellar-population modeling of the LAE populations in small-area
fields where optical/near-IR imaging and \spitzer/IRAC imaging already
exist.  These include results published in \citet{adams11},
\citet{blanc11},  \citet{fink11}, \citet{hagen14}, \ed{and
\citet{chiang15}.} The results of the pilot survey also give us
confidence that we understand the properties of our LAE selection,
including their luminosity function and our ability to select LAEs for
HETDEX with little contamination.

\ifsubmode
\begin{deluxetable}{cccc}
\else
\begin{deluxetable}{cccc}
\fi
\tablecolumns{4}
\tablewidth{0pt}
\tablecaption{Observation Log for IRAC Observations \label{table:obs}}
\tablehead{
\colhead{Observing} & 
\colhead{Observing Dates} & 
\colhead{Position Angles} & 
\colhead{Number} \\ 
\colhead{Epoch} & 
\colhead{(UTC)} & 
\colhead{(deg.\ E of N)} & 
\colhead{of AORS} \\
\colhead{(1)} & 
\colhead{(2)} & 
\colhead{(3)} & 
\colhead{(4)}}
\startdata
1 & 2011-09-28 to 2011-10-10 & $-$110.0 to $-$105.5 & 64 \\
2 & 2012-02-13 to 2012-02-29 & \phs\phn63.1 to \phs\phn65.3 &  63 \\
3 & 2012-09-28 to 2012-10-09 & $-$110.9 to $-$107.4 & 64
\enddata
\ifsubmode
\end{deluxetable}
\else
\end{deluxetable}
\fi

\subsection{Field Location and Ancillary Data}\label{section:survey}

The SHELA field is centered at R.A. = 1$^h$22$^m$00$^s$, Decl. =
+00$^\circ$00$\arcmin$00$\arcsec$ (J2000), (Galactic coordinates, $l =
138.294^\circ$, $b=-62.017^\circ$) and extends approximately $\pm$6.5
deg in R.A. and $\pm$1.25 deg in Decl.  The field was chosen to have
low IR background \citep{schl98} within the SDSS Stripe 82 and DES
fields.  As illustrated in Figure~\ref{fig:field}, the 100~\micron\
background ranges from 1.2 to 1.7~MJy/sr across the field, and with a
mean value of approximately 1.5~MJy/sr.    

Because of its equatorial location,  the SHELA field lies near the
Ecliptic ($\lambda=18.93^\circ$, $\beta=-8.01^\circ$, and ranges in
latitude from $\beta=-4^\circ$ to $-11^\circ$).   Because the primary
component of the background for \spitzer/IRAC is the Zodiacal light,
this results in a higher background than higher (Ecliptic) latitude
fields.   The Ecliptic latitude for SHELA falls between the values
assumed for the ``medium'' and ``high'' background in the  \spitzer\
sensitivity performance estimation tool
(SENS-PET)\footnote{\myhref{http://ssc.spitzer.caltech.edu/warmmission/propkit/pet/senspet}}.
Therefore, it is expected that the SHELA field will suffer higher-than
average Zodiacal backgrounds, which adversely effects the flux
sensitivity of the IRAC data. 

An advantage of the equatorial location is that the SHELA field is
readily observable by current and future optical/IR and radio
telescopes.  The SHELA field is centered on the equator, and overlaps
with the DES optical imaging, and the optical imaging from the deeper
SDSS/Stripe 82 coadd \citep{annis14}.  These data are supplemented
with our own deeper CTIO/DECam $ugriz$ data, which reaches
\ed{$3\sigma$ limiting magnitudes of $u = 26.0$, $g = 25.6$, $r =
25.4$, $i = 25.0$, and $z = 24.5$}  (in $2\arcsec$-diameter
apertures).  In addition, the field is being imaged in the $K_s$ band
down to a $5\sigma$ depth of 22.8 mag using the NEWFIRM camera at Kitt
Peak (PI: S.\ Finkelstein).   \ed{The DECam and NEWFIRM limits are
illustrated in Figure~\ref{fig:fluxlimits}.}  The SHELA field also has 250,
350, and $500~\mu$m images from the SPIRE instrument taken as part of
the {\sl Herschel\/} Stripe 82 Survey \citep[HerS,][]{viero14}, and
X-ray coverage from \chandra\ and \textit{XMM-Newton}
\citep{lama13a,lama13b}.  Finally, the SHELA field has received
microwave observations at 148, 218, and 270~GHz from the Equatorial
Survey of the Atacama Cosmology Telescope \citep[ACT,][]{hass13}.  The
growing amount of multiwavelength data makes SHELA a unique resource
for the study of physical properties of evolution of galaxies as a
function of environment. 

Additional \spitzer/IRAC imaging of the SDSS Stripe 82 field exists
from the \spitzer\ IRAC Equatorial Survey \citep[SpIES, proposal ID
(PID) 90045, PI: G.~Richards;][]{timlin15}.  The SpIES data cover an
additional $\sim$115 deg$^2$ outside the SHELA footprint along SDSS
Stripe 82, with an effective IRAC integration time of 120~s. Scaling
by integration times, the SHELA data are approximately $2.5\log(
\sqrt{270/120}) = 0.44$~mag deeper than SpIES.   The reader is
referred to \citeauthor{timlin15}, for a description of SpIES and its
data products.

\ifsubmode
\begin{figure}[t]
\epsscale{0.7}
\else
\begin{figure}  
\epsscale{1.2} 
\fi 
\plotone{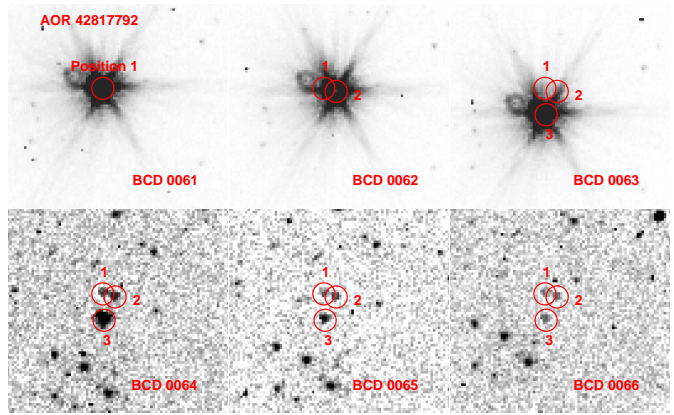}
\caption{Example of persistence in the IRAC 3.6~\micron\ data. The
data are from AOR 42817792 of SHELA.   Each
panel shows the same region of the IRAC detector in a time series of
sequential BCDs, from 0061--0066.  In the first three BCDs
(0061-0063), the bright star (HD 9670, $V=6.9$~mag) falls at position
1, 2, and 3 in this portion of the IRAC detector.  The bright star
leaves a fractionally small persistence effect ($<0.01$\% of the
fluence) in the same x,y pixels in subsequent BCDs (0064--0066), which
decays with an exponential timescale.}\label{fig:persist}
\epsscale{1}
 \ifsubmode
\end{figure}
\else
\end{figure}
\fi
 
\subsection{SHELA IRAC Survey Strategy}

\ed{IRAC \citep{fazio04} observes
simultaneously in channel 1 (at 3.6~\micron) and channel 2 (at
4.5~\micron), with each channel covering  a $5\farcm2 \times 5\farcm2$
field.}  The field centers observed by each channel are separated on
the sky by approximately 6.7 arcminutes, with a gap of about 1.52
arcminutes between the fields.   We designed the mapping strategy for
SHELA to obtain coverage in both channels over approximately the same
area of sky.

Several constraints affected the design of our survey.   We desired
multiple dithers with slightly different position angles to allow
redundancy, to identify cosmic rays, and to guard against image
defects.   We desired observations during several epochs to ease
\spitzer\ scheduling requirements.  The multiple epochs are separated
by long enough periods of time (approximately 4--7 months) to identify
time--variable objects including asteroids (as our observations
are close to the ecliptic we expect asteroids to be detected at higher
rates than higher--latitude extragalactic fields).  We also required
that all astronomical observation requests (AORs) be shorter than the
maximum observing time, about 6 hours for \spitzer. 

We divided the SHELA observations into three epochs, separated by
approximately six months.  There were two, 30-day duration observing
windows each year for \spitzer\ to observe \shela\ at position angles
optimal for our survey strategy.  

 During each epoch we observed the entire SHELA field to one-third of
the total depth, covering approximately $12 \times 2.5$~deg$^2$.  Each
AOR used a three point dither pattern, with 1 $\times$ 30 s frame time
per position (where the array observes with 23.6 s of exposure time for a 30
s frame).   Each AOR obtained a map divided into 8 rows by 10
columns of IRAC pointings, with a step size of 280$\arcsec$ between
each pointing.   The area covered by each AOR is approximately $38\arcmin \times
47\arcmin$, and each epoch tiled the entire SHELA field using 64 AORS
(epochs 1 and 3) or 63 AORs (epoch 2).   A single AOR required
approximately 2.75 hrs of clock time.  As there are 191 AORs, the
total clock time for SHELA required 525 hrs of \spitzer\ observations.
The \spitzer\ observations of SHELA occurred in the three epochs using
these AORS under program PID 80100 (PI: Papovich), with dates listed
in Table~\ref{table:obs}.   The table also gives the  position angles
of IRAC during the observations and the number of AORs observed during
each epoch. 

\section{IRAC Data}\label{section:data}

\ifsubmode
\begin{figure}[t]
\epsscale{1.0}
\else
\begin{figure*}[hp] 
\epsscale{1.05} 
\fi 
\begin{center}
\includegraphics[clip=true,width=1.3\textwidth,angle=90]{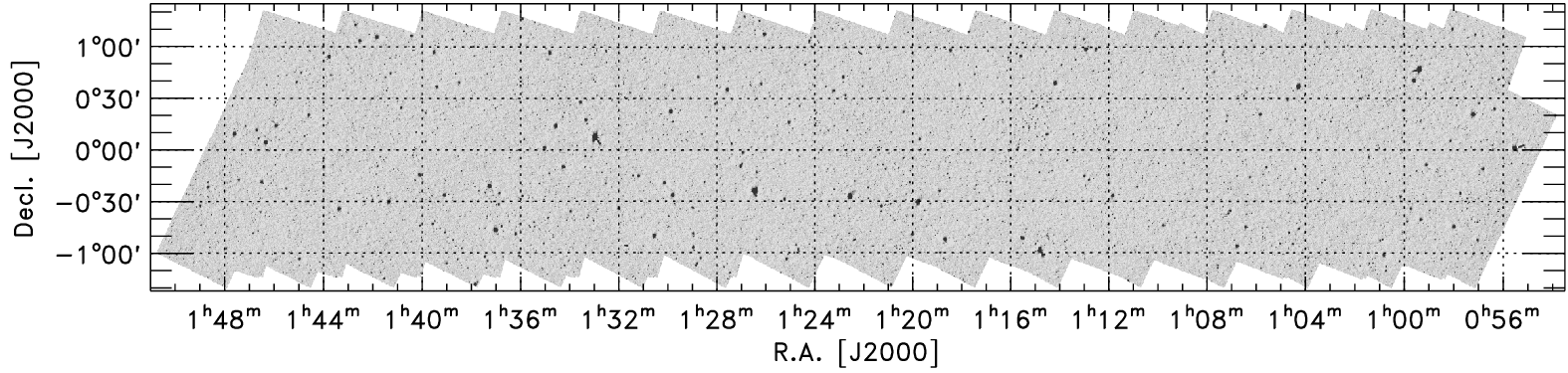}
\includegraphics[clip=true,width=1.3\textwidth,angle=90]{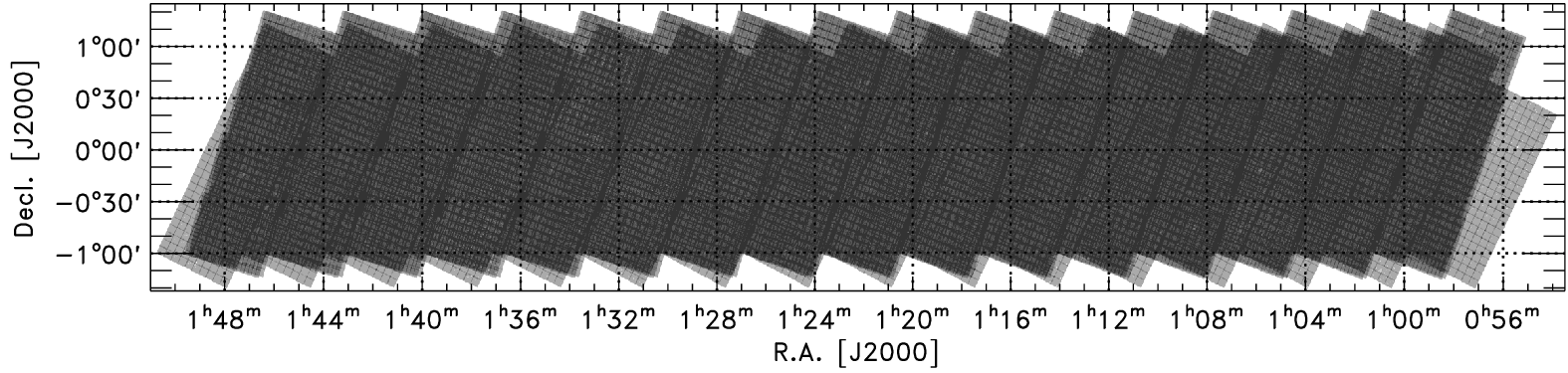}
\end{center}
\caption{Combined, three-epoch mosaic of the SHELA IRAC Channel 1
  (3.6~\micron) data (left) and associated exposure-time
  (weight) map (right). The extreme edges of the image cover nearly 2.5~deg
  $\times$ 13~deg, but the area covered to our 3-epoch depth is $\approx$24~deg$^2$. }\label{fig:mosaic3p6}
\epsscale{1}
 \ifsubmode
\end{figure}
\else
\end{figure*}
\fi
\ifsubmode
\begin{figure} 
\epsscale{1.0}
\else
\begin{figure*}[hp]  
\epsscale{1.05} 
\fi 
\begin{center}
\includegraphics[clip=true,width=1.3\textwidth,angle=90]{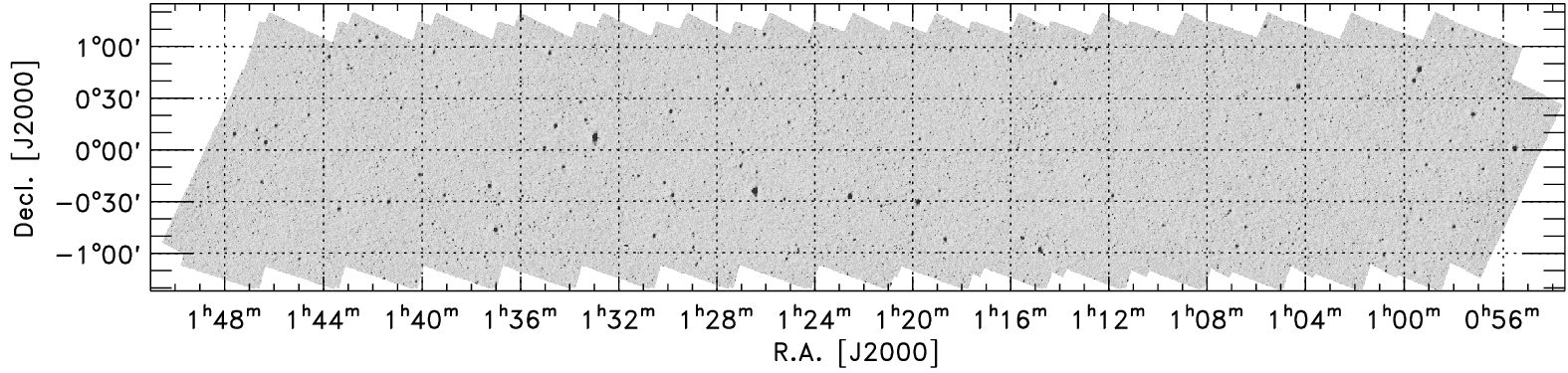}
\includegraphics[clip=true,width=1.3\textwidth,angle=90]{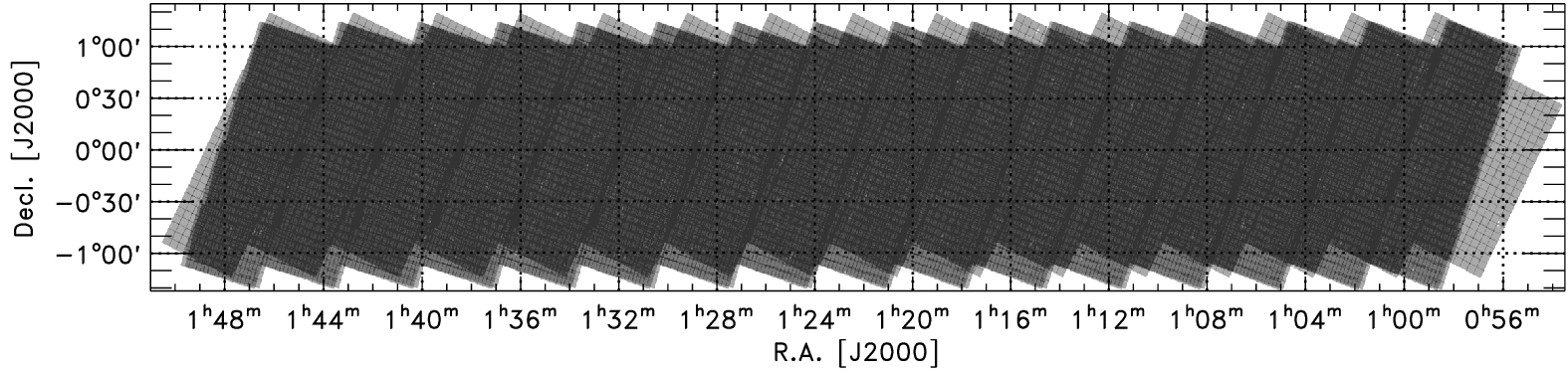}
\end{center}
\caption{Same as Figure~\ref{fig:mosaic3p6}, but for the SHELA IRAC
  Channel 2 (4.5~\micron) data (left) and the associated exposure-time
  (weight) map (right). }\label{fig:mosaic4p5}
\epsscale{1}
 \ifsubmode
\end{figure}
\else
\end{figure*}
\fi

\subsection{Data Reduction}

The SHELA IRAC data reduction began with the data pipeline processing
version S19.0.0  of the IRAC Corrected Basic Calibrated Data (cBCDs)
provided by the \spitzer\ Science Center (SSC).    The data processing
includes a subtraction of the dark current and application of the flat
field, as well as a photometric and astrometric calibration.
Starting with the cBCD products, we next applied several steps to
correct for variations and features in the image backgrounds.   We
corrected each image for column--pulldown effects associated with
bright sources using custom software
(``fixpulldown.pro'')\footnote{\myhref{http://irsa.ipac.caltech.edu/data/SPITZER/docs/dataanalysistools/tools/contributed}}.
We next constructed  a median frame from all cBCDs in a single AOR,
clipping outliers.  In this way we make a sky frame from all images in
an AOR.   We subtracted this image from each individual cBCD to
eliminate structure and residuals.  We then removed additional
striping in the backgrounds by averaging over five columns in each
image (clipping for objects), and subtracted this from each column.
We also excluded the first frame from each series of exposures in a
given AOR sequence to remove any variable instrument bias level
associated with idiosyncrasies of the post-cryogenic IRAC electronics
(the ``first frame effect''\footnote{See the IRAC Instrument Handbook
(2015, version 2.1), available at:
\myhref{http://irsa.ipac.caltech.edu/data/SPITZER/docs/irac/iracinstrumenthandbook}\label{footnote:iracinstrumenthandbook}}). 
%

Our inspection of preliminary reductions showed some instances where
persistence from bright stars produced spurious
sources in the final mosaics.   This occurs even though the
data-reduction pipeline flags for image persistence (and we set
persistence--flagged pixels as ``fatal'' during the mosaic stage, see
below), and suspect this occurs because the pipeline flags only
persistence from the brightest objects (and fainter objects, which
still cause persistence, are missed).      Figure~\ref{fig:persist} shows an
example of the persistence caused by a bright star  (HD 9670,
$V=6.9$~mag) in a consecutive series of cBCDs from one of the SHELA
AORs.    Following our observing strategy, the star is dithered to 3
different positions on the detector, before the IRAC array is stepped
to a new location on the sky.  The persistence from the star is
evident in several subsequent exposures.  The persistence fades with
an exponential timescale (as expected for trapped electron decay
rates) and is mainly a problem in the 3.6~\micron\
images (it is nearly negligible in the 4.5~\micron\
images, see footnote~\ref{footnote:iracinstrumenthandbook}).
%

%
To correct for the strongest persistence residuals, one of us (CL)
inspected visually each  channel 1 cBCD sequentially in the order they
were observed, identifying persistence events.  We then flagged those
pixels with persistence using  the locations of the bright objects in
the previous cBCDs in the observing sequence.   We combined these flag
maps with the mask files produced by the SSC pipeline and included the
masked pixels as fatal bits in the mosaicking steps.   Even so, we
have likely not accounted for all possible persistence in the images.
Persistence can manifest as ``sources'' that vary in the time domain
between observations in different epochs, and users of the catalogs
(especially for time-domain studies or sources detected in a single
channel of an observing epoch) should be wary that some time-variable
sources may be a result of faint persistence missed by our inspection
of the images.  

\subsection{Image Mosaics}\label{section:mosaics}

 We used a combination of the MOPEX software (v18.5.4) provided by the
SSC\footnote{\myhref{http://irsa.ipac.caltech.edu/data/SPITZER/docs/dataanalysistools/tools}}
and SWarp \citep[v2.19.1][]{bertin02} to produce mosaics of the IRAC
data.   Our choice to use SWarp is a result of the fact that the
memory limitations of MOPEX are too stringent for a dataset with the
size of the SHELA data volume.    We first produced a mosaic for each
AOR separately using MOPEX.   MOPEX includes full propagation of
errors for each pixel and masks pixels set to fatal bit patterns
(including pixels we estimate to contain persistence, see above). 

We next used SWarp to mosaic the output from MOPEX for each AOR into
images covering the full SHELA field.    We employed a background
subtraction with \texttt{BACK\_SIZE=128} and
\texttt{BACK\_FILTERSIZE=3} within SWarp to account for (small)
offsets in the backgrounds between AORs.   We combined AORs using a
weighted average (\texttt{COMBINE\_TYPE=WEIGHTED}) from the
exposure-time maps for each AOR, and we resampled the images to a
common field center and pixel scale of $0\farcs8$ pixel$^{-1}$.  We
produced full mosaics of all the data at 3.6~\micron\ and 4.5~\micron.
We also produced mosaics in each channel in each of the 3 observing
epochs separately.  Figures~\ref{fig:mosaic3p6} and
\ref{fig:mosaic4p5} show the combined, three-epoch mosaics at 3.6 and
4.5~\micron, respectively.  

\ifsubmode
\begin{figure} 
\epsscale{1.0}
\else
\begin{figure}  
\epsscale{1.05} 
\fi 
\plotone{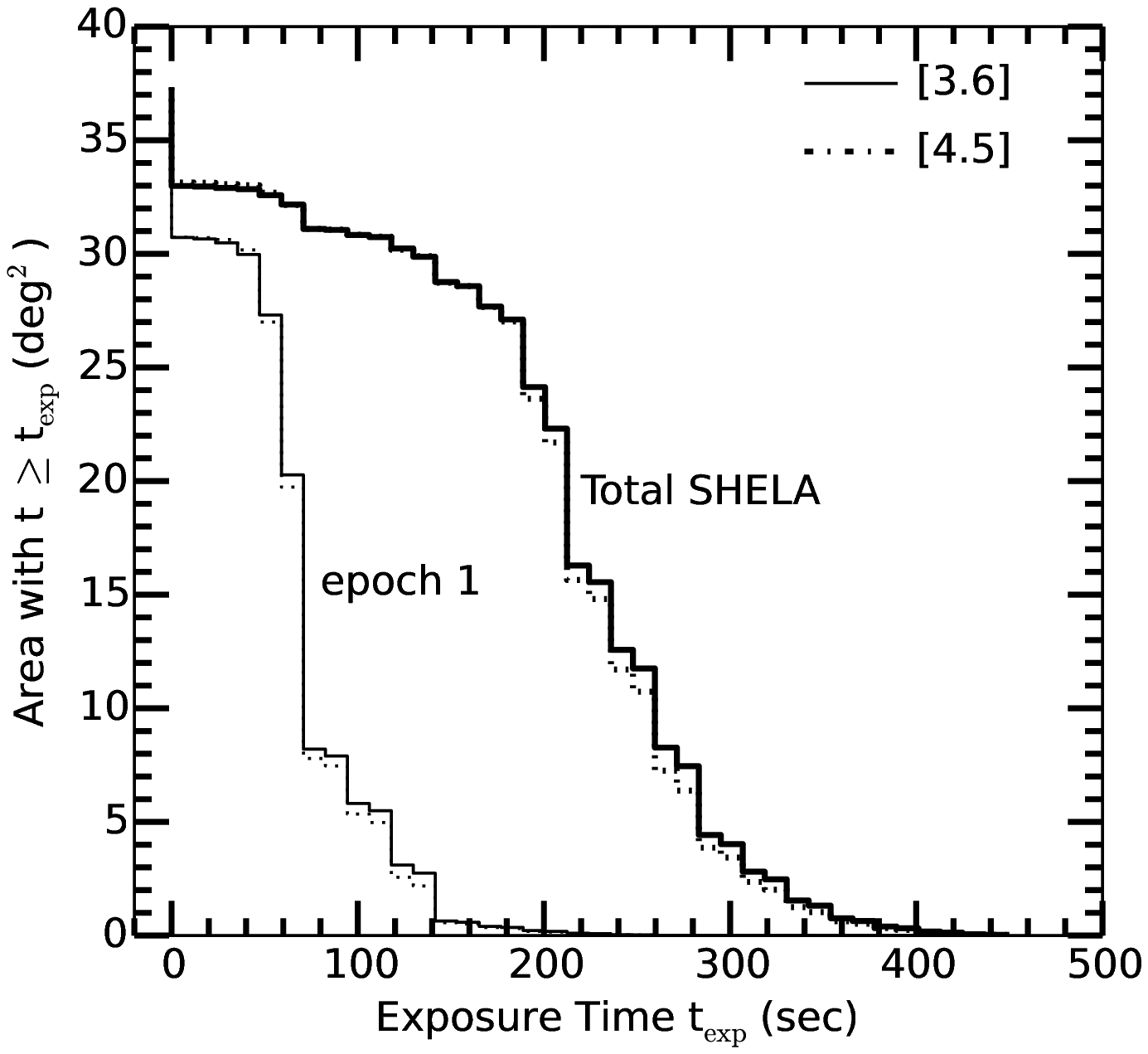}
\caption{The cumulative distribution of area with exposure times $\geq
\texp$.    The thin lines show the distribution only for the first
epoch IRAC data, and the thick lines show the distribution for the
combined three epoch data.  The solid lines show the 3.6~\micron\
distribution and the dotted line shows the 4.5~\micron\ distribution. }\label{fig:whtdist}
\epsscale{1}
 \ifsubmode
\end{figure}
\else
\end{figure}
\fi

We also combined the weight maps using SWarp.
Figures~\ref{fig:mosaic3p6} and \ref{fig:mosaic4p5} show the weight
map coverage for the full mosaic at 3.6 and 4.5~\micron, respectively.
The values in the weight map correspond to the number of IRAC
exposures for each pixel on the sky, and the weight map is therefore
proportional to the effective exposure time, $t_\mathrm{eff} = 23.6$ s
$\times$ $W$, where $W$ is the value of the weight map.
Figure~\ref{fig:whtdist} shows the distribution of area covered to a
given exposure time in the 3.6~\micron\ and 4.5~\micron\ full mosaics
compared to the coverage from epoch 1 only.   A single epoch covers
each region of a 26~deg$^2$ area with three pointings from the dither
pattern ($W=3$), for an effective exposure time,
$t_\mathrm{eff}\simeq$71~s.  \ed{The full mosaic covers an area of 30
deg$^2$ with $W=6$ pointings ($t_\mathrm{eff} = 142$~s depth),  24.2
deg$^2$ with $W=8.5$ pointings ($t_\mathrm{eff}$$>$200~s depth), and
22.4~deg$^2$ with $W>9$ pointings reaching the full survey depth
($t_\mathrm{eff}>212$~s depth).}

\subsection{Astrometric Quality}

In preliminary versions of the SHELA IRAC mosaics, we identified small
astrometric offsets between cBCDs from different AORs.  On subsequent
re-reductions, we corrected for these inter-AOR shifts using multiple
tests.  

We computed coarse astrometric offsets by cross-correlating the
positions of objects in each cBCD with sources detected in the SDSS DR7
\citep{abaz09}  catalog and updating the image headers.  The
astrometric offsets were mostly small, with shifts of up to
$\approx$$0\farcs2$ in both R.A. and Decl.   We then combined all cBCDs
from all mosaics from each epoch, and we again checked the absolute
astrometry of each mosaic, using the newer SDSS DR7
as a reference frame.    Finally, we corrected for the remaining (small)
relative shifts between each individual epoch.   Compared to SDSS DR7 the offsets
of the 3.6 micron images were: $\Delta\alpha = \alpha_\mathrm{SHELA} -
\alpha_\mathrm{DR7} = -140$, $+$180, $-$140~mas, and $\Delta\delta =
\delta_\mathrm{SHELA} - \delta_\mathrm{DR7} = +60$, $-$80, $+$50~mas,
for epochs 1, 2, and 3, respectively.   The offsets to the 4.5 micron
images were slightly different:  $\Delta\alpha = -134$, $+140$,
$-130$~mas; and $\Delta\delta = -140$, $+$110, $-$150~mas, for epochs
1, 2, and 3, respectively.  The origin of the offsets is unclear, but
may be related to the errors measured in the positions of the stars in the
\spitzer\ star trackers for the different spacecraft orientations
(where the errors may be a combination of uncertainties in the proper
motions in the guide star catalog, proper motions of stars in the
2MASS catalog used for the pointing refinement step of the IRAC pipeline,
and intrapixel sensitivity variations that add noise to the measured
star positions).  The orientation of epochs 1 and 3 were approximately
the same, while the spacecraft orientation for epoch 2 was different
by approximately 180 degrees, and indeed, the largest offsets were
between epoch 1 and 2 and epochs 2 and 3 (see above).    We corrected
for these astrometric offsets between each epoch before combining the
data into the final mosaics.  Our tests showed that correcting for
astrometric shifts for each epoch improved the image quality of point
sources in the final mosaic. 

\ifsubmode
\begin{figure} 
\else
\begin{figure*}  
\epsscale{0.85} 
\fi 
\plottwo{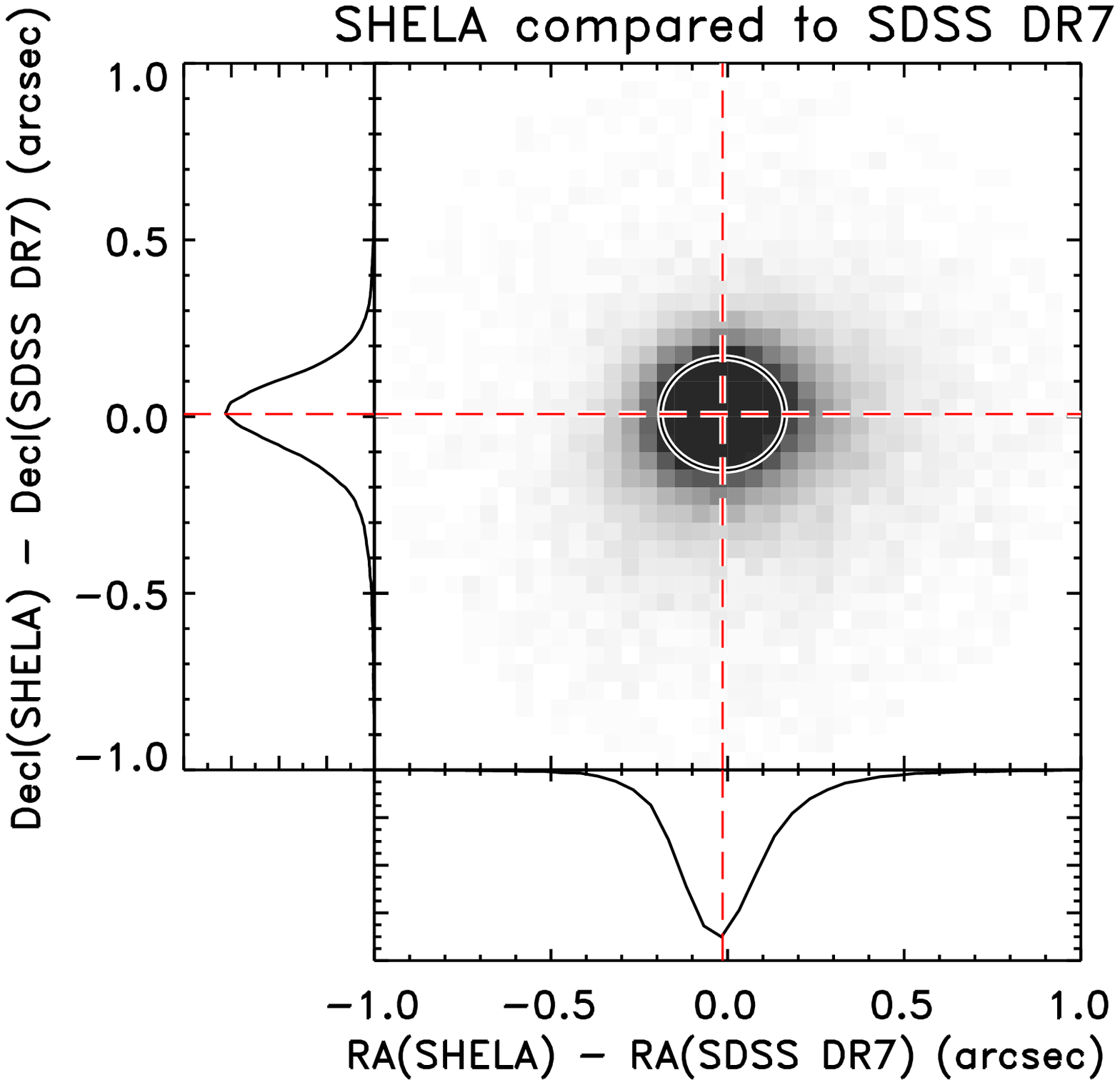}{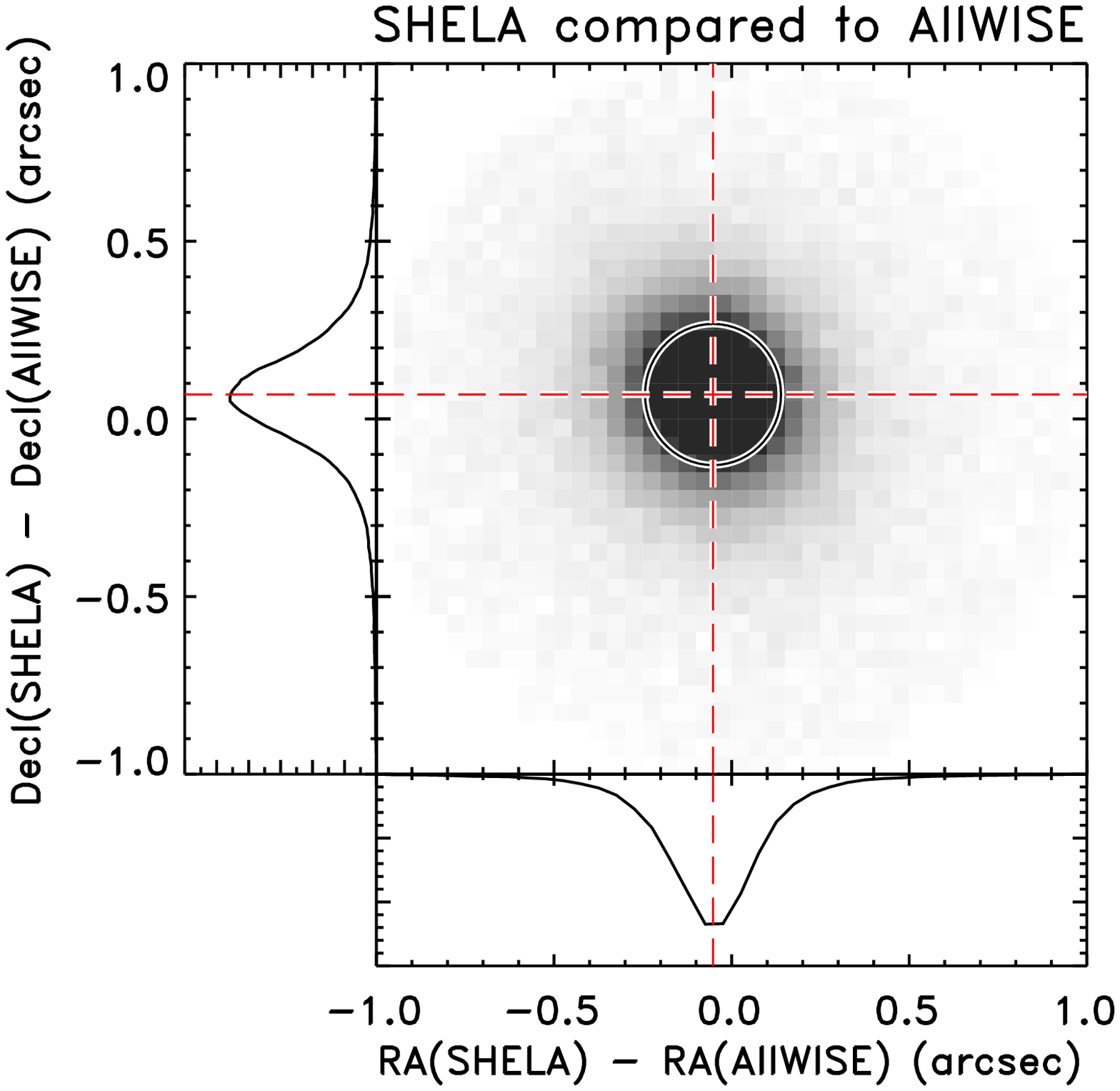}
\caption{Astrometric comparison for point sources between the
SHELA/IRAC and SDSS DR7 (left panel) and between SHELA/IRAC and
AllWISE (right panel).  \ed{In each main plot, the shading is
proportional to the density of  objects in each region of the plot.
The subpanels of each plot show the distributions of the differences
in R.A. and Decl., separately.}    The red, dashed lines show the mean
offset.  The ellipses contain 68\% of the sources.  There are
negligible offsets between SDSS DR7 and SHELA (by construction), and
the scatter is consistent with the SDSS astrometric accuracy of
$0\farcs2$ \citep{abaz09}.  In contrast, the AllWISE astrometry has
a non-negligible offset of $\approx$60--70~mas in both R.A. and Decl.,
with a scatter of 0.2\arcsec\ in each dimension.  }\label{fig:astro}
\epsscale{1}
 \ifsubmode
\end{figure}
\else
\end{figure*}
\fi


%

We remosaicked all the data using the improved astrometric
corrections.   As a result, the final astrometric solutions are very
good compared to SDSS DR7.    Figure~\ref{fig:astro} shows that
offsets between SHELA and DR7 are indeed very small, $\Delta\alpha =
\alpha_\mathrm{SHELA} - \alpha_\mathrm{DR7} = -14$ mas and
$\Delta\delta = \delta_\mathrm{SHELA} - \delta_\mathrm{DR7} = 7$~mas.
For comparison, the scatter is $\sigma(\alpha) = 180$~mas and
$\sigma(\delta) = 160$~mas in each dimension.  The
scatter is comparable to the quoted uncertainty in the SDSS DR7
astrometric solution \citep{abaz09}.   We also rechecked the
astrometry between SHELA and the newer SDSS DR9 coordinates, and
obtained	similar values, with systematic shifts of
$\Delta \alpha = -23$~mas and $\Delta \delta = 18$~mas, and an
overall scatter of $\sigma(\alpha) = 160$~mas and $\sigma(\delta) =
150$~mas. The slight increase in the offset between DR7 and DR9 is
well within the uncertainty in the absolute astrometric calibration of
SDSS \citep{ivezic07}.     The offsets are also small between SDSS DR7
and the IRAC images from each individual epoch.
%

We also compared the astrometry between SHELA and the 2MASS all-sky
point-source catalog \citep{skru06}.  There are very small shifts of
$\Delta \alpha = -8$~mas and $\Delta \delta = -30$~mas, with scatter
$\sigma(\alpha) = 270$~mas and $\sigma(\delta) = 260$~mas in each
dimension. This is larger than the typical positional uncertainty for
$K_s < 14$~mag sources \citep[$\lsim 100$~mas,][]{skru06}, but we have
made no correction for proper motion of stars, and the accuracy is
consistent with that reported in \citet{sand07}, who state an accuracy of
$\sim$200~mas for their \spitzer\ IRAC data.

There are larger shifts between the SHELA IRAC astrometry and the
astrometry of point sources in the AllWISE catalog \citep{cutri13}.
Figure~\ref{fig:astro} shows that the offsets are $\Delta\alpha =
-53$~mas and $\Delta\delta = 69$~mas, with scatter of $\sigma(\alpha)
= 190$~mas and $\sigma(\delta) = 200$~mas in each dimension.  The
scatter is consistent with the astrometric uncertainty of the AllWISE
catalogs \citep{cutri13}, but the larger offsets in the astrometry
(approaching a tenth of an arcsecond) may be non-negligible for some
applications.

\subsection{Point Response Functions}\label{section:prf}

\ifsubmode
\begin{figure} 
\else
\begin{figure}  
\epsscale{1.15} 
\fi 
\plotone{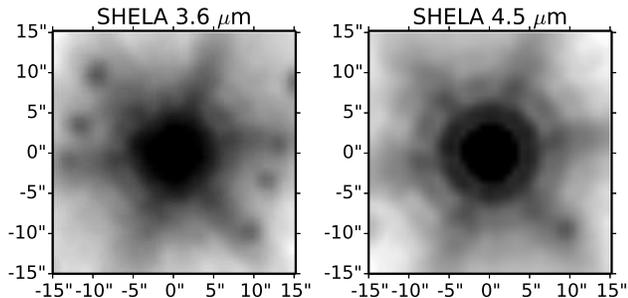}
\caption{Empirical PRFs for the SHELA IRAC 3.6 and 4.5~\micron\ data.
The PRFs are constructed by combining the IRAC fluxes for point
sources from the AllWISE catalog with magnitudes $14 <
W1< 15$ Vega mag.  }\label{fig:psf}
\epsscale{1}
 \ifsubmode
\end{figure}
\else
\end{figure}
\fi

For a variety of quality tests of the data and catalogs, it is useful
to have an empirical point response function (PRF) for the 3.6 and
4.5~\micron\ images.  In each SHELA image, we identified point sources from
the AllWISE catalog brighter than $W1 < 15$~Vega mag with flag value
\texttt{ex}$= 0$.    We kept objects only
in the magnitude range $14 < W1 < 15$ Vega mag as these have high
signal-to-noise, are well away from being saturated, and because
brighter objects are weighted more in the construction of the PRF.   We
constructed average PRFs using the routines provided in
IDLPhot\footnote{\myhref{http://idlastro.gsfc.nasa.gov/ftp/pro/idlphot}},
which is based on the DAOPhot software \citep{stet87}.  Figure~\ref{fig:psf}
shows the PRFs for the 3.6 and 4.5~\micron\ data.\footnote{\ed{FITS
   versions of these PRFs are available through IRSA, see
   \myhref{http://irsa.ipac.caltech.edu/data/SPITZER/SHELA/prfs}.}}

We use the
empirical PRF for tests of object photometric accuracy and
completeness in \S~\ref{section:completeness} and
\S~\ref{section:errors} below.  We measure a full-width at half
maximum (FWHM) of 1$\farcs$97 and 1$\farcs$99 from Gaussian fits to
the 3.6~\micron\ and 4.5~\micron\ PRFs, respectively.  These agree
with the expected values at the native IRAC pixel scale for the IRAC
channel 1 and 2 detectors during the warm mission.\footnote{Data taken
with IRAC during the warm \spitzer\ mission have measured FWHMs for
the PRFs $\approx$15\% larger than for data taken during the the cold
\spitzer\ mission, see the IRAC instrument handbook, link in
footnote~\ref{footnote:iracinstrumenthandbook}.}

We measured a curve-of-growth of the PRFs using circular apertures and
compared those to the flux measured with the fiducial IRAC aperture
(radius $R=12$\arcsec) used to derive the IRAC flux calibration (see
the IRAC Instrument Handbook, link in
footnote~\ref{footnote:iracinstrumenthandbook}).    The curve of
growth provides an estimate of the amount of light lost outside the
photometric aperture.  For large apertures ($R >$2\arcsec) these
corrections are identical to the ones we adopt below
(\S~\ref{section:apcor}), but they differ at the 0.05--0.10 mag level
for apertures  $R<1-2$\arcsec.   
%

\subsection{Photometric Aperture Corrections}\label{section:apcor}

\ifsubmode
\begin{figure}[t]
\epsscale{1}
\else
\begin{figure*}  
\epsscale{1.15} 
\fi 
\plotone{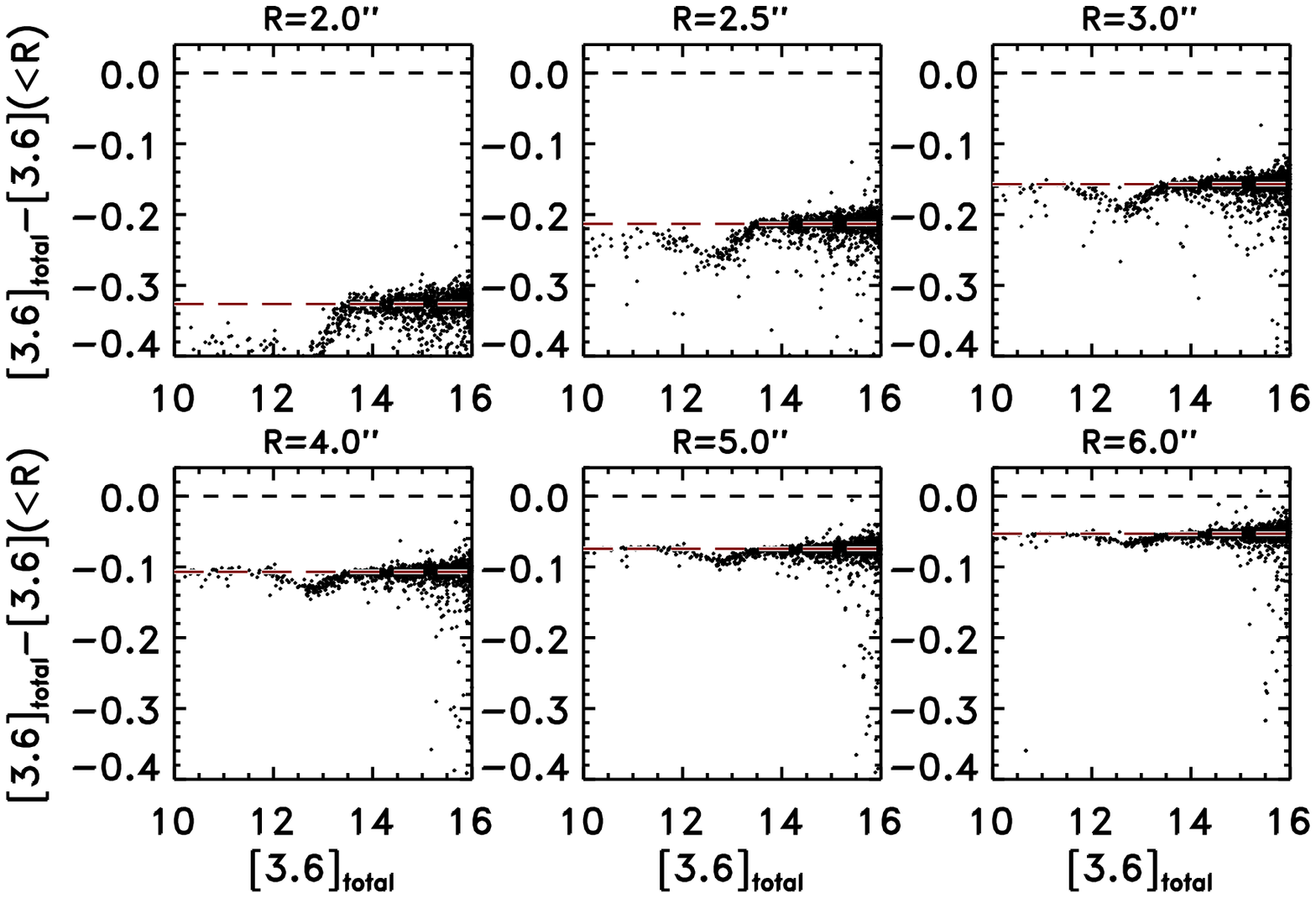}
\caption{Estimate of light lost outside circular apertures of varying
radius in the SHELA IRAC 3.6~\micron\ image.    Each panel shows the
difference between a total magnitude (defined in a
12\arcsec-radius aperture) and the magnitude measured in  a smaller circular aperture of radius
$R$ for point sources from the AllWISE catalog.  Each panel shows a
different aperture radius $R$, given above each panel.    The dashed,
thick red line in each panel shows the median difference for stars
between $13.5 < \mone < 16$~mag, used to derive the aperture
correction.  }\label{fig:magdiff1}
\epsscale{1}
 \ifsubmode
\end{figure}
\else
\end{figure*}
\fi

\ifsubmode
\begin{figure}[t]
\epsscale{1}
\else
\begin{figure*}  
\epsscale{1.15} 
\fi 
\plotone{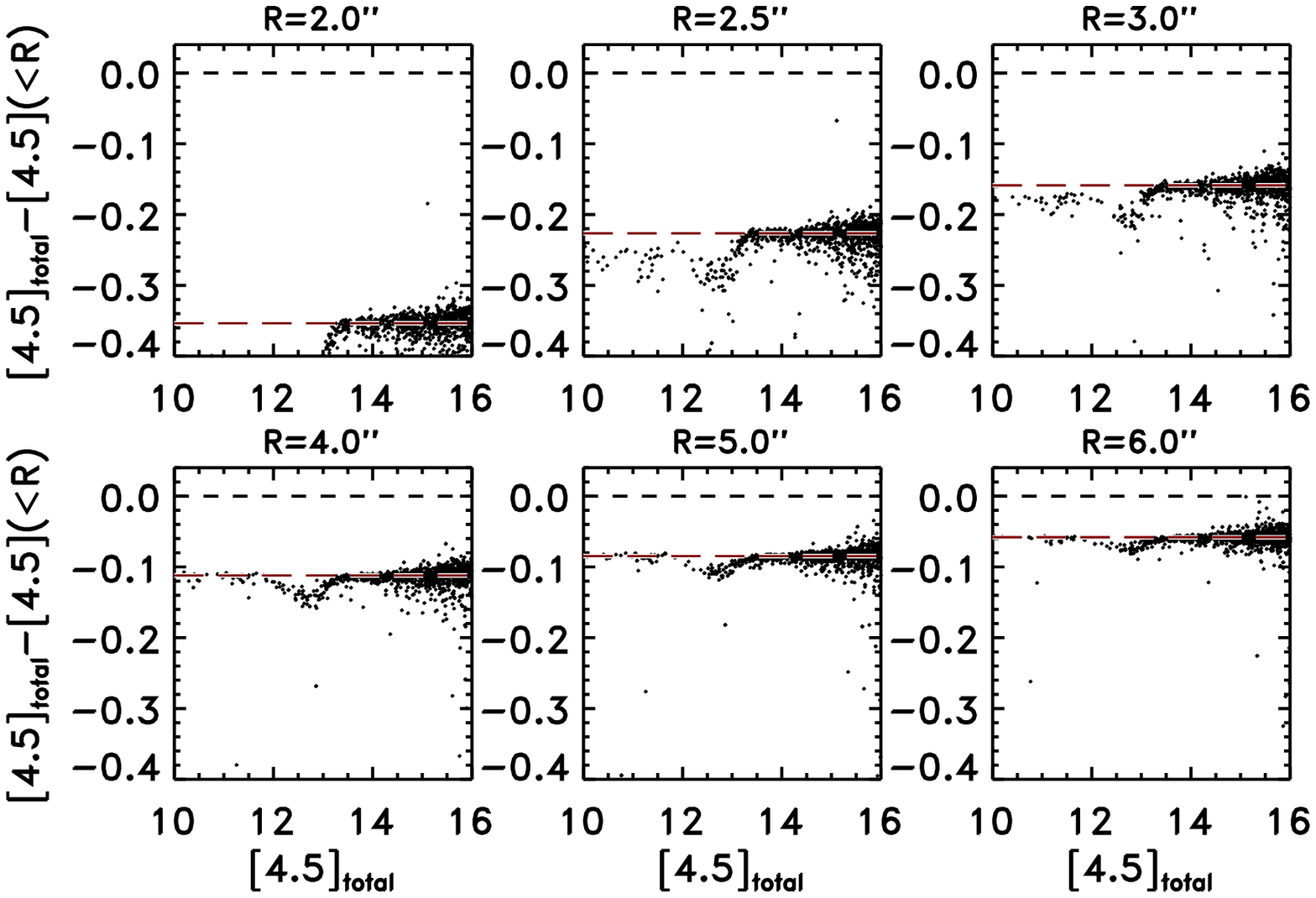}
\caption{Same as Figure~\ref{fig:magdiff1} for the SHELA IRAC
  4.5~\micron\ image.  }\label{fig:magdiff2}
\epsscale{1}
 \ifsubmode
\end{figure}
\else
\end{figure*}
\fi

Because we used the SExtractor software package for object photometry
in our SHELA catalog (\S~\ref{section:catalogs}, below), we found it
advantageous to derive aperture corrections for point-source
photometry from the images themselves using the same photometric
software package.   We used the same bright ($W1< 15$ Vega mag)
objects selected from the AllWISE catalogs used for the construction
of the PRF (see \S~\ref{section:prf}).     We then photometered those
objects in the IRAC images  using SExtractor with the same paramaters
as the source catalog (see \S~\ref{section:catalogs} and and
Table~\ref{table:sextractor}), using the AllWISE point sources as an
associated list with a search radius of 5 pixels (4\arcsec).
SExtractor photometered these sources with circular apertures ranging
in radius from 1\arcsec\ to 12\arcsec, where the 12\arcsec-radius
aperture ``defines'' the total aperture (see \S~\ref{section:prf},
above). 

Figures~\ref{fig:magdiff1} and \ref{fig:magdiff2} compare the total
$R=12\arcsec$ aperture magnitudes of the AllWISE stars in the IRAC 3.6
and 4.5~\micron\ frames to measurements performed in smaller
apertures.   As is clear from the figures,  there are offsets owing to
light lost outside the smaller apertures.   We measure aperture
corrections based on the median $m(<R) - m(<12\arcsec)$ magnitude for
stars with magnitudes between 13.5 and 16 (AB) mag.   (These median
offsets are denoted by the long-dashed red lines in the figures.)  At
brighter magnitudes, $\lsim$14~mag, the effects of saturation cause
the offsets to increase sharply, and become function of magnitude.
We caution against using small aperture magnitudes in this regime, as
they are unreliable.

\ifsubmode
\begin{figure}[t]
\epsscale{0.75}
\else
\begin{figure}  
\epsscale{1.15} 
\fi 
\plotone{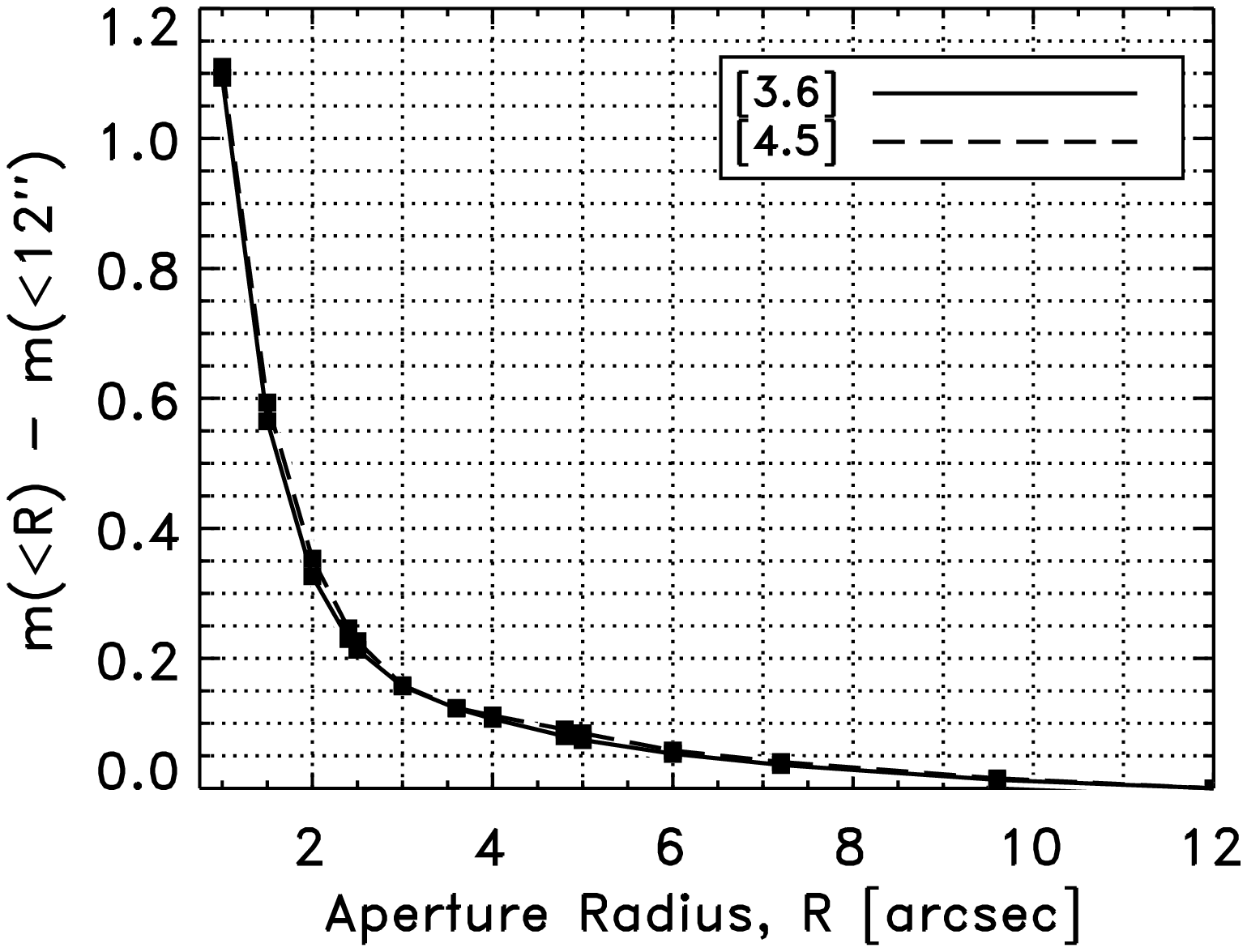}
\caption{Aperture corrections for the IRAC images.   The plot shows
the difference between the IRAC photometric magnitude measured for
point sources with magnitude between 13.5 and 16 mag in circular
apertures of radius $R$ and a total magnitude measured in an
$R=12\arcsec$ aperture.  The aperture corrections are measured in
apertures with discrete radii, as indicated by the solid black
squares, and interpolated linearly between those points.   The solid
(dashed) line shows the 3.6~\micron\ (4.5~\micron) data, as labeled in
the plot legend.  }\label{fig:apcor}
\epsscale{1}
 \ifsubmode
\end{figure}
\else
\end{figure}
\fi

\begin{deluxetable}{cccc}
\tablecolumns{4}
\tablewidth{0pt}
\tablecaption{Aperture Corrections for IRAC Data\label{table:apcor}}
\tablehead{
\colhead{$R$} & 
\colhead{$R$} & 
\colhead{\mone($<$$R$) $-$ \mone($<$12)} &
\colhead{\mtwo($<$$R$) $-$ \mtwo($<$12)} \\
\colhead{(pix)} & 
\colhead{(arcsec)} & 
\colhead{(mag)} &
\colhead{(mag)} \\
\colhead{(1)} & 
\colhead{(2)} & 
\colhead{(3)} & 
\colhead{(4)}}
\startdata
      2.500  &  2.0   &   0.326 &    0.353 \\
      3.125  &  2.5   &   0.213 &     0.226 \\
      3.750  &  3.0   &   0.157 &    0.159 \\
      5.000  &  4.0   &   0.107 &    0.112 \\
      6.250  &  5.0   &  0.074 &    0.085 \\
      7.500  &  6.0   &  0.053 & 0.058 
\enddata
\tablecomments{The aperture correction is the difference between the magnitude
  measured in a circular aperture of radius $R$ and the magnitude
  measured in a circular aperture of radius 12\arcsec.}
\end{deluxetable}
  
Figure~\ref{fig:apcor} shows the aperture corrections for the IRAC 3.6
and 4.5~\micron\ data measured for point sources with magnitude
between 13.5 and 16 mag in circular apertures, assuming an
$R=12\arcsec$ aperture encompasses the total light of a point
source. Table~\ref{table:apcor} lists the aperture corrections.   Our
measurements are consistent with those of the IRAC Instrument Handbook
(see footnote~\ref{footnote:iracinstrumenthandbook}) and those derived
in the literature \citep[\eg,][]{ashby09,ashby13b} with differences at
the $<0.05$~mag level.  These differences likely depend on the method
of photometry.  We advocate the use of the aperture corrections
derived here as they use the same photometric parameters as the source
catalog.   These corrections are accurate to better than 0.03 mag
based on our comparison of the IRAC photometry to flux measurements
from AllWISE at $W1$ (3.4~\micron) and $W2$ (4.6~\micron) in
\S~\ref{section:irac2wise}, below.

\subsection{Photometric Quality:\\Comparison between SHELA IRAC and AllWISE}\label{section:irac2wise}

\ifsubmode
\begin{figure}[tp]
\epsscale{0.7}
\else
\begin{figure*}  
\epsscale{1.15} 
\fi 
\plottwo{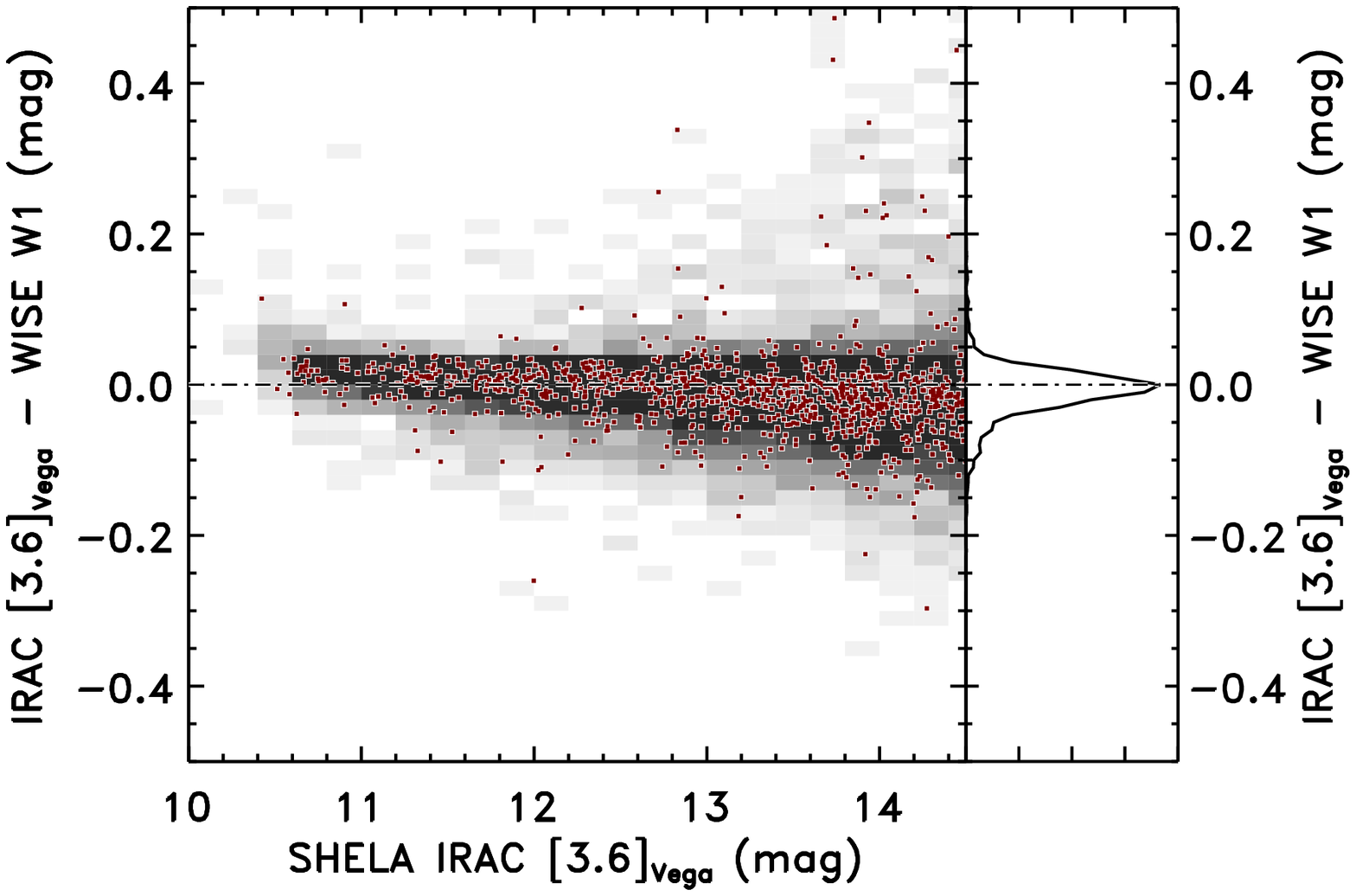}{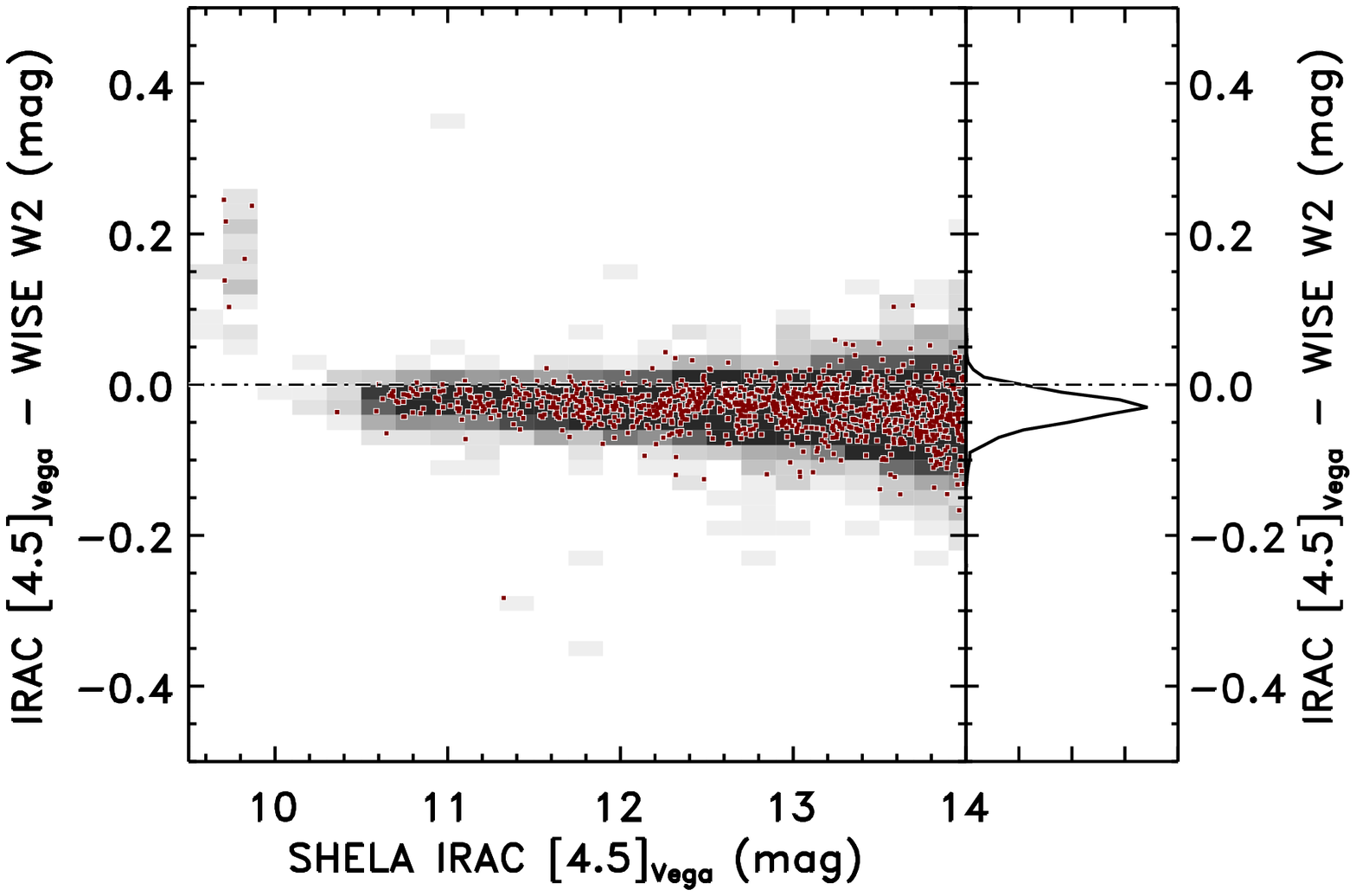}
\caption{The magnitudes measured for point sources in the
AllWISE catalog at 3.4~\micron\ ($W1$) and 4.6~\micron\ ($W2$)
compared with those in the SHELA IRAC 3.6 and 4.5~\micron\ data.
\textit{Note that in this plot all magnitudes are relative to Vega-type stars.}   In both
plots the gray shading shows all point sources, where darker regions
correspond to a higher density of points falling in that region.  The
red points show a random subsample of the data.   In each plot, the
right panel shows the distribution of the magnitude difference between
the IRAC and WISE image.   The left plot
compares the IRAC 3.6~\micron\ and WISE W1 data.  The
photometric offset is negligible.  The right plot compares the
IRAC 4.5~\micron\ and WISE W2 data.  There is
a small offset, $-0.028$ mag.     Stars brighter
than $\lsim$10 (11) Vega mag appear saturated in the IRAC 3.6
(4.5)~\micron\ data.}\label{fig:compWise}
\epsscale{1}
 \ifsubmode
\end{figure}
\else
\end{figure*}
\fi

\ifsubmode
\begin{figure}[t]
\epsscale{0.75}
\else
\begin{figure} 
\epsscale{1.25} 
\fi 
\plotone{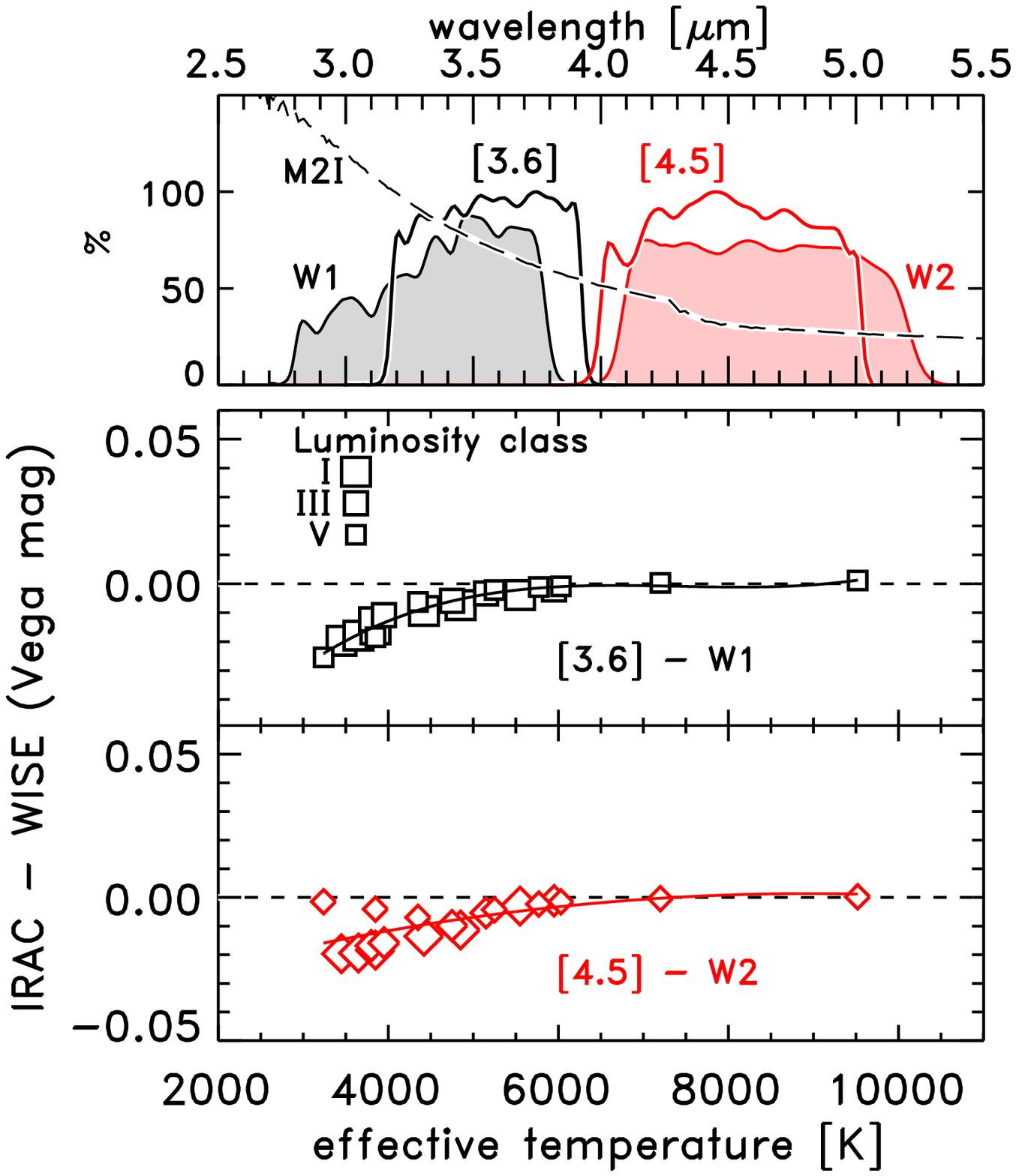}
\caption{Expected differences between the IRAC and WISE photometry due
to color variations in stars of different spectral types.  The top
panel compares the transmittance of the IRAC \mone\ and \mtwo\ filters
to the WISE W1 and W2 filters, \ed{as labeled.  The dashed line shows
the spectrum of an M2I stellar template \citep{kurucz93}.   The bottom
panels show synthesized colors between IRAC 3.6~\micron\ and WISE
$W1$, and between IRAC 4.5~\micron\ and WISE  $W2$, as labeled.}   The
data points correspond to \citet{kurucz93} models for dwarfs
(luminosity class V), giants (class III), and supergiants (class I)
over a range of effective temperature (spectral type).   Because the
WISE instrumental filters are wider, they can include bandhead
absorption features in late-type stars, affecting the IRAC--WISE color
up to 0.03 mag. }\label{fig:irac2wise}
\epsscale{1}
 \ifsubmode
\end{figure}
\else
\end{figure}
\fi

A test of the photometric accuracy of the SHELA IRAC data is possible
by comparing the IRAC photometry to that measured by WISE at
3.4~\micron\ ($W1$) and 4.6~\micron\ ($W2$).   For this test, we use
the same point sources selected from the AllWISE catalog matched to
the IRAC data discussed in \S~\ref{section:apcor}, above, \ed{while
  applying an additional color cut, $-0.1 \leq \mone_\mathrm{Vega} -
  \mtwo_\mathrm{Vega} \leq 0.0$ to minimize potential color terms in
  the stellar atmospheres\footnote{see 
\myhref{http://wise2.ipac.caltech.edu/docs/release/allwise/expsup/sec2\_3a.html}\label{footnote:allwise}}.}
Figure~\ref{fig:compWise} shows the magnitude difference between the
IRAC \mone\ and WISE $W1$ photometry and between IRAC \mtwo\ and WISE
$W2$ photometry.  The AllWISE $W1$ and $W2$ catalogs have $5\sigma$
sensitivity limits $\gsim$16 mag (Vega, Cutri et al.\ 2013) so our
comparison is for stars well above this limit and biases should be
minimal.        The difference between $\mone - W1$ is negligible.
The figure shows this distribution, and a Gaussian fit, which gives a
mean $\mu$=0.001 mag and standard deviation, $\sigma$=0.023 mag.

Figure~\ref{fig:compWise} shows the distribution of
$\mtwo - W2$, where a Gaussian fit gives a mean of $\mu = -0.028$ mag with
$\sigma = 0.021$~mag.   While this offset is small ($-0.028$ mag), its
origin is unclear.   The offset is $1.4\sigma$ times the scatter, and
corresponds to a flux ratio of 1.8\%.  \citet{jarr11} find similar
offsets and conclude these are consistent with the differences in the
IRAC and WISE bandpasses (the ``relative system response'' curves),
and the uncertainty in the calibration.  A similar offset is found by
\citet{cutri13} between $W2 - \mtwo$ (see footnote~\ref{footnote:allwise})
This offset persists regardless of the size of the
photometric aperture (once corrected to total), so a systematic offset
in the aperture correction seems unlikely.   It therefore seems to be
consistent with differences in the WISE $W2$ and IRAC \mtwo\ spectral
response curves.

We furthermore considered (and rejected) the possibility that there
are additional color terms between the IRAC and WISE photometry.
Figure~\ref{fig:irac2wise} shows the expected color between the IRAC
and WISE bands for different stellar types, using models from
\citet{kurucz93} over a range of luminosity class and spectral type.
The  IRAC--WISE colors are zero (relative to Vega) for early-type
(i.e., Vega-analogs) main-sequence stars.  This is expected as these
stars are used for the calibration of the instruments.   \ed{ However,
the IRAC and WISE filters have different central wavelengths and
filter widths.  Therefore, there will be color terms depending on the
source spectral energy distributions between the IRAC and WISE bands.
These color terms include color-dependent transformations from the
Vega to AB magnitude system.  Nominally, the WISE AB-to-Vega system
conversion constants are $W1_{\rm AB} - W1_{\rm Vega} = 2.699$ and
$W2_{\rm AB} - W2_{\rm Vega} = 3.338$ \citep{jarr11}.    When compared
to the conversions for IRAC, this implies an IRAC--WISE color offset of
$\sim 0.1$~mag for Vega-like stars on the AB system \citep[see also
][]{rich15}.}

%

\ed{Furthermore, because the WISE bands are broader in wavelength than
the IRAC bands, they are more sensitive to the molecular absorption in
later-type stellar atmospheres  (e.g., H$_2$O, HCN, C$_2$H$_2$, CO),
especially in the red giants, whose contribution to the SHELA field
star counts may be substantial.  As illustrated in
Figure~\ref{fig:irac2wise} late-type supergiants should have a
IRAC--Wise photometric offset of of $\mtwo - W2 \approx -0.02$~mag, a
value consistent with that observed in the data and that given in the
literature \citep[e.g.,][]{jarr11}.  The offset between $\mone - W1$
is similarly small, though the fact that the bandpass contains of very
different set of absorption features --- H$_2$O, C$_2$H$_2$+HCN bands
at $3.8~\mu$m as opposed to CO bands at $5.0~\mu$m
\citep[see][]{mats05,mats14} --- calls this interpretation into
question.  Therefore, while the color terms likely can explain the
\textit{scatter} in the IRAC--AllWISE colors
(Figure~\ref{fig:compWise}), they likely are not the cause of the
systematic offset between the IRAC 4.5~\micron\ and AllWISE
4.6~\micron\ photometry discussed above}.

Regardless of its origin, the offset is small, and is within the
uncertainty of the absolute IRAC calibration \citep{reach05}.  \citet{jarr11} argue
the offset likely results from a combination of absolute calibrations,
aperture corrections, and/or color corrections.  The photometry is
sufficiently accurate for most science applications, although those
requiring better than 2\% absolute photometric accuracy should be
aware of this systematic. 

\section{SHELA IRAC Catalogs}\label{section:catalogs}

\begin{deluxetable}{lc}
\tablecolumns{2}
\ifsubmode
\else
\tablewidth{3in}
\fi
\tablecaption{SHELA SExtractor Parameter Settings\label{table:sextractor}}
\tablehead{\colhead{SExtractor Parameter} & \colhead{Value} \\
\colhead{(1)} & \colhead{(2)} 
}
\startdata
\texttt{DETECT\_MINAREA} & 3~pixels \\
\texttt{DETECT\_THRESH} & 1.5 \\
\texttt{ANALYSIS\_THRESH} & 1.5 \\
\texttt{FILTER\_NAME} &  gauss\_2.0\_5x5\tablenotemark{a} \\
\texttt{WEIGHT\_TYPE} & MAP\_WEIGHT \\
\texttt{DEBLEND\_NTHRESH} & 64 \\
\texttt{DEBLEND\_MINCONT} & 0.0005 \\
\texttt{MAG\_ZEROPOINT} & 20.9555\tablenotemark{b} \\
\texttt{PIXEL\_SCALE} & 0.80~arcsec\\ 
\texttt{BACK\_TYPE} & AUTO \\
\texttt{BACK\_SIZE} & 256~pixels \\
\texttt{SEEING\_FWHM} & 1.7~arcsec
\enddata
\tablecomments{SExtractor was run using the weighted sum of the 3.6 and
  4.5~\micron\ images for detection, and using the images separately
  for photometry. All SExtractor parameters are identical for both
  images.   All other SExtractor parameters are
  set to the program defaults (for SExtractor v.2.19.5).}
\tablenotetext{a}{This is a Gaussian kernel with $\sigma$=2 pixels and
  size $5 \times 5$ pixel$^2$ used to filter the image for
  source detection.} 
\tablenotetext{b}{The AB magnitude zeropoint for the images,
  converting from the \spitzer\ default of MJy sr$^{-1}$ to $\mu$Jy
  pixel$^{-1}$ at the $0\farcs8$ pixel$^{-1}$ scale.} 
\end{deluxetable}

We used Source Extractor \citep[SExtractor v.\ 2.19.5;][]{bertin96} to detect and to
photometer sources in the IRAC images.    To detect sources, we
constructed a detection image as the weighted sum of the 3.6 and
4.5~\micron\ images.   The detection image, $D$,  is then,
\begin{equation}\label{eqn:detect}
D = \frac{W_1 \times I_1 + W_2 \times I_2}{W_1 + W_2}
\end{equation} where $W_1$ and $W_2$ are the weight maps (proportional
to the exposure time) for the IRAC channel 1 and 2 images,
respectively, and $I_1$ and $I_2$ are the science (flux) images for
channel 1 and 2, respectively.    We then ran SExtractor in ``double
image mode'' using the detection image and science images with the
parameters listed in Table~\ref{table:sextractor}.
%
%
Tables~\ref{table:preamblecatalog} and \ref{table:fullcatalog} provide
all the information from the full-mosaic catalogs.

We also constructed catalogs for the mosaics from each of the three
observing epochs individually.  For each epoch, we used the detection
for the combined images (see eq.~\ref{eqn:detect} above).  In this
way, sources detected in the combined epoch, 3.6+4.5~\micron\ image
are photometered in each image from each epoch.  We used the identical
SExtractor parameters as for the full-mosaic catalogs
(Table~\ref{table:sextractor}).  Tables
\ref{table:ep1catalog}--\ref{table:ep3catalog} provide the
photometry from the individual epoch data.

\ifsubmode
\begin{deluxetable}{lcccc}
\else
\begin{deluxetable}{lcccc}
\fi
\tablecolumns{5}
\tablewidth{0pt}
\tablecaption{Completeness and Error Estimates for \shela\ IRAC data\label{table:completeness}}
\tablehead{\colhead{AB mag} & \colhead{Raw Completeness} &
  \colhead{Completeness} & \colhead{$\sigma_{3.6}$} &
  \colhead{$\sigma_{4.5}$} \\
\colhead{(1)} & 
\colhead{(2)} & 
\colhead{(3)} & 
\colhead{(4)} & 
\colhead{(5)}
}
\startdata
18.0 & 0.98 & 0.98 & 0.02 & 0.02 \\
18.5 & 0.97 & 0.97 & 0.02 & 0.02 \\
19.0 & 0.98 & 0.98 & 0.02 & 0.02 \\
19.5 & 0.97 & 0.97 & 0.03 & 0.03 \\
20.0 & 0.95 & 0.95 & 0.05 & 0.05 \\
20.5 & 0.94 & 0.94 & 0.07 & 0.07 \\
21.0 & 0.90 & 0.90 & 0.11 & 0.12 \\
21.5 & 0.86 & 0.86 & 0.17 & 0.17 \\
22.0 & 0.79 & 0.79 & 0.25 & 0.25 \\
22.5 & 0.57 & 0.56 & 0.37 & 0.38 \\
23.0 & 0.20 & 0.19 & 0.49 & 0.51 \\
23.5 & 0.05 & 0.04 & 0.76 & 0.77 \\
24.0 & 0.01 & 0.00 & 0.90 & 0.87
\enddata \tablecomments{(1) Magnitude bin, (2) ratio of the number of
recovered fake sources to the total number of fake sources in this
magnitude bin, (3) completeness corrected for ``false positives'',
fake sources recovered in the detection image even when no fake
sources were added, (4) estimate of the photometric uncertainty for
point sources in \mone, (5) estimate of the photometric uncertainty
for point sources in \mtwo.  The photometric uncertainty estimates are
the standard deviation between the input magnitudes and the measured
magnitudes (measured in $2\arcsec$--radii [i.e., $4\arcsec$--diameter] apertures, corrected to
total using the values in Table~\ref{table:apcor}).}
\end{deluxetable}

\subsection{Completeness Simulations}\label{section:completeness}

\ifsubmode
\begin{figure*} 
\epsscale{0.5}
\else
\begin{figure}  
\epsscale{1.15} 
\fi 
\plotone{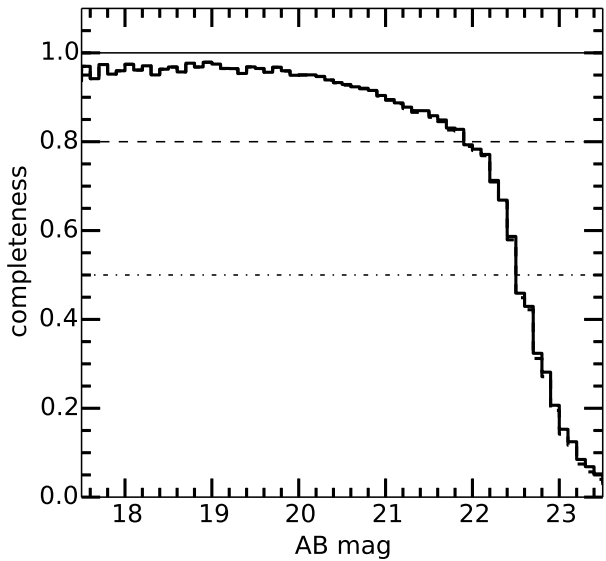}
\caption{Completeness for point-sources in the SHELA data as a
function of the input source magnitude.   The plot shows the recovery
fraction as a function of magnitude for simulated point sources, which
are added to both the 3.6 and 4.5~\micron\ images with the same AB
magnitude. The solid--line histogram shows the raw completeness fraction.  The
dashed--line histogram shows the completeness corrected for ``false positives''
(sources that are ``recovered'' in the image, in which no simulated
sources are added), which shows only slight differences with respect
to the raw completeness fraction. The solid, dashed, and dot-dashed horizontal lines
show 100\%, 80\%, and  50\%
completeness.  }\label{fig:completeness}
\epsscale{1}
 \ifsubmode
\end{figure*}
\else
\end{figure}
\fi

We performed simulations to estimate the completeness in the \shela\
IRAC catalogs following the method in \citet{papo15}.   We inserted
fake point sources into the 3.6 and 4.5 images using the empirical
PRFs derived above (\S~\ref{section:prf}).  We inserted each fake source at the same
($\alpha$, $\delta$) location in the 3.6 and 4.5~\micron\ images,
where the source has the same total brightness (AB magnitude) in each
channel.   Fake sources were assigned magnitudes chosen randomly from a wide
distribution (17--24th magnitude), and the sources are located
randomly in the images.  In this way fake sources may fall within the
isophotes of real objects in the image, and therefore our completeness
simulations include the effects from blended objects.  We reconstructed
the detection image as the weighted sum of the 3.6 and 4.5~\micron\
images and reran SExtractor.   This latter step was computationally
expensive given the size of the images (see above; Equation~1).  We
repeated the simulation only 15 times, where we inserted
into each simulated image 10,000 fake sources ($\approx$0.4\% the
total number of real sources).  In this way we sampled the full range
of source magnitude using a minimum investment of resources.  

We computed the completeness as the ratio of the number of recovered
(detected) fake sources to the number of input fake sources in bins of
source magnitude.  Figure~\ref{fig:completeness} shows the completeness,
where the 50\% (80\%) completeness limit is 22.6 (22.0) AB mag.
Table~\ref{table:completeness} gives these as the ``raw'' completeness
as a function of source magnitude in the detection image.   We also
added to the completeness a correction for  ``false positives'',
sources at the location of the fake that are ``recovered'' even when
no sources are added to the image.   Table~\ref{table:completeness}
gives these as the ``completeness''.   As illustrated in
Figure~\ref{fig:compFlux} the difference between the raw completeness
and the completeness corrected for false positives is small,
accounting for only 1\% of  recovered sources down to the 50\%
completeness limit. 

\ifsubmode
\begin{figure} 
\epsscale{1.0}
\else
\begin{figure*}  
\epsscale{1.12} 
\fi 
\plottwo{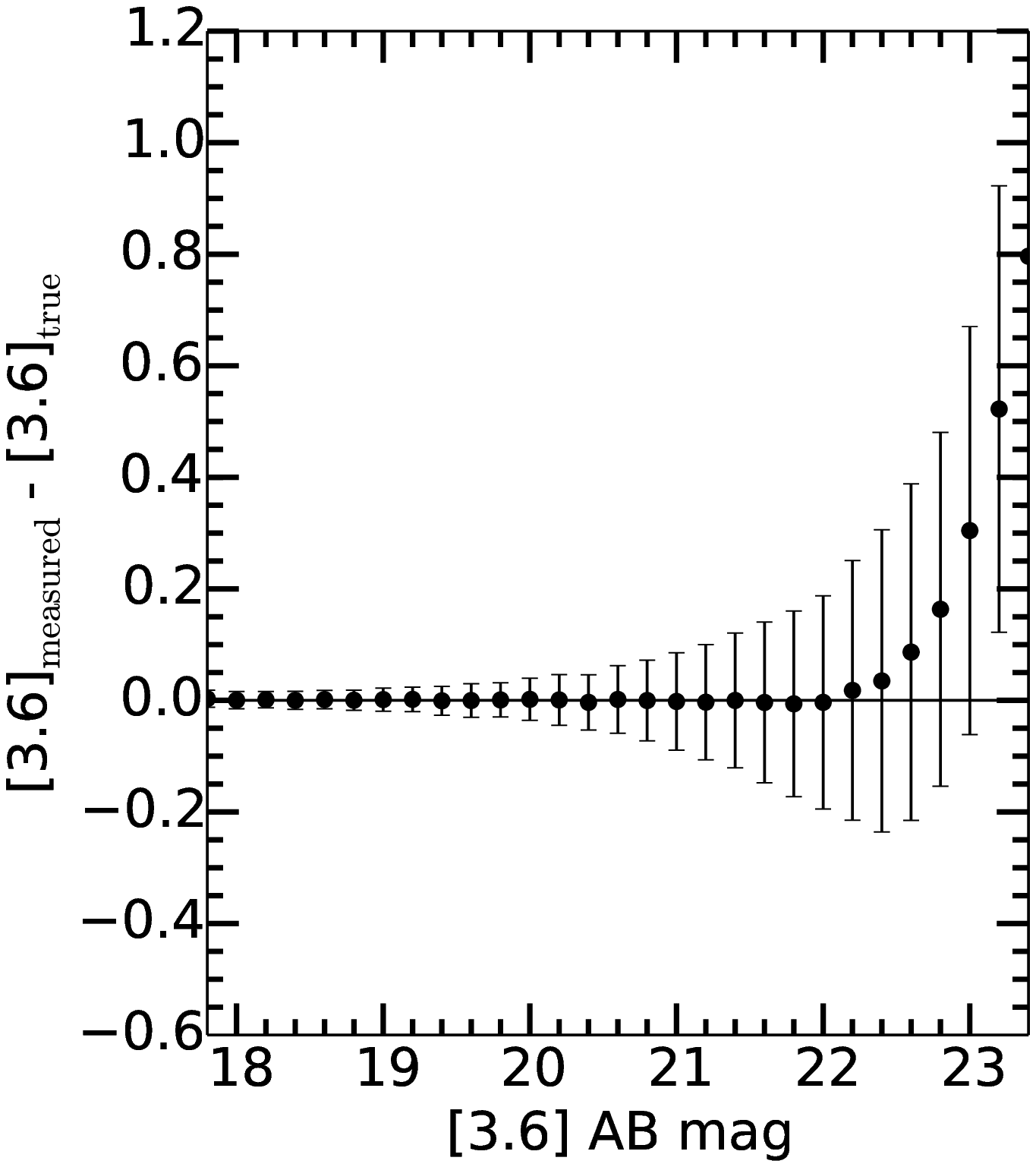}{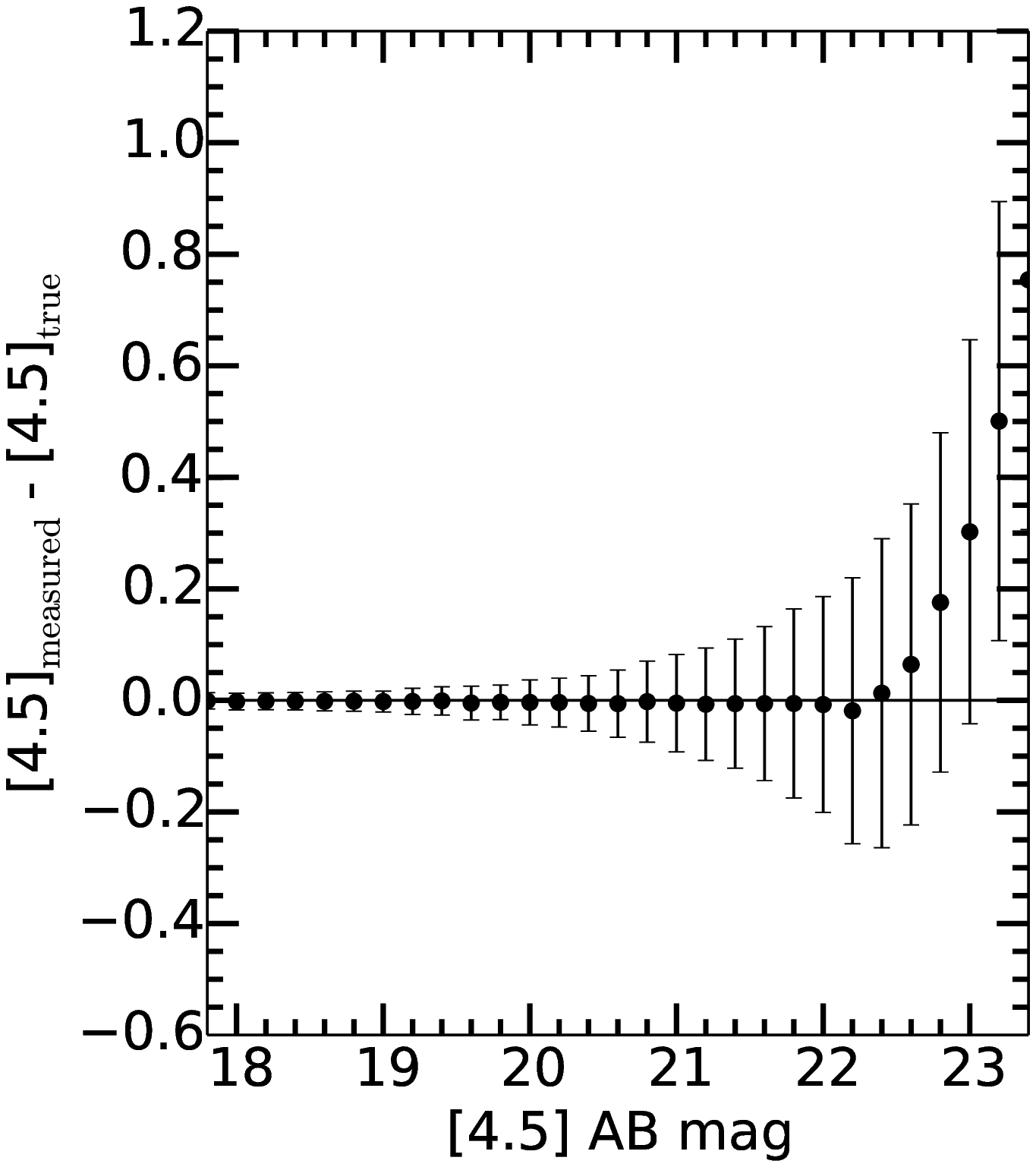}
\caption{Comparison between the ``true'' (input) magnitude for
  simulated sources and the measured magnitude as a function of
  sources magnitude.   The left panel shows the results for
  3.6~\micron, the right panel shows the results for 4.5~\micron.
}\label{fig:compFlux}
\epsscale{1}
 \ifsubmode
\end{figure}
\else
\end{figure*}
\fi

\ifsubmode
\begin{figure} 
\epsscale{1.0}
\else
\begin{figure*}  
\epsscale{1.15} 
\fi 
\plottwo{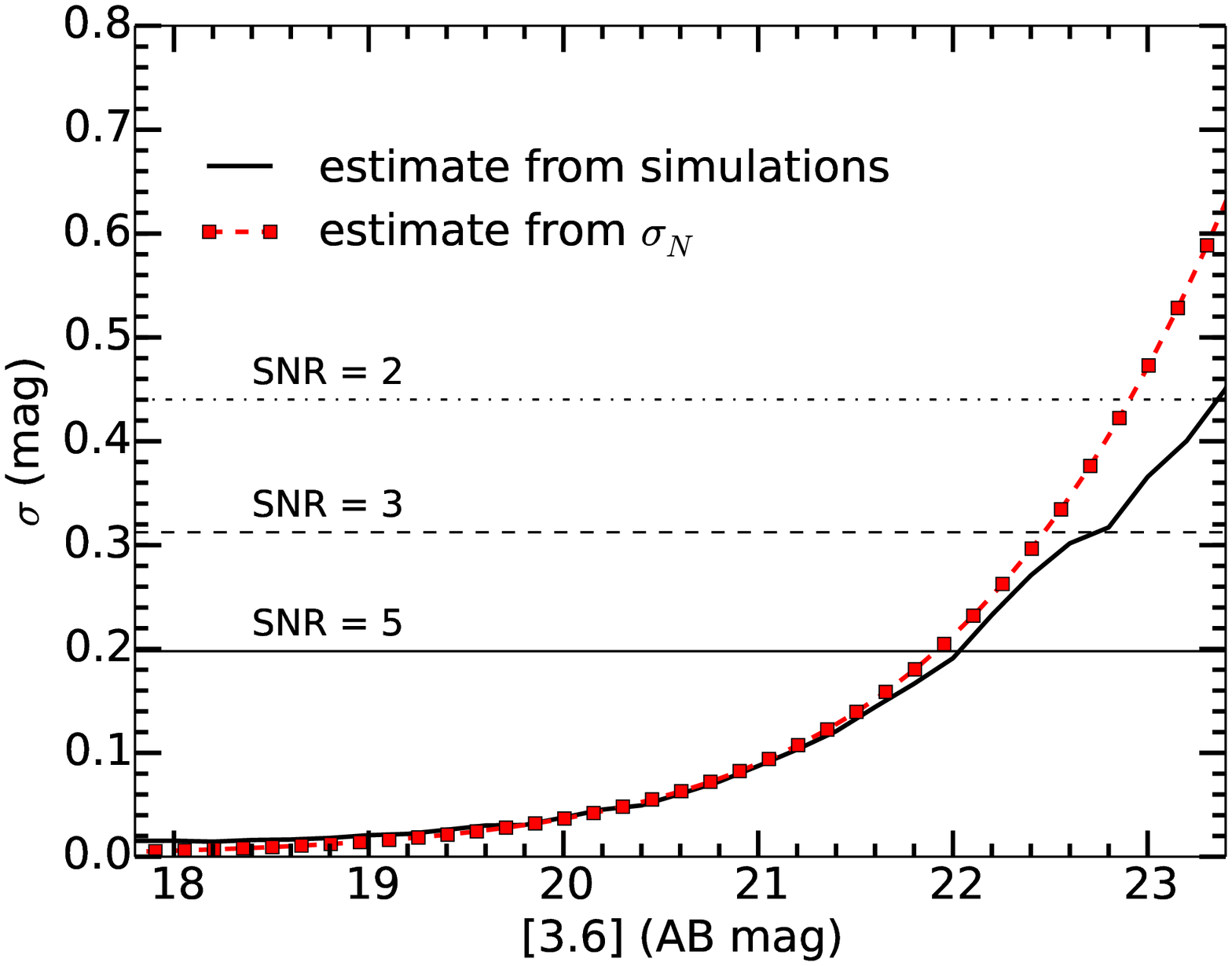}{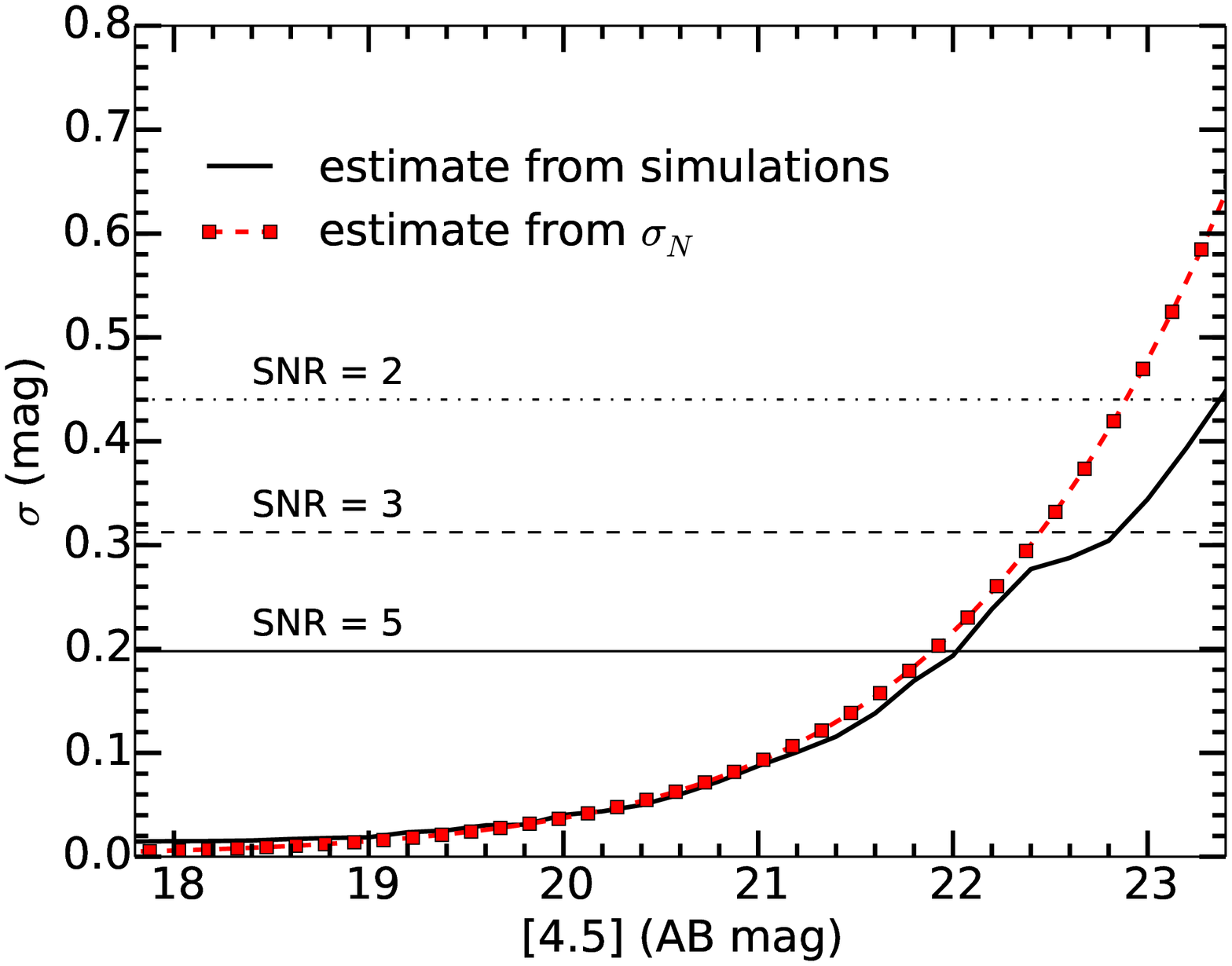}
\caption{Estimates of uncertainties for the SHELA IRAC data.    The
left panel shows the results for the IRAC 3.6~\micron\ data.  The
right panel shows the results for the 4.5~\micron\ data.   In each
panel, the solid-line curve shows the estimated uncertainty measured from a
comparison of the recovered
magnitudes to the input magnitudes for fake sources added to the
images.      The points connected by the dashed-line curve show the estimates
derived from $\sigma_N$ for $2\arcsec$-radii apertures, scaled to
total magnitudes.   The horizontal lines show the equivalent magnitude
uncertainty for a source with \snr=5, 3, and 2, as labeled.  }\label{fig:sigmaFlux}
\epsscale{1}
 \ifsubmode
\end{figure}
\else
\end{figure*}
\fi

\subsection{Error Estimates}\label{section:errors}

We estimate uncertainties for sources in the IRAC catalogs using two
methods.  We first used the simulations from
\S~\ref{section:completeness} to estimate the uncertainty for point
sources of a given magnitude.     In each of the IRAC 3.6 and
4.5~\micron\ images, we computed the difference between the input
(``true'') magnitude and measured magnitude from SExtractor in
{$R=2$\arcsec} radii (corrected to total using the aperture
corrections in Table~\ref{table:apcor}).  Figure~\ref{fig:compFlux}
shows the median and inter-68-percentile of the distributions of these
differences as a function of \mone\ and \mtwo magnitude.   The mean
offset is near zero down to $\approx$22.2 mag (below the 80\%
completeness limit).   In each bin of magnitude, we compute the ratio
$\sigma / F_\nu$ (the inverse of the SNR) as a
measure of the relative error for sources of that magnitude.  The
solid black line of Figure~\ref{fig:sigmaFlux} shows this ratio as a
function of \mone\ and \mtwo.   This yields a limiting SNR=5 at
22.0~AB mag, or a $1\sigma$ limit of 1.1~\ujy\ for both \mone\ and
\mtwo. 

This $1\sigma$ flux-density limit is consistent with estimates from
the \spitzer\ SENS-PET for fields with
higher background.  SENS-PET gives for the ``warm'' \spitzer\ mission a
$1\sigma$ limit for point sources of 0.9--1.2~\ujy\ and 1.1--1.6~\ujy\
for 3.6 and 4.5~\micron\ observations for ``medium'' and ``high'' background
(where as noted above, the medium and highs backgrounds in SENS-PET
assume a sightline with latitude $\beta=40^\circ$ and 0$^\circ$ from the
Ecliptic); therefore the values we derive are reasonable. 

Figure~\ref{fig:sigmaFlux} shows a ``kink'' in the error estimated from
the simulations for both \mone\ and \mtwo\ below $\gsim 22.5$~mag
(black, solid-lined curves in each panel of the figure).
This is likely a bias owing to incompleteness of recovered sources.
The 50 and 80\% completeness limits are 22.6 and 22.0~mag,
respectively.  Figure~\ref{fig:compFlux} shows that in this magnitude
range, the median difference between recovered and input photometry is
biased to positive values because fainter sources are missed in the
catalog.  Therefore, at these magnitudes, the distribution is clipped,
and the inter-68 percentile range is biased smaller.  This means the
errors estimated from the simulations underestimate the true
photometric uncertainty for sources with magnitudes below about the
80\% completeness limit.   Partly for this reason we will adopt the
alternative method to estimate errors, described in the rest of this
subsection below.

\ifsubmode
\begin{deluxetable}{lcccccc}
\else
\begin{deluxetable*}{lcccccc}
\fi
\tablecolumns{7}
\tablewidth{0pc}
\tablecaption{Coefficients for Error Estimates using $\sigma_N$\label{table:errorCoeffs}}
\tablehead{
  \colhead{Channel} & 
  \colhead{Epoch} &
  \colhead{$\sigma_1$/$\mu$Jy} & 
  \colhead{$\alpha$} &
  \colhead{$\beta$} &
  \colhead{$\gamma$} &
  \colhead{$\delta$} \\
\colhead{(1)} & 
\colhead{(2)} & 
\colhead{(3)} & 
\colhead{(4)} &
\colhead{(5)} &
\colhead{(6)} &
\colhead{(7)} }
\startdata
3.6~\micron\ & Combined 3 Epochs & 0.106 & 0.959 & 0.67 &
$8.80\times10^{-4}$ & 1.77 \\
  & Single Epoch & 0.178 & 0.944  & 0.69 & $1.98\times 10^{-3}$ &
  1.88 \\\hline
4.5~\micron\ & Combined 3 Epochs & 0.103 & 0.981 & 0.67 & $3.49\times
10^{-4}$ & 2.28 \\
  & Single Epoch & 0.169 & 0.893  & 0.70 & $1.95\times 10^{-6}$ &
  2.59
\enddata
\ifsubmode
\end{deluxetable}
\else
\end{deluxetable*}
\fi
 
Second, we derived error estimates from the noise in the images in
apertures of \ed{increasing number of pixels, $N$, where $N \propto
{A}$, the  area of the photometric aperture.   The flux
uncertainty within an aperture has a contribution from photon
statistics.  The theoretical uncertainty in an aperture with $N$
pixels would then scale as $\sigma_N = \sigma_1 \times \sqrt{N}$,} where
$\sigma_1$ is the standard deviation of background pixels.   This
relation assumes that the pixel values are independent (uncorrelated).
In practice, multiple effects are expected to introduce some
pixel--to--pixel correlation, such as image alignment and mosaicking,
sky subtraction, the extended wings of bright sources, and the flux
from undetected objects.   The limiting case
of perfect correlations between pixels implies that the uncertainty in
an aperture of \ed{$N$ pixels should scale as $\sigma_N = \sigma_1 \times
N$ \citep{quadri07b}}.   Generalizing, we expect the uncertainty to
scale with \ed{$N^\beta$ with $0.5 < \beta < 1$}, between the limiting
cases of uncorrelated pixels and perfectly correlated pixels
\citep[see also][]{labbe03,gawi06,blanc08,whit11}.

\ifsubmode
\begin{figure} 
\epsscale{0.5}
\else
\begin{figure}  
\epsscale{1.05} 
\fi 
\plotone{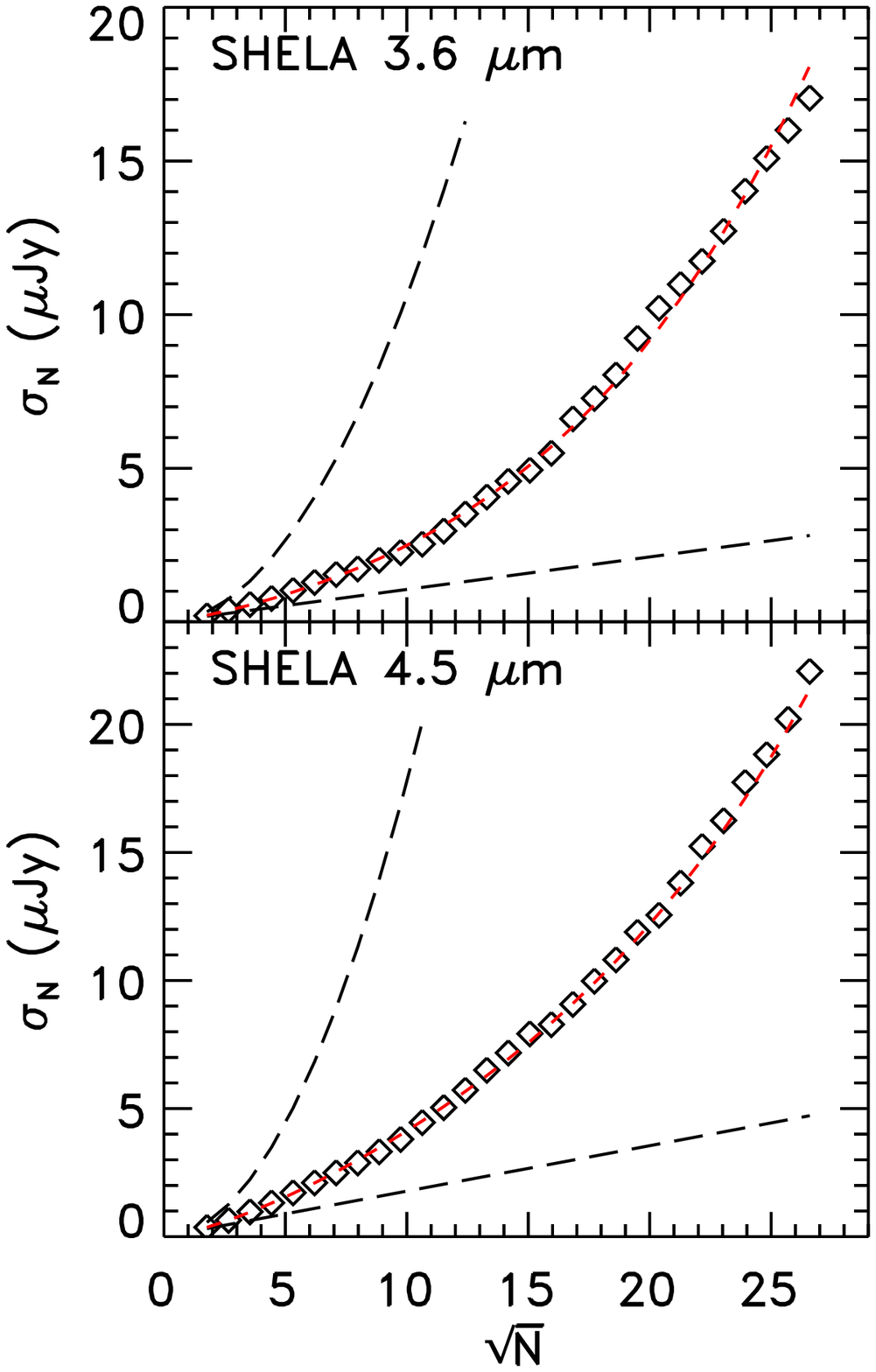}
\caption{Scaling relation between the measured noise in the SHELA IRAC
images and the \ed{square root of the number of pixels, $\sqrt{N}$, in the area of the
photometric aperture}.    Both panels show the measured noise,
$\sigma_N$, in each aperture of $N$ pixels.  The top panel shows the
3.6~\micron\ data and the bottom panel shows the 4.5~\micron\ data.
In each panel, the bottom-most dashed line shows the theoretical
relation assuming uncorrelated pixels in the Gaussian limit, $\sigma_N
\sim \sqrt{N}$.  The top-most dashed line shows the relation for perfectly
correlated pixels \citep[$\sigma_N \sim N$,][]{quadri07b}.   The red,
short-dashed line shows the parameterized fit to the data, which we
use to define the flux uncertainties measured in different-sized
apertures.}\label{fig:noise}
\epsscale{1}
 \ifsubmode
\end{figure}
\else
\end{figure}
\fi

We estimated the noise as a function of pixels by measuring
the sky counts in circular apertures of varying size in $\approx$5000
randomly placed regions in the  SHELA IRAC images, ensuring that apertures do
not overlap, and excluding regions containing objects. We then computed
the standard deviation of the distribution of aperture fluxes from the
normalized median absolute deviation, $\sigma_\mathrm{nmad}$ \citep{beers90}, as an
estimate for $\sigma_N$ for each aperture with $N$ pixels.  
Figure~\ref{fig:noise} shows the measured relation of $\sigma_N$ as a
function of $\sqrt{N}$ for the 3-epoch IRAC \mone\ and \mtwo\ images.

Following the suggestion in \citet{labbe03}, we fit a parameterized
function to estimate the noise in an arbitrary aperture of linear size
$N$,
\begin{equation}\label{eqn:sigma}
\sigma_N = \sigma_1 \left(  \alpha N^\beta + \gamma N^\delta \right)
\end{equation} 
where $\sigma_1$ is the pixel-to-pixel standard deviation in the sky
background, and $\alpha$, $\beta$, $\gamma$, and $\delta$ are free
parameters.  We required that  $\alpha$ and $\gamma$ be non-negative,
that $0.5 < \beta < 1$, and we placed  no restrictions on $\delta$.   In this
way the first term of Equation~\ref{eqn:sigma} represents the expected
noise for partially correlated pixels.  The second term includes an
additional correction that better reproduces the noise in large
apertures \citep[see also,][]{labbe03}.  Table~\ref{table:errorCoeffs}
lists the parameters for the fits in Equation~\ref{eqn:sigma} for the
combined, 3 epoch IRAC 3.6 and 4.5~\micron\ data.  The table also
includes fits for the individual IRAC epoch data (where our tests
showed each individual epoch had noise properties consistent with
being the same, so we combined the random apertures of all individual
images for a single fit).    

The product of $\sigma_1 \alpha$  reflects the pixel-to-pixel rms, and
these decrease roughly with the square-root of the exposure time such
that the value of $\sigma_1$ for the 3-epoch combined image is roughly
$\sqrt{3}$ lower than that for an individual epoch.    The fitted
values for the slope in the first term,  $\beta \sim 0.6-0.7$, are
consistent with partially correlated pixels, as found in other imaging
surveys \citep[\eg,][]{gawi06,quadri07b}.   The values for $\gamma$ are
relatively small (the ratio of the coefficients is $\alpha / \gamma
\sim 5 \times 10^2 - 5 \times 10^5$),  implying there is a small, but
increasing correction to the noise model for apertures with larger
numbers of pixels.

\ifsubmode
\begin{deluxetable}{lllcc}
\else
\begin{deluxetable*}{lllcc}
\fi
\tablecolumns{5}
\tablewidth{0pc}
\tablecaption{Column Definitions in SHELA IRAC Catalog\label{table:columnDefs}}
\tablehead{
  \colhead{Catalog Column Name\tablenotemark{a}} & 
  \colhead{Table Column Name\tablenotemark{b}} & 
  \colhead{Description} &
  \colhead{units} & 
  \colhead{data type}
}
\startdata
ID & \ldots & Unique ID number from SExtractor for each source
  in the IRAC catalogs & \ldots & long int \\
X\tablenotemark{c} & \ldots & $x$-pixel coordinate in IRAC image & pixel & float \\
Y\tablenotemark{d} & \ldots & $y$-pixel coordinate in IRAC images & pixel & float \\
RA & RA(J2000) & Right Ascension (J2000) of IRAC source & deg & double
\\
DEC & DEC(J2000) & Declination (J2000) of IRAC source & deg & double
\\
ISOAREA & Isophotal Area &  isophotal area of source in detection
image & arcsec$^2$ & float \\ 
A & $a$ & source semimajor axis & arcsec & float \\
E & $e$ & source ellipticity, $e = 1 -  b/a$, where $b$ is the
semiminor axis & \ldots & float \\
THETA & $\theta$ & position angle of the semi-major axis, degrees east from
celestial north & deg & float \\
W3P6 & W(3.6) & value of the 3-epoch 3.6~\micron\ weight map at the object's
center position\tablenotemark{e} & \ldots & float \\
W4P5 & W(4.5) & value of the 3-epoch 4.5~\micron\ weight map at the object's
center position\tablenotemark{e} & \ldots & float \\
W3P6\_1 & W(3.6)$_1$ & value of the epoch 1, 3.6~\micron\ weight map at the object's
center position\tablenotemark{e} & \ldots & float \\
W4P5\_1 & W(4.5)$_1$ & value of the epoch 1, 4.5~\micron\ weight map at the object's
center position\tablenotemark{e} & \ldots & float \\
W3P6\_2 & W(3.6)$_2$ & value of the epoch 2, 3.6~\micron\ weight map at the object's
center position\tablenotemark{e} & \ldots & float \\
W4P5\_2 & W(4.5)$_2$ & value of the epoch 2, 4.5~\micron\ weight map at the object's
center position\tablenotemark{e} & \ldots & float \\
W3P6\_3 & W(3.6)$_3$ & value of the epoch 3, 3.6~\micron\ weight map at the object's
center position\tablenotemark{e} & \ldots & float \\
W4P5\_3 & W(4.5)$_3$ & value of the epoch 3, 4.5~\micron\ weight map at the object's
center position\tablenotemark{e} & \ldots & float \\
FLAGS3P6 & Flags (3.6\micron) & SExtractor flags for photometry of
3.6~\micron\ image\tablenotemark{f} & \ldots & integer \\
FLAGS4P5 & Flags (4.5\micron) & SExtractor flags for photometry of
4.5~\micron\ image\tablenotemark{f} & \ldots & integer \\
F3P6\_ISO & $\mathbf{f^\mathbf{(3.6)}_{\mathbf\nu,ISO}}$ &  Isophotal flux
  density for sources in the  3.6~\micron\ image\tablenotemark{f} & $\mu$Jy & float \\
F3P6ERR\_ISO & $\mathbf{\sigma^{(3.6)}_{\mathbf\nu,ISO}}$ &  Error on
the 3.6~\micron\ isophotal flux density\tablenotemark{f}
& $\mu$Jy & float \\ 
 F4P5\_ISO & $\mathbf{f^\mathbf{(4.5)}_{\mathbf\nu,ISO}}$ &  Isophotal flux
  density for sources in the  4.5~\micron\ image\tablenotemark{f} & $\mu$Jy &
  float \\
F4P5ERR\_ISO & $\mathbf{\sigma^{(4.5)}_{\mathbf\nu,ISO}}$ &  Error on
the 4.5~\micron\ isophotal flux density\tablenotemark{f}
& $\mu$Jy & float \\ 
F3P6\_AUTO & $\mathbf{f^\mathbf{(3.6)}_{\mathbf\nu,AUTO}}$ &  Total
flux measured in the Kron aperture for sources in the  3.6~\micron\ image\tablenotemark{f} & $\mu$Jy & float \\
F3P6ERR\_AUTO & $\mathbf{\sigma^{(3.6)}_{\mathbf\nu,AUTO}}$ &  Error on
the 3.6~\micron\ total flux density\tablenotemark{f}
& $\mu$Jy & float \\ 
 F4P5\_AUTO & $\mathbf{f^\mathbf{(4.5)}_{\mathbf\nu,AUTO}}$ &  Total
 flux measured in a Kron aperture for sources in the 4.5~\micron\ image\tablenotemark{f} & $\mu$Jy &
  float \\
F4P5ERR\_AUTO & $\mathbf{\sigma^{(4.5)}_{\mathbf\nu,AUTO}}$ &  Error on
the 4.5~\micron\ total flux density\tablenotemark{f}
& $\mu$Jy & float \\ 
F3P6\_4ARCS & $\mathbf{f^\mathbf{(3.6)}_{\mathbf\nu,4\arcsec}}$ & Flux
density measured at 3.6~\micron\ for sources measured in
4\arcsec-diameter apertures\tablenotemark{f,g} & $\mu$Jy & float \\  
F3P6ERR\_4ARCS & $\mathbf{\sigma^{(3.6)}_{\mathbf\nu,4\arcsec}}$ &
Error on the flux density at 3.6~\micron\ measured in the
4\arcsec-diameter apertures\tablenotemark{f} & $\mu$Jy & float \\
F4P5\_4ARCS & $\mathbf{f^\mathbf{(4.5)}_{\mathbf\nu,4\arcsec}}$ & Flux
density measured at 4.5~\micron\ for sources measured in
4\arcsec-diameter apertures\tablenotemark{f,g} & $\mu$Jy & float \\  
F4P5ERR\_4ARCS & $\mathbf{\sigma^{(4.5)}_{\mathbf\nu,4\arcsec}}$ &
Error on the flux density at 4.5~\micron\ measured in the
4\arcsec-diameter apertures\tablenotemark{f} & $\mu$Jy & float \\
F3P6\_6ARCS & $\mathbf{f^\mathbf{(3.6)}_{\mathbf\nu,6\arcsec}}$ & Flux
density measured at 3.6~\micron\ for sources measured in
6\arcsec-diameter apertures\tablenotemark{f,g} & $\mu$Jy & float \\  
F3P6ERR\_6ARCS & $\mathbf{\sigma^{(3.6)}_{\mathbf\nu,6\arcsec}}$ &
Error on the flux density at 3.6~\micron\ measured in the
6\arcsec-diameter apertures\tablenotemark{f} & $\mu$Jy & float \\
F4P5\_6ARCS & $\mathbf{f^\mathbf{(4.5)}_{\mathbf\nu,6\arcsec}}$ & Flux
density measured at 4.5~\micron\ for sources measured in
6\arcsec-diameter apertures\tablenotemark{f,g} & $\mu$Jy & float \\  
F4P5ERR\_6ARCS & $\mathbf{\sigma^{(4.5)}_{\mathbf\nu,6\arcsec}}$ &
Error on the flux density at 4.5~\micron\ measured in the
6\arcsec-diameter apertures\tablenotemark{f} & $\mu$Jy & float
\enddata
\tablenotetext{a}{Column name in binary FITS tables}
\tablenotetext{b}{Column name in
  Tables~\ref{table:preamblecatalog}-\ref{table:ep3catalog}, if
  different from column name in binary FITS table}
\tablenotetext{c}{\texttt{PIX\_X} in the catalog on IRSA.}
\tablenotetext{d}{\texttt{PIX\_Y} in the catalog on IRSA.}
\tablenotetext{e}{The weight
  map values are proportional
to the effective exposure time, with a constant of proportionality
$t_\mathrm{eff} = 23.6$ s $\times$ Weight.}
\tablenotetext{f}{These column names exist in each of the catalogs
  (combined 3 epoch, and individual epochs) with the same column names.}
\tablenotetext{g}{The flux densities for sources measured in circular
  apertures have been corrected to total using the aperture
  corrections in Table~\ref{table:apcor}.}
\ifsubmode
\end{deluxetable}
\else
\end{deluxetable*}
\fi

\setcounter{table}{11}

\ifsubmode
\begin{deluxetable}{lllcc}
\else
\begin{deluxetable*}{lllcc}
\fi
\tablecolumns{5}
\tablewidth{0pc}
\tablecaption{Column Definitions in SHELA/SDSS-matched Catalog\label{table:SDSScolumnDefs}}
\tablehead{
  \colhead{Catalog Column Name\tablenotemark{a}} & 
  \colhead{Table Column Name\tablenotemark{b}} & 
  \colhead{Description} &
  \colhead{units} & 
  \colhead{data type}
}
\startdata
SHELA\_ID & ID & Unique ID from the SHELA catalog in
Table~\ref{table:preamblecatalog} & \ldots & long int \\
SDSS\_ID & SDSS ID & ID of object in the SDSS catalogs & \ldots & 64-bit long
  int \\
SDSS\_RA & SDSS RA &  Right ascension (J2000) of object in the SDSS
catalog & deg & double \\
SDSS\_DEC & SDSS DEC & Declination (J2000) of object in the SDSS
catalog & deg & double \\
TYPE & \ldots & Object type from SDSS catalog\tablenotemark{c} &
\ldots & int \\
FLAGS & SDSS FLAGS & SDSS Flags for the object & \ldots & 64-bit long
int \\
U\tablenotemark{d} & $\mathbf{\sdssu}$ & SDSS \sdssu\ total AB magnitude corrected for Galactic
  extinction & mag & float \\
UERR & $\mathbf{\sigma_{\sdssu}}$ & uncertainty on SDSS \sdssg\ AB
magnitude & mag & float \\
G\tablenotemark{d} & $\mathbf{\sdssg}$ & SDSS \sdssg\ total AB magnitude corrected for Galactic
  extinction & mag & float \\
GERR & $\mathbf{\sigma_{\sdssg}}$ & uncertainty on SDSS \sdssg\ AB
magnitude & mag & float \\
R\tablenotemark{d} & $\mathbf{\sdssr}$ & SDSS \sdssr\ total AB magnitude corrected for Galactic
  extinction & mag & float \\
RERR & $\mathbf{\sigma_{\sdssr}}$ & uncertainty on SDSS \sdssr\ AB
magnitude & mag & float \\
I\tablenotemark{d} & $\mathbf{\sdssi}$ & SDSS \sdssi\ total AB magnitude corrected for Galactic
  extinction & mag & float \\
IERR & $\mathbf{\sigma_{\sdssi}}$ & uncertainty on SDSS \sdssi\ AB
magnitude & mag & float \\
Z\tablenotemark{d} & $\mathbf{\sdssz}$ & SDSS \sdssz\ total AB magnitude corrected for Galactic
  extinction & mag & float \\
ZERR & $\mathbf{\sigma_{\sdssz}}$ & uncertainty on SDSS \sdssz\ AB
magnitude & mag & float
\enddata
\tablenotetext{a}{Column name in binary FITS tables}
\tablenotetext{b}{Column name in
  Table~\ref{table:sdsscatalog}, if
  different from column name in binary FITS table}
\tablenotetext{c}{The most common type values are TYPE=3 for
  Galaxy or TYPE=6 for Star.}
\tablenotetext{d}{The source SDSS magnitudes have the extension \texttt{\_MAG} in
  the catalogs on IRSA.}
\ifsubmode
\end{deluxetable}
\else
\end{deluxetable*}
\fi

\setcounter{table}{13}

\begin{deluxetable*}{ccccccc}
\tablecolumns{7}
\tablewidth{0pt}
\tablecaption{Number of SHELA/IRAC Sources Detected in Different Channel Combinations\label{table:combo}}
\tablehead{
\colhead{} & 
\multicolumn{6}{c}{Channel Combination} \\
\colhead{Aperture} & 
\colhead{\snr$_1$$\ge$3} & 
\colhead{\snr$_2$$\ge$3} & 
\colhead{\snr$_1$$\ge$3 $\lor$ \snr$_2$$\ge$3} & 
\colhead{\snr$_1$$\ge$3 $\land$ \snr$_2$$\ge$3} & 
\colhead{\snr$_1$$\ge$3 $\land$ \snr$_2$$<$3} & 
\colhead{\snr$_1$$<$3 $\land$ \snr$_2$$>=$3} \\
\colhead{(1)} & 
\colhead{(2)} & 
\colhead{(3)} & 
\colhead{(4)} & 
\colhead{(5)} & 
\colhead{(6)} & 
\colhead{(7)}}
\startdata
4ARCS     &    1,729,650     &   1,701,167    &  1,982,997                         &  1,447,820                            &  207,981                            &  181,398 \\
6ARCS     &    1,329,326     &   1,290,800    &  1,564,399                         &  1,055,727                            &  214,091                            &  176,862 \\
ISO          &    1,677,349     &   1,618,215    &  1,946,029                         &  1,349,535                            &  254,104                            &  197,157 \\ 
AUTO      &    1,019,486     &      960,500    &  1,267,282                         &     712,704                            &  255,712                            &  194,841
\enddata
\tablecomments{Column (1) ``Aperture'' , gives the name of the extension of
  each aperture as it appears in Table~\ref{table:columnDefs}.
  Columns (2--7) \snr$_1$\ and \snr$_2$\ are the signal--to--noise
  ratios in the IRAC 3.6~\micron\ and 4.5~\micron\ flux density for
  each aperture, respectively.  $\land$ and $\lor$ are the logical AND
and OR operators, respectively.}
\end{deluxetable*}
 
The red points in Figure~\ref{fig:sigmaFlux} show the magnitude
uncertainty calculated for $2\arcsec$-radius apertures (scaled up to
the total aperture) as a function of \mone\ and \mtwo.    There is
generally good agreement between the estimated uncertainties from
$\sigma_N$ and those from the simulations described above.  For
objects with $<$22 AB mag, there is a slight offset, where the
estimates from $\sigma_N$ are lower at about the 0.02 mag level
compared to the estimates from the completeness simulations.  This
could arise from several effects, including the fact that the
completeness simulations allow fake objects to fall on image regions
that contain other (real) galaxies.  As this will tend to increase the
average difference between the input and recovered magnitude, an
offset is not unexpected.   \ed{We include an additional 0.02 mag
systematic uncertainty into our estimates to account for this effect
(see below).}

\ed{For the IRAC catalogs, we computed errors using
Equation~\ref{eqn:sigma} for the number of pixels $N$ in the aperture
used to measure the object, scaled up to the total aperture.    We add
these errors in quadrature with an additional error
$\sigma_{\mathrm{sys}}=0.02$~mag, to account for systematics derived
from the completeness simulations (see \S~\ref{section:completeness}
and Table~\ref{table:apcor}).  The total photometric error,
$\sigma_{i,c}$ on each source $i$ in IRAC channel $c$ is then given
by,}
\begin{equation}
\sigma^2_{i,c} = 
\frac{\sigma^2_{N,c}}{(w_{i,c}/w_{\mathrm{med},c})} +
0.921\ \sigma_{\mathrm{sys}} \times F_{i,c},
\end{equation}
\ed{where $F_{i,c}$ is the flux density of each object in each
  channel, $\sigma^2_{N,c}$ is given by Equation~\ref{eqn:sigma} for
  each channel, 
  $w_{i,c}$ is the value of the weight map at the location of each object, and
  $w_{\mathrm{med},c}$ is the median value of the weight map in each channel.}
 We opted to use these uncertainty estimates as they can be scaled to
arbitrarily sized apertures (unlike the errors on the simulations,
which are otherwise valid only for point sources).   

\subsection{Catalogs}

With this \textit{Paper}, we publish the full SHELA photometric
catalog.  The catalogs include the IRAC fluxes measured in multiple
apertures (4\arcsec\ and 6\arcsec\ diameter circular apertures,
corrected to total magnitudes, isophotal magnitudes, and the ``total''
(\texttt{MAG\_AUTO}) magnitudes from SExtractor).  Errors are
estimated from Equation~\ref{eqn:sigma} and
Table~\ref{table:errorCoeffs} for the number of 
pixels for each object/aperture.    In addition, we include a catalog
with photometry for the IRAC sources from the SDSS Stripe 82 coadd
field \citep{annis14} in $\sdssu\sdssg\sdssr\sdssi\sdssz$, where
sources in the SHELA catalog have been matched to the astrometric
positions of sources in the SDSS Stripe 82 catalogs using a 1\arcsec\
search radius.  Only SDSS sources matched to SHELA sources are
included in the catalog, and we include only the closest source in the
cases where multiple SDSS sources are located within 1\arcsec\ of a
given SHELA source.


Table~\ref{table:columnDefs} provides a description of each column
name in the tables and the binary FITS tables.
Table~\ref{table:preamblecatalog} contains object astrometry and
quantities measured from the detection image and weight maps.
Tables~\ref{table:fullcatalog}--\ref{table:ep3catalog} present the
SHELA catalogs for the data release.  This includes a catalog for the
full, combined 3-epoch data (Table~\ref{table:fullcatalog}) and
catalogs for each individual epoch
(Tables~\ref{table:ep1catalog}-\ref{table:ep3catalog}).     The full
catalogs are provided as binary tables in Flexible Image Transport
System \citep[FITS,][]{hani01} format. 
 
The SExtractor flags (\texttt{FLAGS (3.6\micron)} and
\texttt{FLAGS (4.5\micron)}) are stored as bits and coded in decimal as
the sum of powers of 2 
($2^\mathrm{bit}$) for bits that are flagged.  Common
flag bit values are:
\begin{description}\itemsep1pt \parskip0pt \parsep0pt
\item[bit 1] The object has bright neighbors that may bias the photometry, or
  the object has more than 10\% of its pixels flagged as bad
  or have zero weight;
\item[bit 2] The object was deblended from another object. 
\end{description}
Neither flag bit value is fatal, but users may require a more thorough
vetting of these sources depending on their needs. 
Other, higher (very uncommon in the SHELA catalogs) bit values denote
objects whose photometry is dubious.  These objects should likely be
excluded from use.   These bits are available in the SExtractor User's Manual
(v2.13).

The column descriptions for the binary FITS table and table for the
merged SHELA--SDSS Stripe 82 catalog are listed in
Table~\ref{table:SDSScolumnDefs}. Table~\ref{table:sdsscatalog}
presents the photometric data for the merged SHELA--SDSS Stripe 82
catalog. 

In the catalogs, objects with no coverage (in a given wavelength
and/or epoch) will have weight = 0 and zero flux density and error.
These objects also have SExtractor bit=1 set in their flag values.
\ed{The catalogs and images are available as part of this publication
  through IRSA, see footnote~\ref{footnote:IRSA}}
%
%
%

\ed{Because sources are detected in the weighted sum combination of
  the IRAC 3.6 and 4.5~\micron\ data, the catalogs contain different
  numbers of sources detected in 3.6~\micron\ and 4.5~\micron\
  individually.   Table~\ref{table:combo} provides the number of
  sources detected in different combinations of the 3.6 and
  4.5~\micron\ bands based on the requirement that sources be detected in
  \textit{at least} one of the two channels with significance $\snr
  \ge 3$.    The table shows the number of sources for the \snr\
  defined in different apertures.   Clearly, the 4\arcsec--diameter
  apertures provide $\snr \ge 3$ for the most sources.  This is
  expected as this aperture encompasses $\sim$75\% of the light for
  unresolved objects, while containing the fewest low \snr\ pixels.
  Therefore for objects whose light is well
  contained with 4\arcsec--diameters, we recommend this catalog as it
  maximizes the \snr.  Catalog users can determine if the
  4\arcsec--diameter aperture is too small by comparing the flux
  densities measured in this aperture and those in larger apertures
  (the 6\arcsec--diameter aperture and the Kron, $\_$AUTO aperture,
  for example), and make decisions about aperture choice for their
  specific requirements.}

\section{A Modicum of Science}

\subsection{Number Counts}\label{section:counts}

\begin{deluxetable}{lcccc}
\tablecaption{SHELA IRAC Number Counts\label{table:counts}}
\tablehead{
\colhead{} & \multicolumn{2}{c}{IRAC 3.6~$\mu$m} &
\multicolumn{2}{c}{IRAC 4.5~$\mu$m} \\ 
\colhead{$m_\mathrm{AB}$} & \colhead{$dN/dm$} &
  \colhead{Error} & \colhead{$dN/dm$} &
  \colhead{Error} \\
\colhead{(mag)} & \colhead{(mag$^{-1}$ deg$^{-2}$)} & 
\colhead{(mag$^{-1}$ deg$^{-2}$)} & 
\colhead{(mag$^{-1}$ deg$^{-2}$)} & 
\colhead{(mag$^{-1}$ deg$^{-2}$)} \\
\colhead{(1)} & \colhead{(2)} & \colhead{(3)} & \colhead{(4)} & \colhead{(5)}}
\startdata
12.0 & 7.3 & 1.1 & 3.8 & 0.8 \\
12.2 & 7.0 & 1.0 & 3.5 & 0.7 \\
12.4 & 9.9 & 1.2 & 6.7 & 1.0 \\
12.6 & 13.8 & 1.4 & 7.8 & 1.1 \\
12.8 & 20.8 & 1.8 & 14.1 & 1.5 \\
13.0 & 24.3 & 1.9 & 22.6 & 1.9 \\
13.2 & 31.5 & 2.2 & 19.3 & 1.7 \\
13.4 & 36.9 & 2.4 & 21.4 & 1.8 \\
13.6 & 35.4 & 2.3 & 22.9 & 1.9 \\
13.8 & 44.4 & 2.6 & 31.0 & 2.2 \\
14.0 & 52.0 & 2.8 & 34.2 & 2.3 \\
14.2 & 61.2 & 3.1 & 40.7 & 2.5 \\
14.4 & 67.2 & 3.2 & 47.7 & 2.7 \\
14.6 & 78.6 & 3.5 & 56.4 & 2.9 \\
14.8 & 93.5 & 3.8 & 62.0 & 3.1 \\
15.0 & 106.2 & 4.0 & 77.2 & 3.4 \\
15.2 & 117.2 & 4.2 & 84.7 & 3.6 \\
15.4 & 140.0 & 4.6 & 95.8 & 3.8 \\
15.6 & 168.2 & 5.1 & 116.6 & 4.2 \\
15.8 & 175.8 & 5.2 & 134.7 & 4.5 \\
16.0 & 215.1 & 5.7 & 152.7 & 4.8 \\
16.2 & 243.3 & 6.1 & 186.2 & 5.3 \\
16.4 & 287.1 & 6.6 & 219.3 & 5.8 \\
16.6 & 335.4 & 7.1 & 250.2 & 6.2 \\
16.8 & 394.6 & 7.7 & 306.9 & 6.8 \\
17.0 & 469.0 & 8.4 & 358.1 & 7.4 \\
17.2 & 561.0 & 9.2 & 429.1 & 8.1 \\
17.4 & 693.5 & 10.0 & 533.6 & 9.0 \\
17.6 & 853.4 & 11.0 & 643.2 & 9.9 \\
17.8 & 1096.0 & 13.0 & 809.5 & 11.0 \\
18.0 & 1417.0 & 15.0 & 1032.0 & 13.0 \\
18.2 & 1839.0 & 17.0 & 1324.0 & 14.0 \\
18.4 & 2427.0 & 19.0 & 1697.0 & 16.0 \\
18.6 & 3138.0 & 22.0 & 2260.0 & 19.0 \\
18.8 & 4005.0 & 25.0 & 2952.0 & 21.0 \\
19.0 & 5088.0 & 28.0 & 3881.0 & 24.0 \\
19.2 & 6322.0 & 31.0 & 5012.0 & 28.0 \\
19.4 & 7523.0 & 34.0 & 6372.0 & 31.0 \\
19.6 & 8904.0 & 37.0 & 7931.0 & 35.0 \\
19.8 & 10390.0 & 40.0 & 9481.0 & 38.0 \\
20.0 & 11780.0 & 42.0 & 11180.0 & 41.0 \\
20.2 & 13040.0 & 45.0 & 12680.0 & 44.0 \\
20.4 & 14240.0 & 47.0 & 14110.0 & 46.0 \\
20.6 & 15330.0 & 48.0 & 15310.0 & 48.0 \\
20.8 & 16760.0 & 50.0 & 16620.0 & 50.0 \\
21.0 & 18400.0 & 53.0 & 18080.0 & 52.0 \\
21.2 & 19810.0 & 55.0 & 19630.0 & 55.0 \\
21.4 & 20950.0 & 56.0 & 20900.0 & 56.0 \\
21.6 & 21420.0 & 57.0 & 21210.0 & 57.0 \\
21.8 & 21360.0 & 57.0 & 21100.0 & 57.0 \\
22.0 & 20740.0 & 56.0 & 20730.0 & 56.0 \\
22.2 & 19680.0 & 55.0 & 19610.0 & 55.0 \\
22.4 & 17840.0 & 52.0 & 18120.0 & 52.0 \\
22.6 & 15140.0 & 48.0 & 15670.0 & 49.0 \\
22.8 & 11690.0 & 42.0 & 12490.0 & 44.0 \\
23.0 & 8166.0 & 35.0 & 9101.0 & 37.0 
\enddata \tablecomments{(1) magnitude of number count bin, (2) number
counts at 3.6~\micron, (3) Poisson error on 3.6~\micron\ number
counts, (4) number counts at 4.5~\micron, (5) Poisson error on
4.5~\micron\ number counts.  Note that the counts are not corrected
for completeness.  To do so requires dividing by the
magnitude-dependent completeness corrections in
Table~\ref{table:completeness}.}
\end{deluxetable}

Galaxy number counts provide tests of galaxy evolution and cosmology
\citep[e.g.,][]{peeb93}.  The number counts are the integral over
the luminosity function and distance (redshift), containing the
total contribution of galaxies of a given luminosity and distance to
the cosmic background emission.     The galaxy number counts in the
mid-IR are particularly useful as they contain information about
stellar-mass growth, 
dust--obscured populations, and AGN.   Galaxy number counts with
\spitzer\ have demonstrated the abundance of faint sources attributed
to (rest-frame) near-IR and mid-IR emission from distant galaxies and
their contribution to the IR background
\citep[\eg,][]{fazio04b,papo04a,dole04,sand07,ashby09,ashby13b,maud12}.  

Figure~\ref{fig:counts} shows the IRAC 3.6 and 4.5~\micron\ number
counts for the full SHELA data.   The raw counts (uncorrected for
completeness, see \S~\ref{section:completeness}) are provided in
Table~\ref{table:counts}.    The SHELA counts agree well with previous
measurements \citep[e.g.,][]{fazio04b,sand07}.   At bright magnitudes
(AB $\lsim$ 18) the counts follow roughly the expected contribution
from Galactic stars \citep[\eg,][and references
therein]{fazio04b,ashby13b}.   The SHELA counts show a slight excess of
bright counts compared to the data in Fazio et al.\ and Sanders et al.
This is likely a result of the different Galactic sight lines among the surveys,
and is therefore not too surprising.  

The counts provide an independent measure of the completeness of the
SHELA IRAC catalogs.   \ed{As illustrated in Figure~\ref{fig:counts},
  the SHELA counts agree with those of the deeper imaging in 
\citet{fazio04b} and \citet{sand07} to better than 10\% near the peak
of the distribution.   Similarly, when corrected for incompleteness
(\S~4.1), the SHELA data are consistent with the counts in Fazio et
al.\ and Sanders et al.\ at least down to the 50\% completeness level
of $m_\mathrm{AB}\sim 22.5$~mag}.  At fainter
magnitudes, the completeness corrections for the SHELA data are
significantly higher and the uncertainties on the completeness
corrections dominate the counts  (and Eddington--type biases are most
severe).  Therefore, we have confidence in the completeness
corrections and the number counts down to the 50\% completeness
limit. 

\ifsubmode
\begin{figure} 
\epsscale{1.1}
\else
\begin{figure*}  
\epsscale{1.05} 
\fi 
\plottwo{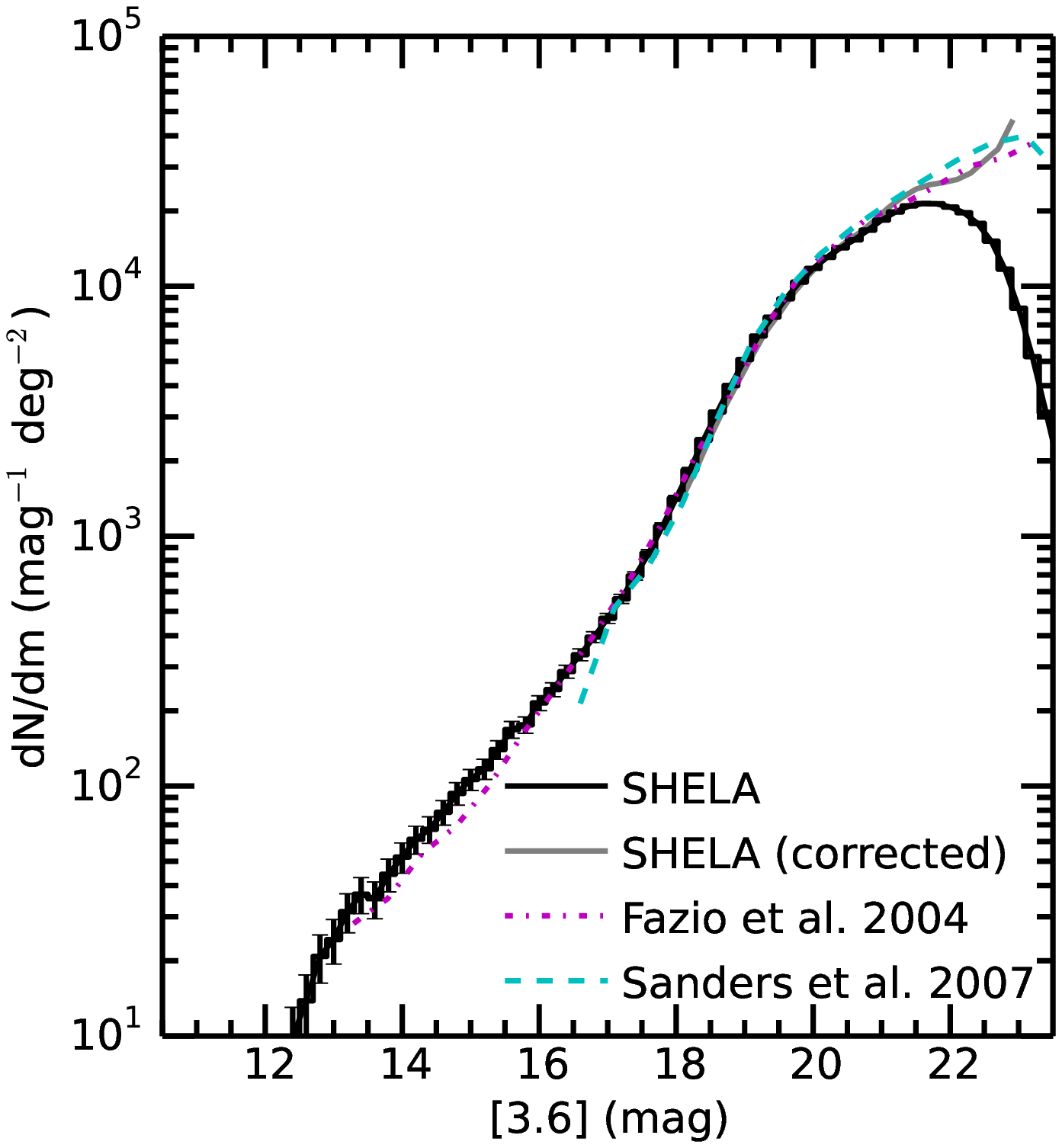}{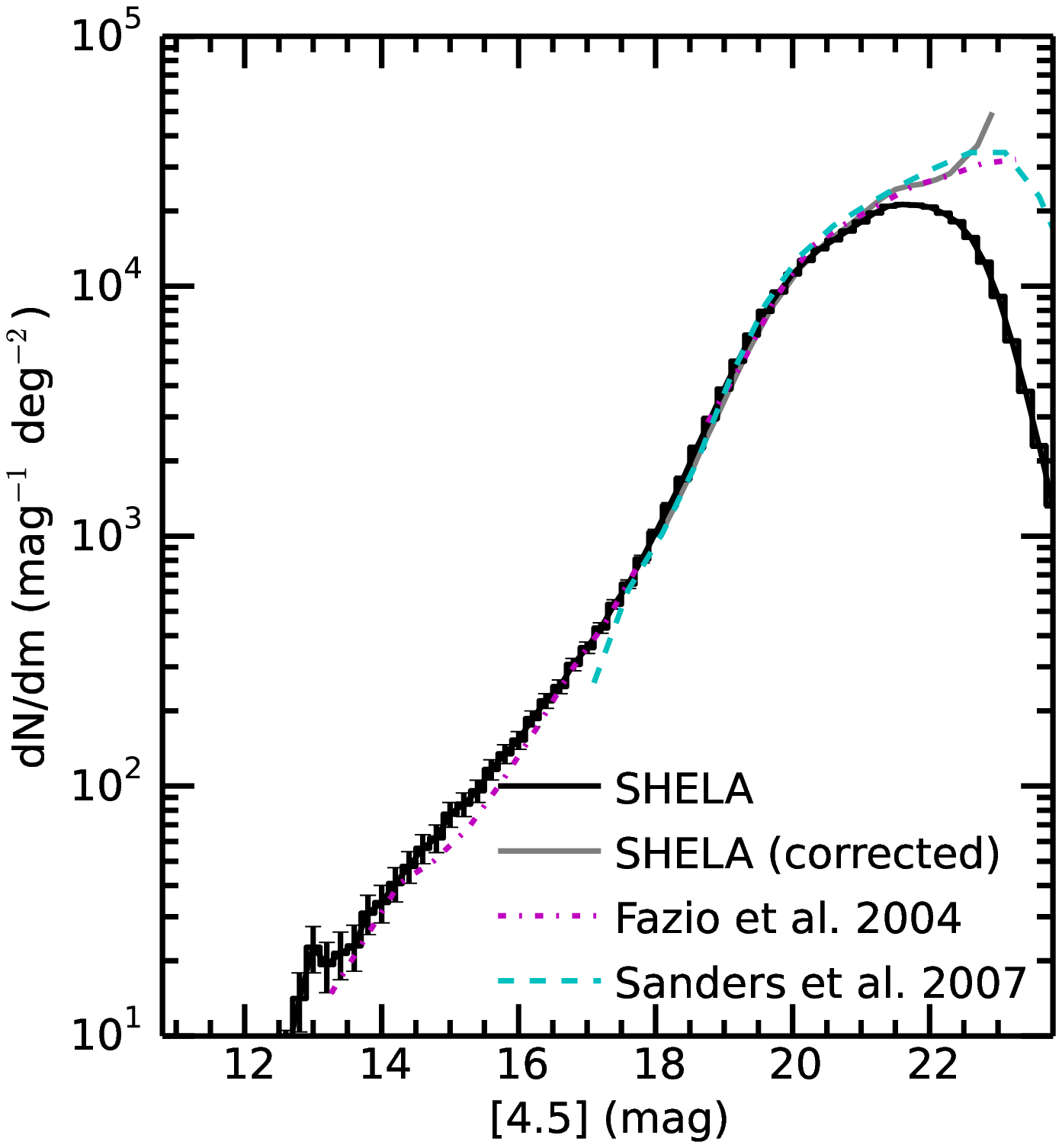}
\plottwo{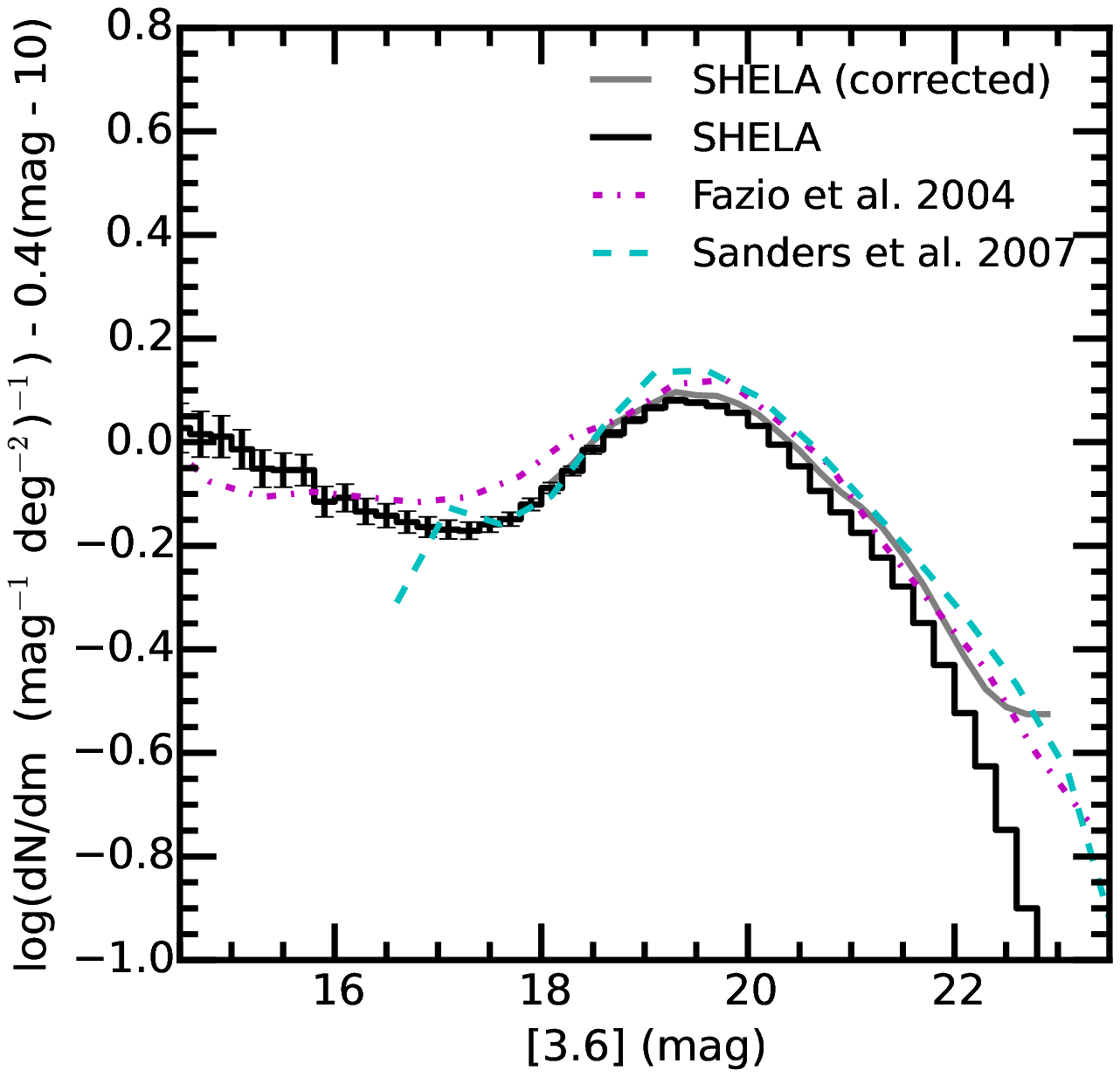}{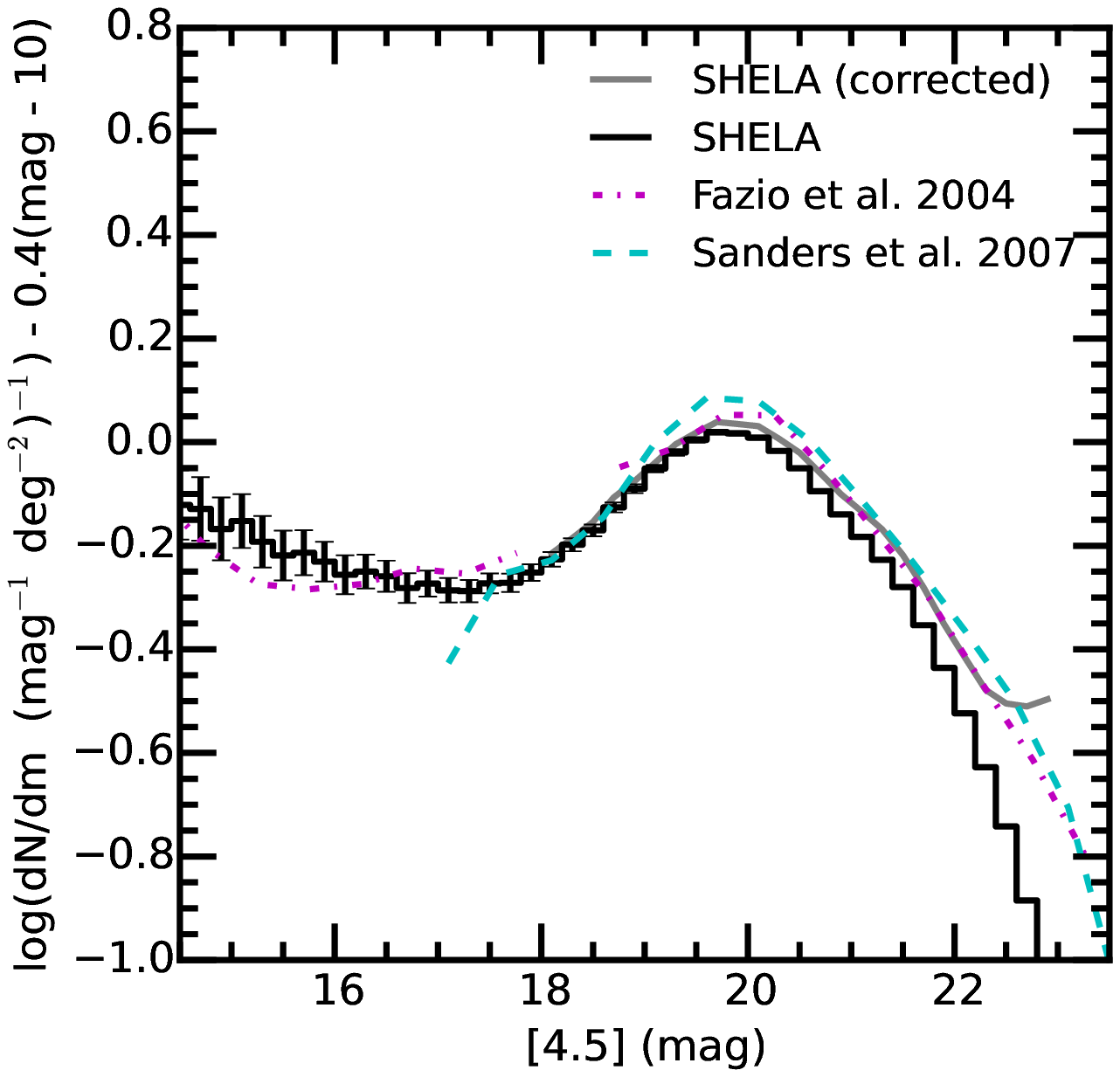}
\caption{\ed{The top panels show the differential number counts of
    IRAC sources in the SHELA field in bins of 0.2 mag.    The top-left panel shows the results for the IRAC
3.6~\micron\ data.  The top right panel shows the results for the
4.5~\micron\ data.   The bottom panels show the differential number
counts normalized to a Euclidean slope with an arbitrary offset
applied. }   In each panel the heavy black  histogram shows
the IRAC counts with no correction for completeness.   The error bars
show Poisson uncertainties on the number counts.   The gray line
shows the counts corrected for incompleteness.  For comparison the
dashed line shows counts from S-COSMOS \citep{sand07} and the
dot-dashed line shows counts from the IRAC GTO data
\citep{fazio04b}. \ed{The counts near the ``peak'' of the emission (in
  the lower panels) are consistent to better than 10\%.}}\label{fig:counts}
\epsscale{1}
 \ifsubmode
\end{figure}
\else
\end{figure*}
\fi
 
\subsection{Time-Variable Objects}

The combination of the large area and multi-epoch nature of the SHELA dataset allows
for the identification of sources whose brightness varies across the
$\sim$6 and 12 month baselines in time.   This includes rare sources that show
large changes in brightness,  and sources with high proper
motion \citep[see also,][]{ashby09}.   

\ifsubmode
\begin{figure} 
\epsscale{1.}
\else
\begin{figure*}  
\epsscale{1.0} 
\fi 
\plottwo{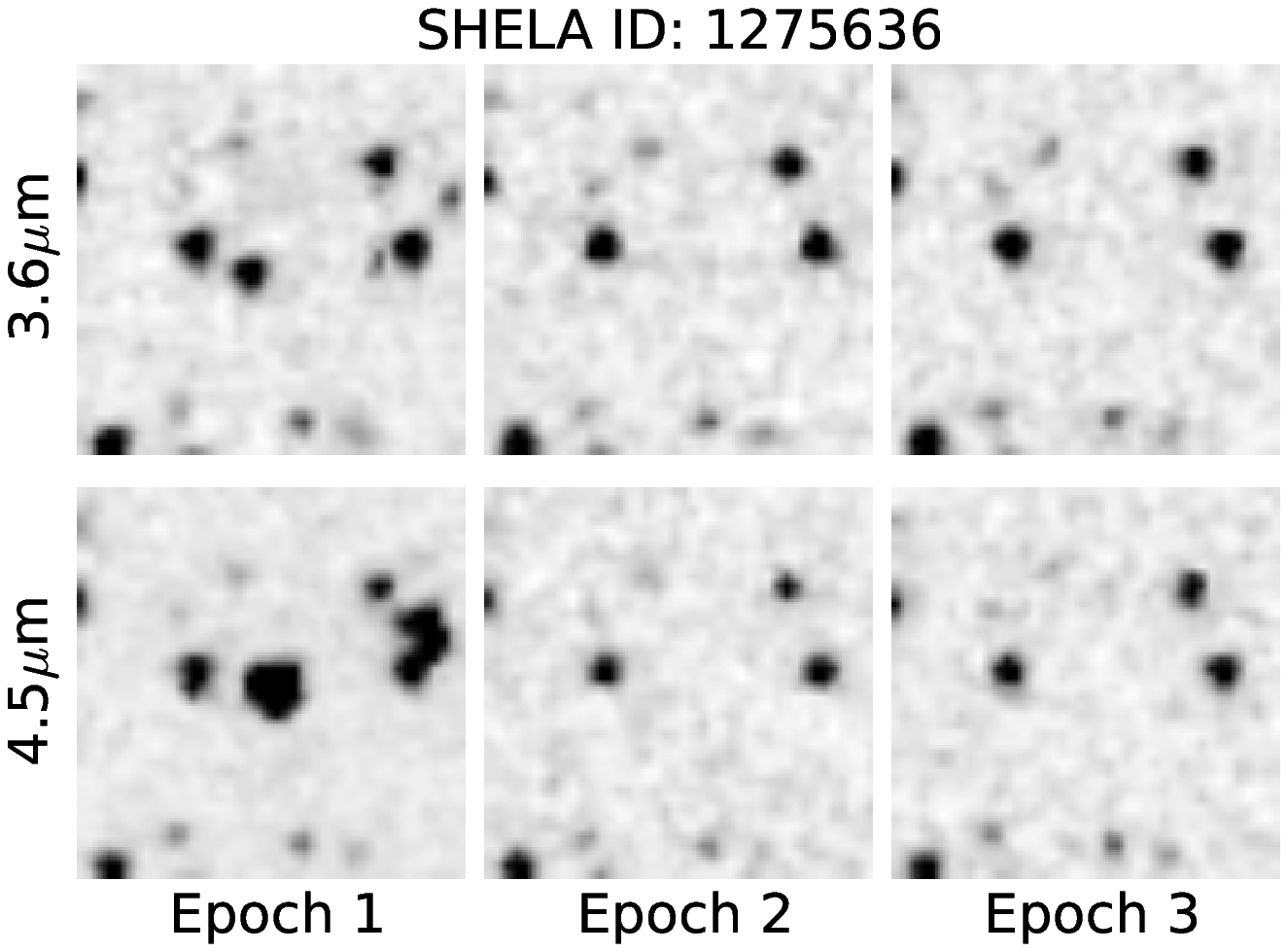}{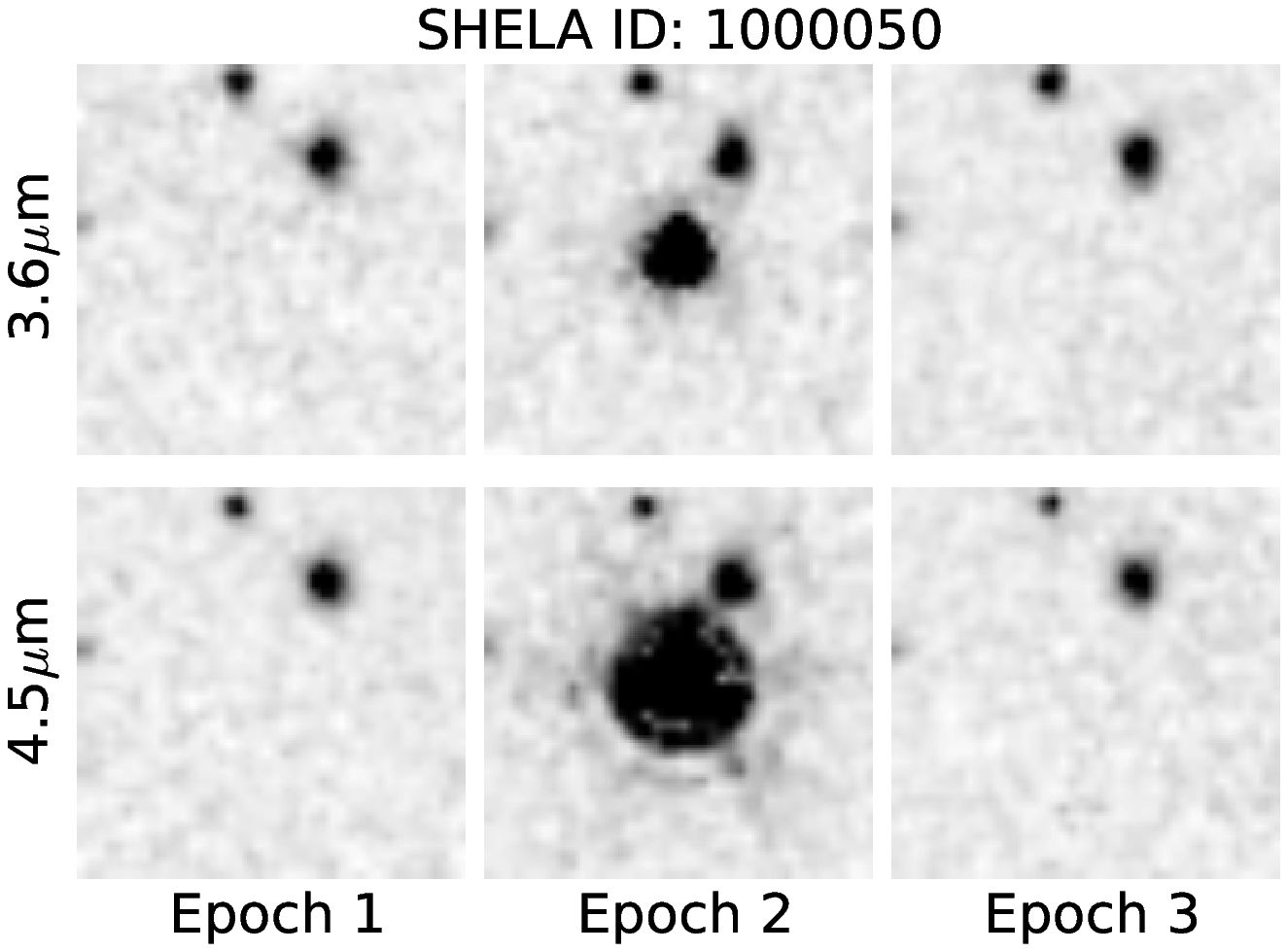}
\plottwo{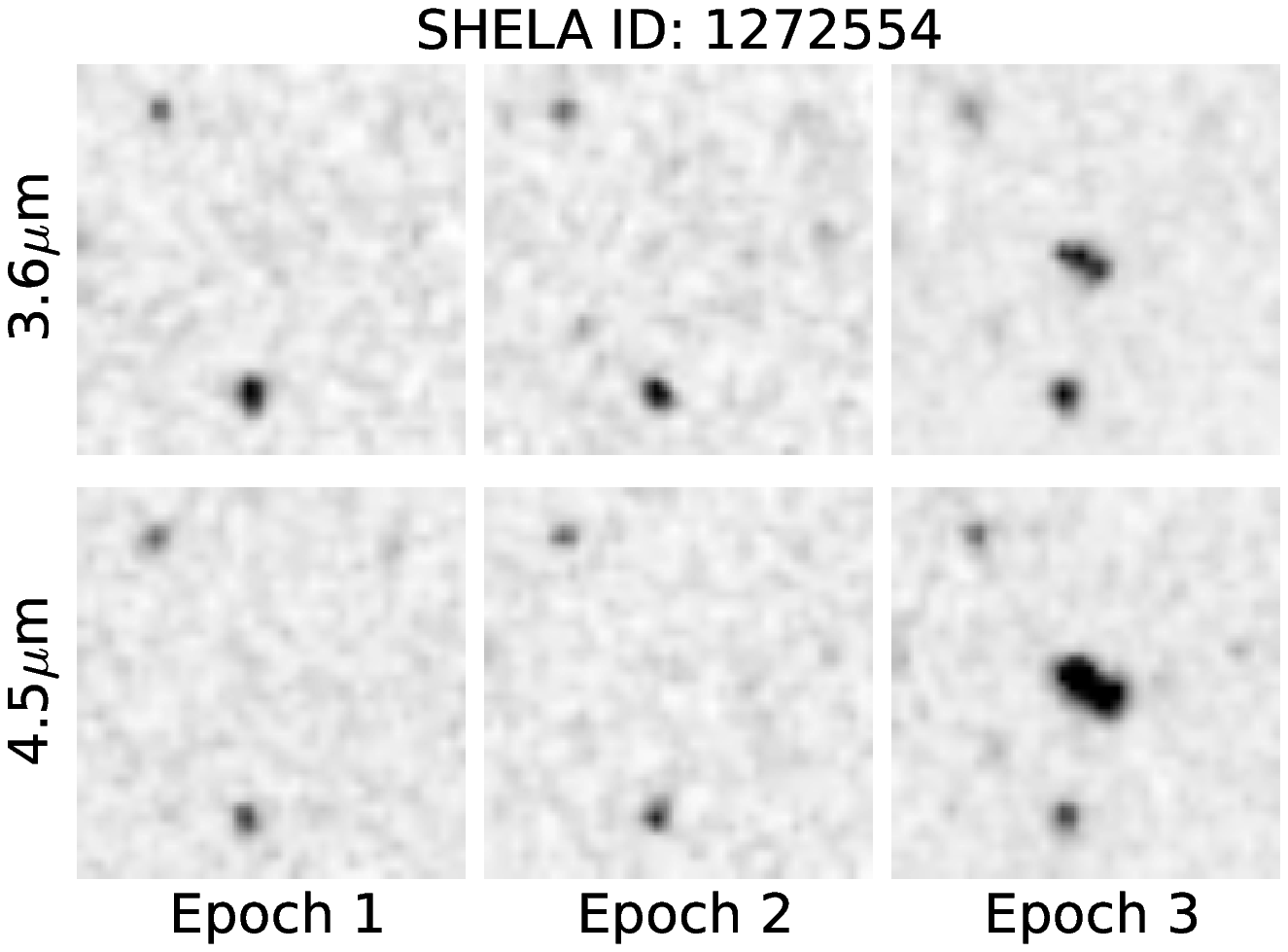}{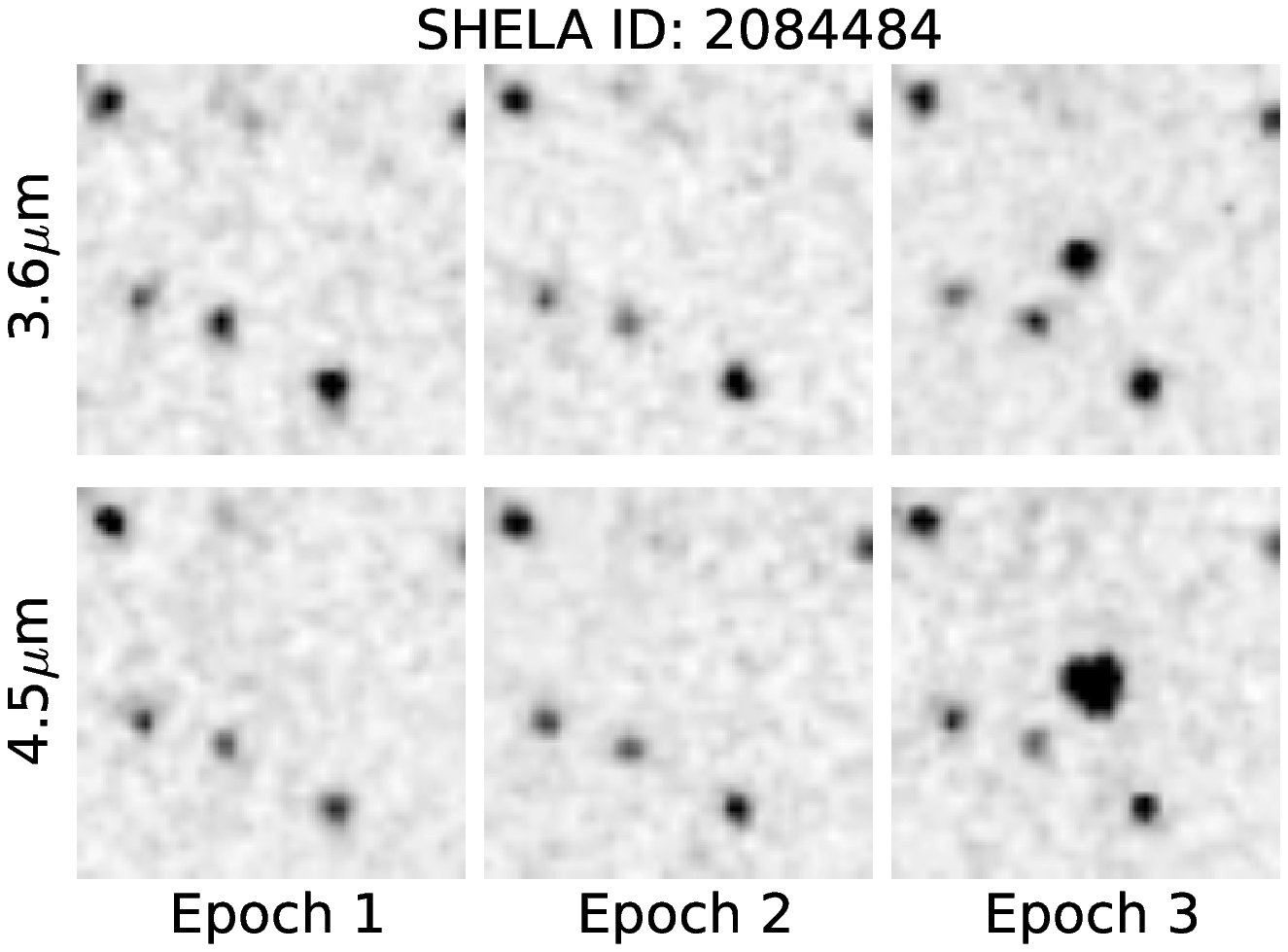}
\caption{Examples of sources that vary in brightness by more than 2.5
  magnitudes between the different SHELA observing epochs.  Each set
  of panels shows the 3.6 and 4.5~\micron\ images for epochs 1, 2, and
3 for 4 sources in the SHELA catalog.  The object catalog IDs are
given as the title for each set of plots. }\label{fig:timevary}
\epsscale{1}
 \ifsubmode
\end{figure}
\else
\end{figure*}
\fi
 
As an illustration, we selected objects from the SHELA catalog that
are detected in both the combined 3.6 and
4.5~\micron\ data, have coverage in all three observing epochs,
but vary by more than 2.5 mag (a factor of 10 in brightness) between
any two observing epochs.    There are 291 objects in the SHELA field
that satisfy these requirements with $[3.6] \leq 20.5$~AB mag or
$[4.5] \leq 20.5$~AB mag.  An inspection of these objects shows they
are all consistent with point sources, with several ``double''
(resolved, or multiple component) objects
and some objects that appear to show astrometric centroid shifts
between the 3.6 and 4.5~\micron\ image (which would imply very high proper
motions, indicative of asteroids).  Figure~\ref{fig:timevary} shows four objects that
appear in only a single observing epoch.   Because such objects make
it into the final, three-epoch, combined catalog, care must be taken
when creating object sets that require no (significant) temporal variations.

\subsection{The Relation between the Scale Radius and Mass of Dark-Matter Halos}

The ACT survey includes the SHELA IRAC footprint, and its catalog
includes SZ emission from distant ($z < 1$) clusters in the Stripe 82 field \citep{hass13}.    The
thermal SZ effect is a decrement in the emission from the CMB owing to the
presence of a massive (virialized) galaxy cluster along the line of
sight.  The hot ($T \sim 10^7-10^8$~K) ICM gas associated with the galaxy cluster causes
inverse Compton scattering of the CMB photons, leaving a distortion in
the direction of the cluster.  The strength of the distortion is
proportional to the line-of-sight integral of the thermal pressure
(the Compton $y$ parameter), which correlates with the total mass ($M_{500}$)
associated with the galaxy cluster \citep[\eg,][]{vikh09,marr12,sifon13}.

\ifsubmode
\begin{figure} 
\epsscale{1.1}
\else
\begin{figure*}  
\epsscale{1.2} 
\fi 
%
\epsscale{1.0}
\plottwo{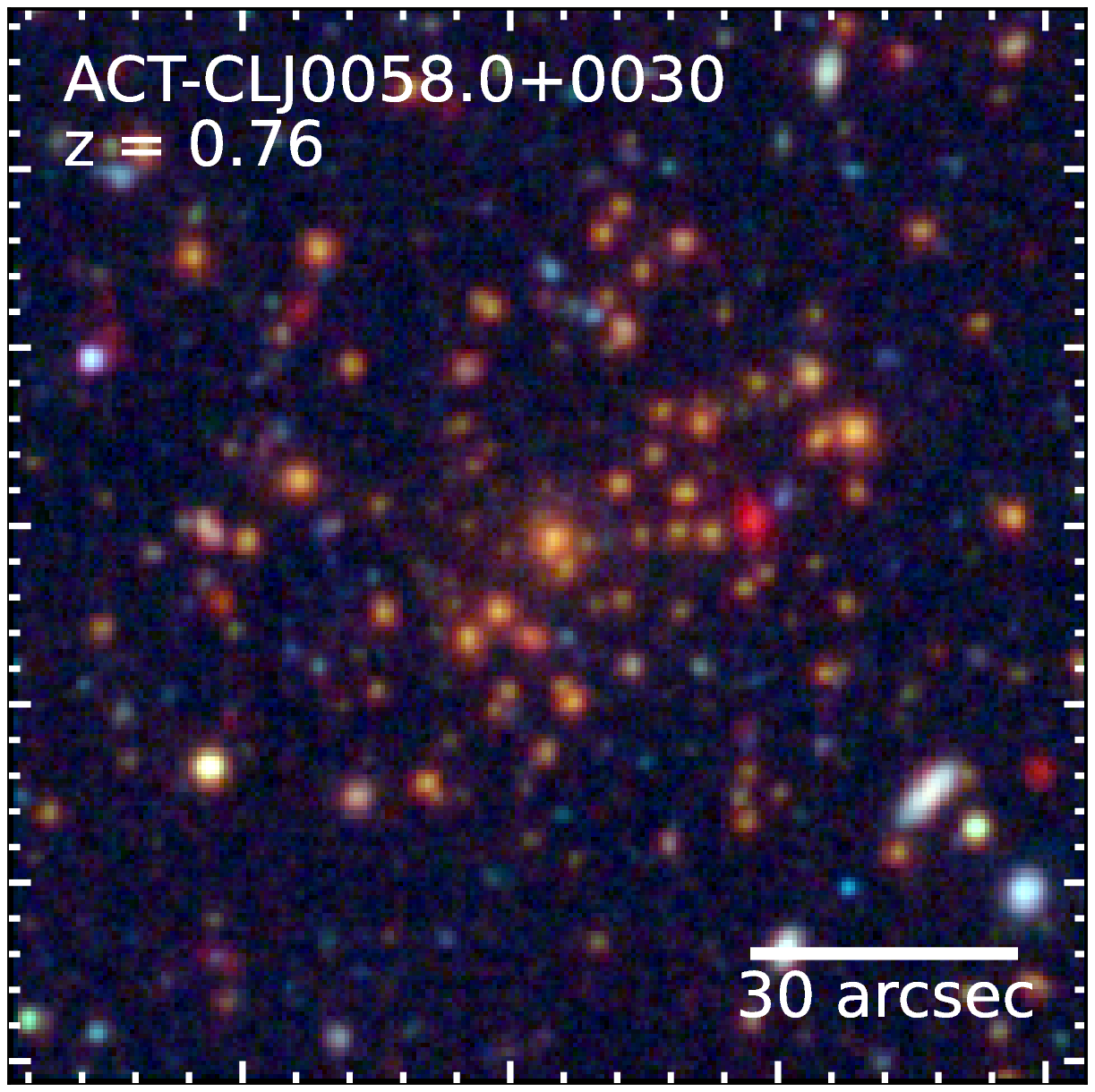}{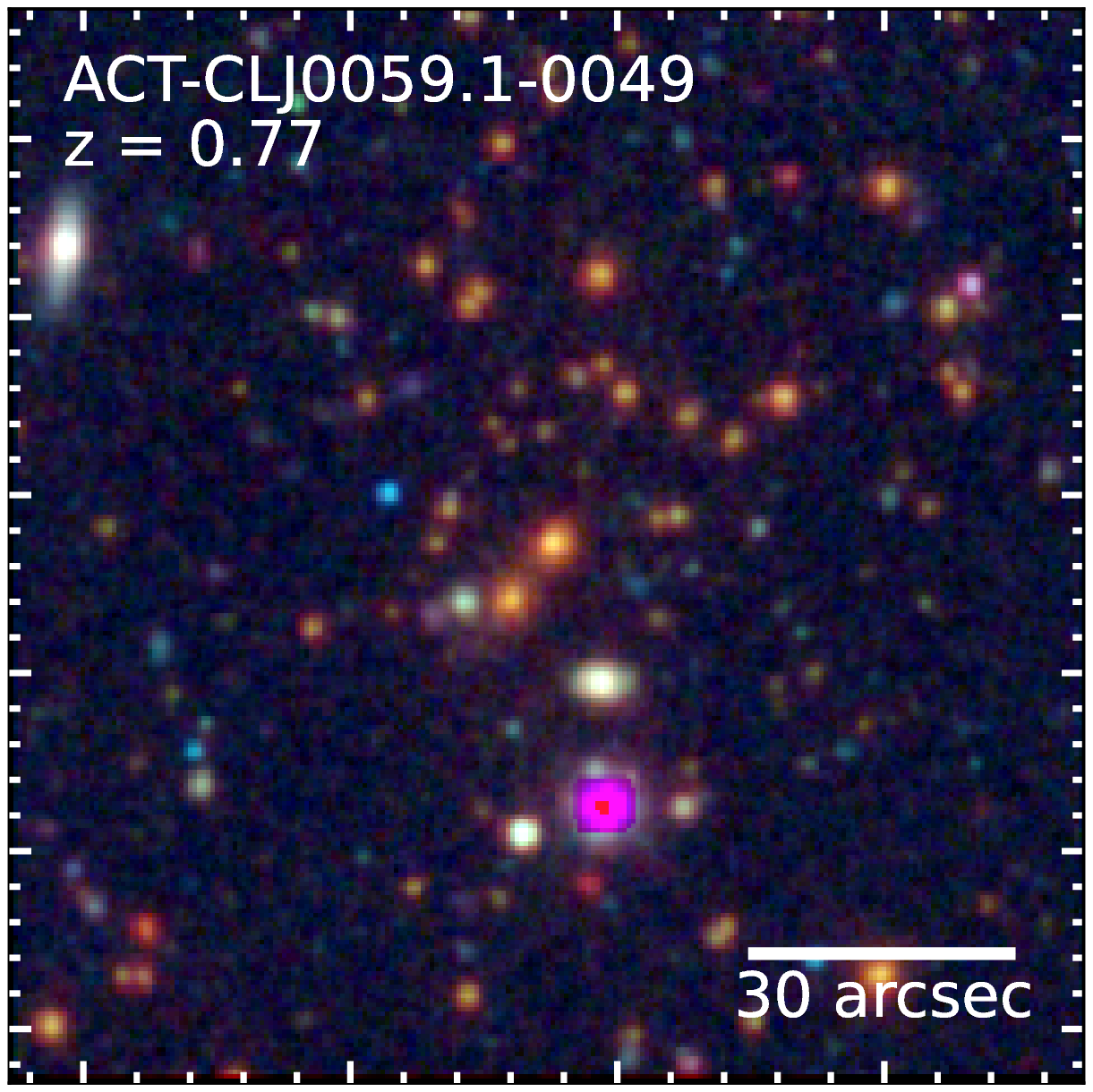}
\plottwo{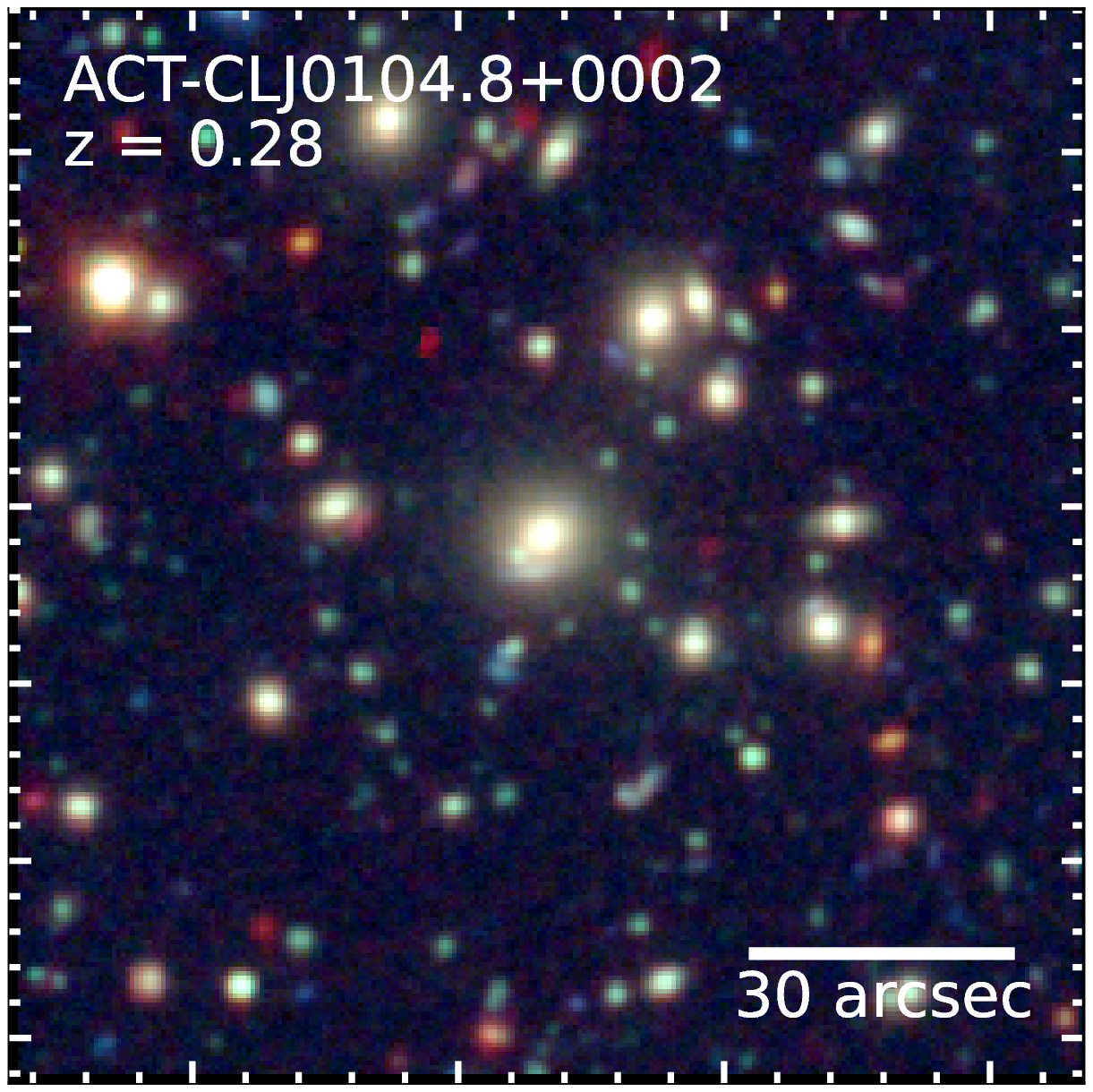}{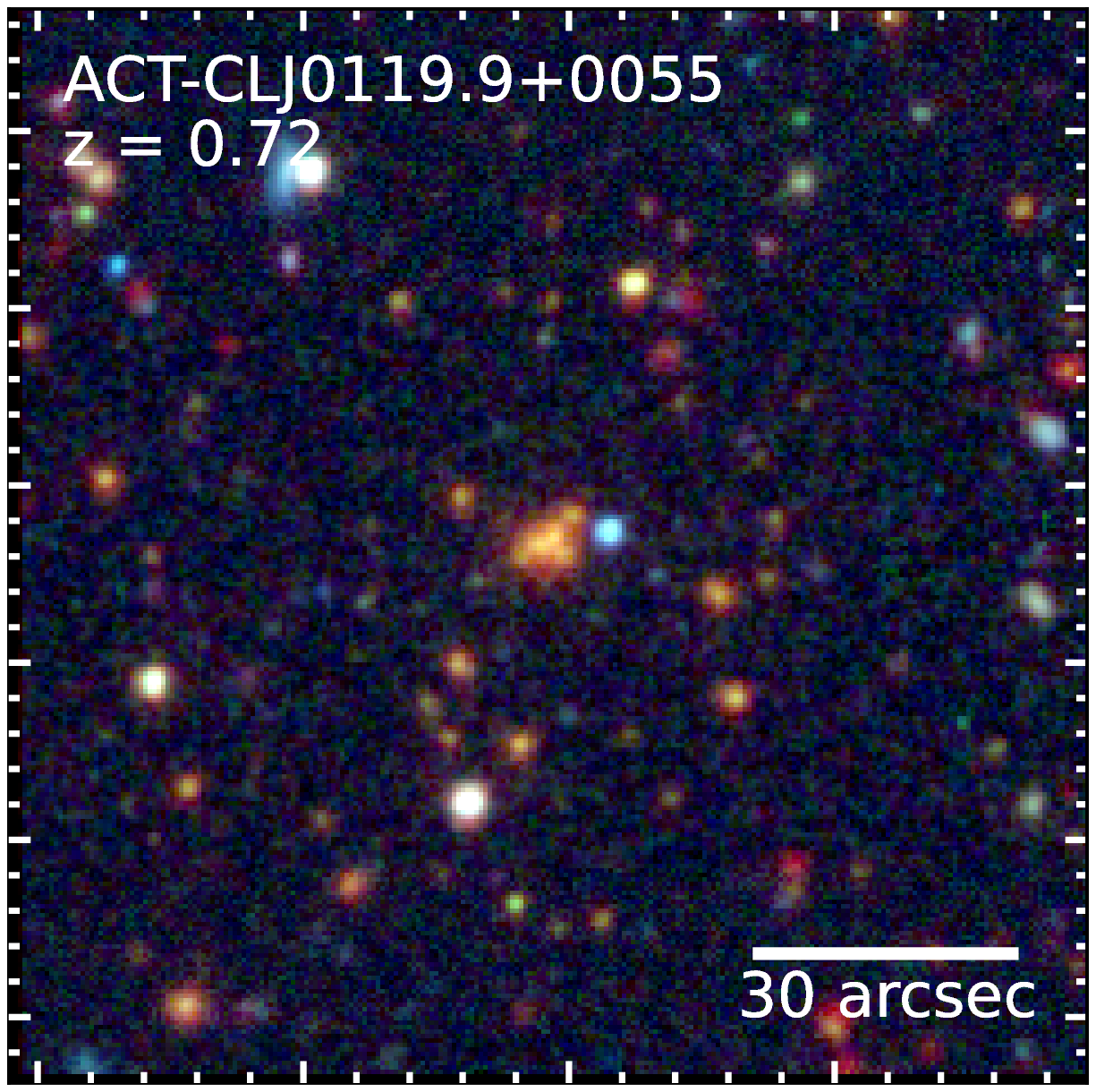}
\epsscale{0.55}
\plotone{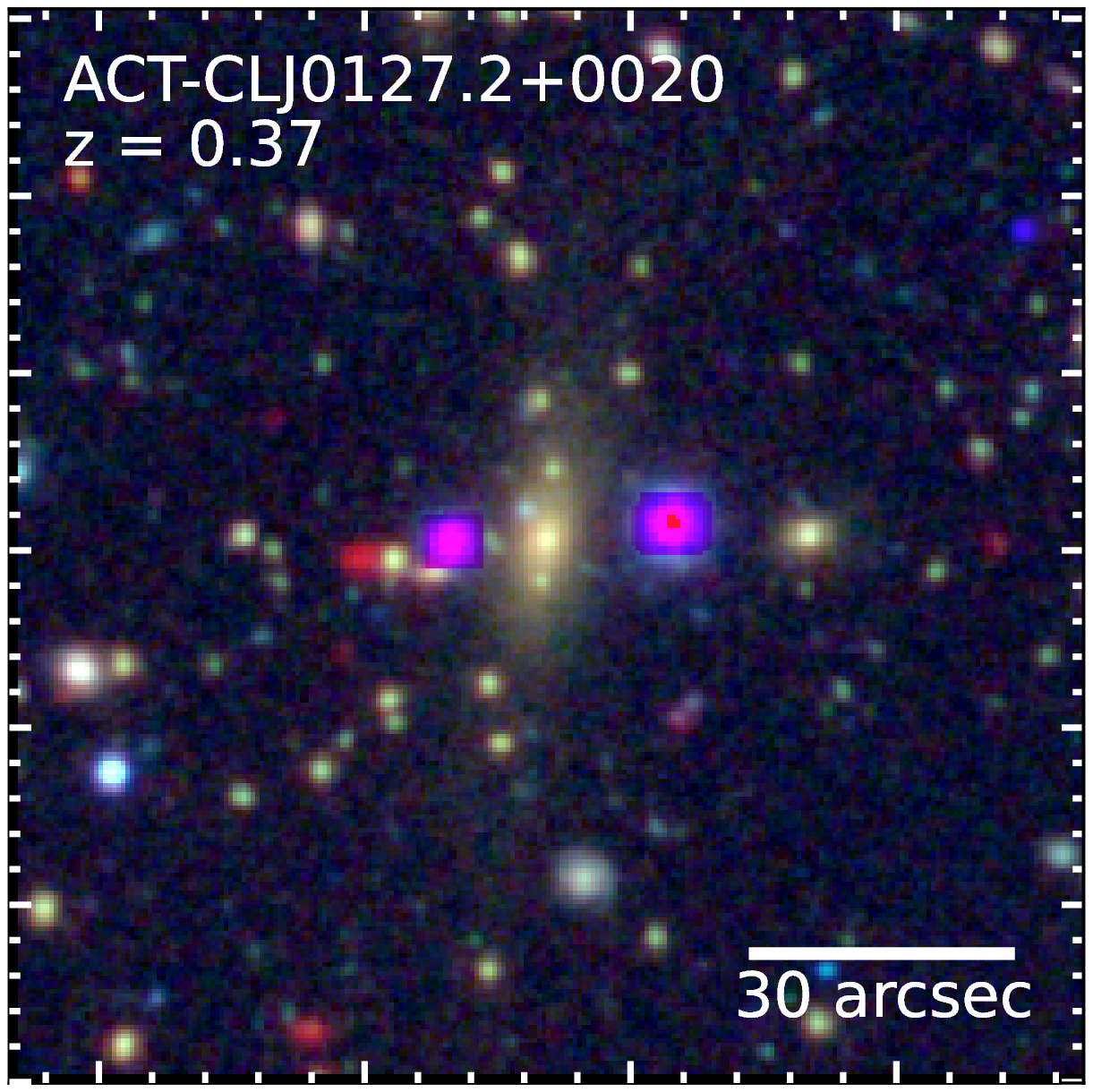}
\caption{\ed{False-color RGB images of the five SZ--selected galaxy
clusters detected in ACT that fall in the SHELA field.  Each image
shows a $2\arcmin \times 2\arcmin$ field centered on the astrometric
position of each ACT cluster (with North up and East to the left).
In each  panel, the red, green, and blue colors correspond to the
image from the SHELA 3.6~\micron\ image, and DECam $i$, and $g$--bands,
respectively. (The ``magenta'' objects are stars whose  data are
saturated in the DECam $i$-band images and have been masked.)}
}\label{fig:act_rgb}
\epsscale{1}
 \ifsubmode
\end{figure}
\else
\end{figure*}
\fi

\ifsubmode
\begin{figure} 
\epsscale{1.1}
\else
\begin{figure}[th]  
\epsscale{1.2} 
\fi 
\plotone{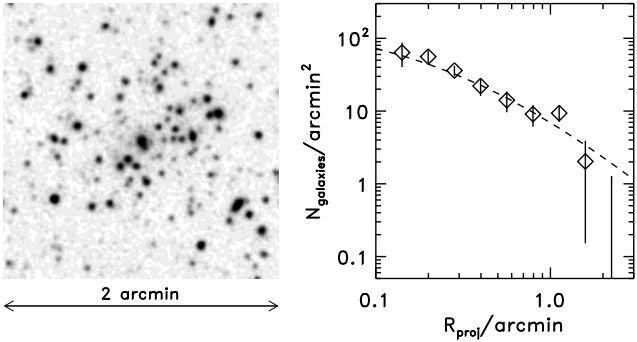}
\caption{The left panel shows the SHELA IRAC 3.6~\micron\ image of the
ACT SZ-selected cluster CLJ0058.0+0030 at $z=0.76$ \citep{hass13}.
The right panel shows the projected surface-density distribution of
galaxies centered on the peak of the SZ signal of the cluster.   The surface density
is the number of galaxies in the SHELA IRAC data measured in
concentric annuli centered on the cluster, corrected for the average
density of galaxies in random apertures in the IRAC image.  The
surface density is consistent with a projected NFW profile with scale
radius $0.51\pm 0.14$~arcmin.}\label{fig:cl0058}
\epsscale{1}
 \ifsubmode
\end{figure}
\else
\end{figure}
\fi

\ifsubmode
\begin{deluxetable}{lcccccc}
\else
\begin{deluxetable*}{lcccc}
\fi
\tablecolumns{5}
\tablewidth{\textwidth}
\tablecaption{Measures of Scale Radii of NFW profiles in ACT SZ Clusters in SHELA\label{table:rs}}
\tablehead{
  \colhead{Cluster} & 
  \colhead{z} &
  \colhead{$M_{500}$} & 
  \colhead{$\theta_s$} & 
  \colhead{$r_s$} \\
  \colhead{} & 
  \colhead{} &
  \colhead{($10^{14} M_\odot)$} & 
  \colhead{(arcmin)} & 
  \colhead{(kpc)} \\
\colhead{(1)} & 
\colhead{(2)} & 
\colhead{(3)} & 
\colhead{(4)}  & 
\colhead{(5)} }
\startdata
ACT CLJ0104.8+0002    &  0.28 & 2.6 $\pm$     0.9 &      1.1\phn $\pm$ 0.4\phn &  280  $\pm$ 100  \\
ACT CLJ0127.2+0020   &   0.37  &     3.3   $\pm$    0.9  & 0.67 $\pm$ 0.24   & 205 $\pm$ \phn72 \\
ACT CLJ0119.9+0055   &  0.72 &  3.3   $\pm$    0.8   &   0.72 $\pm$ 0.22 & 311 $\pm$ \phn96 \\
ACT CLJ0058.0+0030    &  0.76 &  3.2   $\pm$    0.8   &  0.51 $\pm$ 0.15 &  223 $\pm$ \phn65 \\
ACT CLJ0059.1$-$0049   & 0.77 &  5.2 $\pm$   0.9   &   0.72 $\pm$ 0.19 &  318 $\pm$ \phn83
\enddata
\tablecomments{(1) Cluster designation \citep[from][]{hass13}, (2) estimated redshift \citep[from][]{hass13}, (3) total cluster mass derived from the SZ signal \citep[from][]{hass13}, (4) angular scale radius $\theta_s$ of the projected NFW profile fit to the background-corrected surface density of galaxies in the SHELA IRAC data centered on each cluster, (5) scale radius converted to physical units assuming $h=0.7$, $\Omega_m=0.3$, and $\Omega_\Lambda=0.7$.}
\ifsubmode
\end{deluxetable}
\else
\end{deluxetable*}
\fi

\ifsubmode
\begin{figure} 
\epsscale{1.1}
\else
\begin{figure}  
\epsscale{1.2} 
\fi 
\plotone{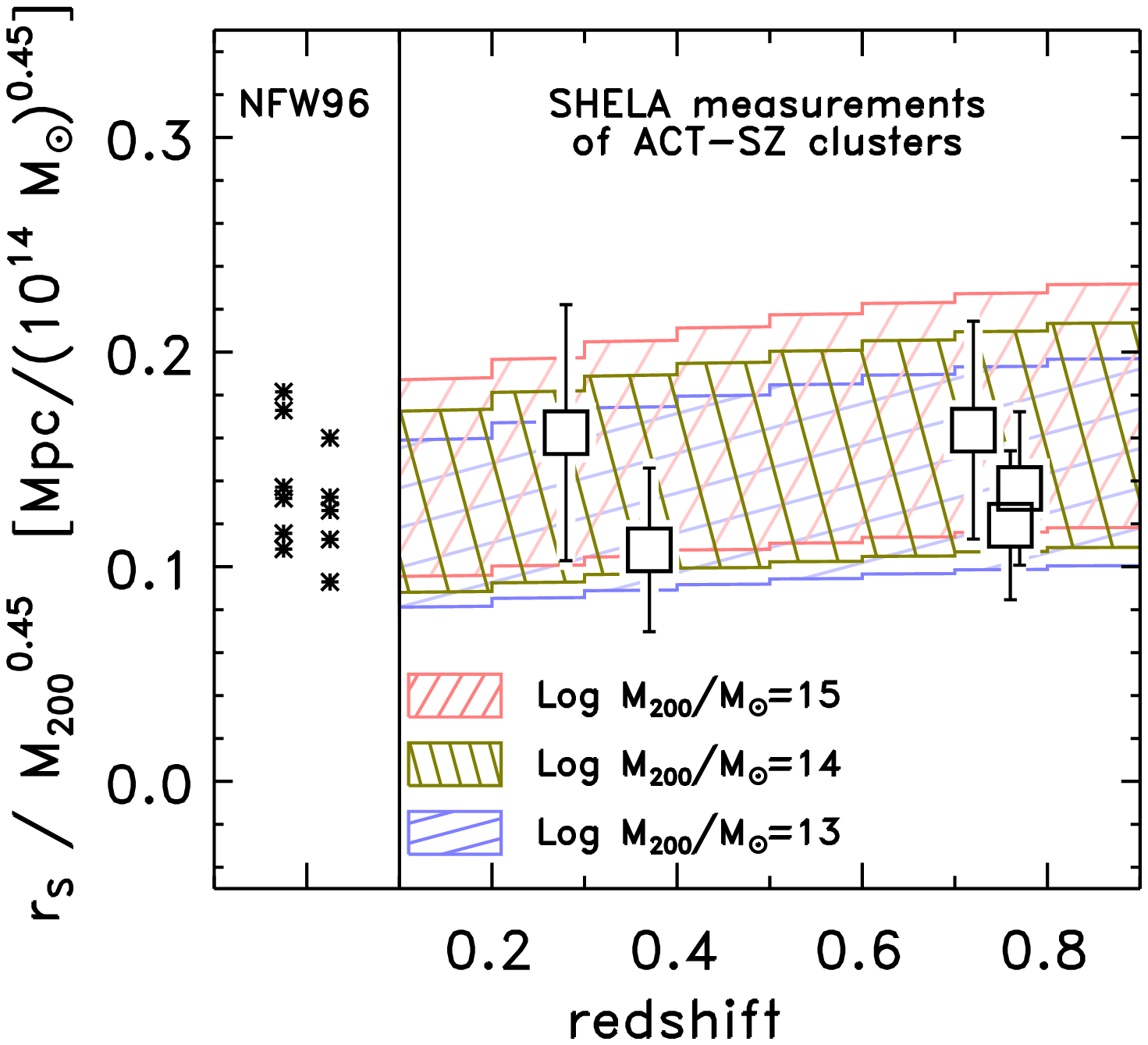}
\caption{The evolution of the ratio between the NFW-profile scale
radius, $r_s$ and halo mass, $M_{200}$.  The boxes and error bars show
values derived for the ACT-selected clusters in SHELA.  The scale
radii are measured by fitting projected NFW profiles to the surface
density of galaxies in each ACT cluster.   The asterisks show the
predicted ratio of $r_s/M_{200}^{0.45}$ for galaxy halos from NFW96.
The hatched swaths show the expected ratio for halos of $\log
M_{200}/\msol = 13$, 14, and 15 (as labeled in the figure legend)
including the scatter in halo concentration using the relations in
\citet{diem15}.  The data are consistent with a near unevolving ratio
of $r_s/M^{0.45}$ over a large baseline in redshift, with values
consistent with DM halo scaling relations as expected from the
distribution of predicted ratios for a CDM--type cosmology.
}\label{fig:zrm}
\epsscale{1}
 \ifsubmode
\end{figure}
\else
\end{figure}
\fi
 
Five of the ACT clusters from \citet{hass13}  fall in the SHELA/IRAC
footprint (ACT CLJ0059.1-0049, ACT CLJ0127.2+0020,  ACT
CLJ0119.9+0055,  ACT CLJ0058.0+0030,  ACT CLJ0104.8+0002).    The IRAC
data probe the amount of starlight associated with the galaxies  in
these clusters, and measure the galaxy spatial distribution.  The
combination of IRAC and ACT data therefore allows us to study the
structural size of the dark matter halo (as traced by the galaxies in
the IRAC image) and compare it with the halo mass as estimated from the SZ
signal.  \ed{Figure~\ref{fig:act_rgb} shows false-color images of the five
clusters in the SHELA IRAC 3.6~\micron\ image (red-color) combined
with $i$-- and $g$--band images from our DECam imaging (Wold et al.,
in preparation). }

Studies have shown that the surface density of satellites roughly
trace the distribution of dark matter
\citep[\eg,][]{tal12,kawi14,vanderburg15}  predicted by the density
profile of the dark matter from numerical simulations \citep[\eg,][NFW
hereafter]{nava96}.  The Spitzer/IRAC data allow the measurement of
the radial distribution of galaxies, and therefore a tracer of the
dark matter density distribution.  Following the method of
\citet{kawi14}, we counted the number of SHELA galaxies with $17 <
m_{3.6} < 22$~mag in concentric annuli centered on each cluster.  To
correct for the background, we measured the average (median) number of
galaxies in each annulus for 10$^4$ randomly placed apertures around
the SHELA image (taking care to avoid the image edges).     We then
measured the radial profile, and fit the projected NFW profile
\citep{bart96} using two parameters,  the NFW scale angular radius,
$\theta_s$, and a normalization.    The results from these fits for
the NFW scale radii for each cluster are given in
Table~\ref{table:rs}.     

Figure~\ref{fig:cl0058} shows the distribution of galaxies centered on
one of the ACT clusters in SHELA, ACT CLJ0058.0+0030 at $z=0.76$.  The
galaxy distribution has been corrected statistically
for the galaxies associated with the field as discussed above.     The figure shows that
the surface density of galaxies follows a projected NFW profile
with a best fit scale radius, $\theta_s =
0.51\pm 0.14$~arcmin, which corresponds to a physical scale radius of
$r_s = 223 \pm 65$~kpc (for $h=0.7$, $\Omega_m=0.3$, and
$\Omega_\Lambda = 0.7$).     

The combination of data on the radial distribution of cluster galaxies
and measures of the total mass of the clusters  is a potentially
powerful way  to study properties of dark matter halos.  For an NFW
profile, the scale radius is expected to increase with halo mass as
$r_s \sim M_{200}^{0.45}$ \citep{nava96}, where $M_{200}$ is the mass
within a radius where the density is 200 times the critical density.
Simulations predict that this relation should be constant constant with
redshift \citep[\eg,][]{bull01,eke01,diem15}.

Figure~\ref{fig:zrm} shows the ratio of $r_s / M_{200}^{0.45}$ for the
five ACT SZ clusters in the SHELA field as a function of their
redshift.  The $r_s$ values come from the projected NFW profile fits
to the radial distributions of galaxies for each cluster from the IRAC
catalogs (as described for CLJ0058.0+0030 in the previous paragraph).
The  $M_{200}$ values come from the estimates of $M_{500}$ from the SZ
$y$-parameter measurements \citep{hass13}, where we have adjusted the
$M_{500}$ values by 0.1 dex to convert them to $M_{200}$.
Figure~\ref{fig:zrm} also shows the values of $r_s/M_{200}^{0.45}$ for
the simulated model halos from \citet{nava96}, for halos ranging from
$M_{200} = 10^{13}-10^{15}$~\msol.  The values range from
$r_s/M_{200}^{0.45} \simeq 0.1-0.2$ Mpc / ($10^{14}$ \msol)$^{0.45}$,
and agree remarkably well with the observations for the 5 ACT-SZ
clusters.  These are consistent with the expected evolution of more
modern simulations for a halo of mass $10^{13}-10^{15}$~\msol\
\citep{diem15}.  The observations show indications that the shape of
the dark--matter profiles has only a weak dependence on mass and
redshift in accordance with predictions from $\Lambda$CDM
\citep{bull01}.

\section{Summary}\label{section:summary}

We presented the \spitzer\ IRAC imaging at 3.6 and 4.5~\micron\ of the
SHELA survey, a \spitzer\ Exploratory program which covers a
$\approx$24 deg$^2$ field within the footprint of HETDEX.   This field
has a rich set of multiwavelength data, including optical imaging from
SDSS Stripe 82 and  CTIO/DECam, near-IR imaging from NEWFIRM $K$-band,
far-IR imaging from \herschel, and X--ray observations from Chandra
and $XMM$-Newton.

The HETDEX survey will obtain redshifts in this field for $\sim
200,000$ galaxies at $1.9 < z < 3.5$ based on Lyman-$\alpha$ emission
(covering a volume of 0.5 Gpc$^3$), and redshifts for an additional
$\sim 200,000$ galaxies at $z < 0.5$ based on their [\ion{O}{2}]
emission.  The SHELA IRAC data are sensitive to galaxies with stellar
masses down to $\simeq 2\times 10^{10}$~\msol\ through the redshift
range of Lyman-$\alpha$ probed by HETDEX.   Thus, the combination of
the HETDEX spectroscopy data, ground-based optical/near-IR imaging,
and the SHELA IRAC data allow the study of the relationship between
structure formation, galaxy stellar mass, dark halo mass,  and
environment during over a large range of cosmic history. 

In this \textit{Paper} we discussed the properties of the SHELA IRAC
data, including the data acquisition, reduction, validation, and
source catalogs.    The imaging includes three observing epochs
separated by approximately 6 months between epochs.   The combined
three-epoch dataset covers 24.2 deg$^2$ with an exposure time of
at least $\approx$200~s.  Our tests show the images and catalogs are 80\%
(50\%) complete to limiting magnitudes of 22.0 (22.6) AB  mag in the
detection image, which is constructed from the weighted-sum of the IRAC 3.6 and
4.5~\micron\ images.  The catalogs reach limiting ($1\sigma$)
sensitivities of 1.1~\ujy\ in each IRAC channel.    The photometric
accuracy is consistent with AllWISE with essentially no different
between the \mone\ and W1 bands and a possible 0.02~mag offset between
\mtwo\ and W2 bands.  

The astrometric solution of SHELA is tied to SDSS DR7, where
the astrometric uncertainty is $<$$0\farcs2$, comparable to the
uncertainty in the SDSS catalogs.   The astrometric solutions are
accurate compared to 2MASS, but show a (possibly) non-negligible
offset compared to AllWISE approaching one-tenth of an arcsecond.  

The IRAC data enable a broad range of scientific explorations,
including studies of galaxy and AGN evolution, and the formation of
large-scale structure.  As a demonstration of science, we present IRAC
number counts, examples of highly temporally variable sources, and
galaxy surface density profiles of rich galaxy clusters.    At faint
magnitudes, the source number counts are consistent with other IRAC
datasets, which provides confidence in our estimated completeness
corrections.   At bright magnitudes we observe a possible excess of
counts, which we attribute to variations in the surface density of
Galactic stars. 

We use a sample of five ACT SZ--selected galaxy clusters between $0.2 < z <
0.8$ to study the relation between cluster mass
(traced by the SZ Compton parameter) and the scale radius, $r_s$, of
the cluster halos as traced by the surface distribution of galaxies as
measured from the SHELA IRAC data.   All clusters show galaxy surface
densities in agreement with a projected NFW halo, with a ratio of
$r_s/M_{200}^{0.45}$ that is consistent with simulations of dark
matter halos and is unevolving in redshift, as predicted by
$\Lambda$CDM models.  

In the spirit of \spitzer\ Exploratory programs we provide all images
and catalogs as part of this \textit{Paper}.  We describe the source IRAC
catalogs and imaging products released for the SHELA data, which will
be available through IRSA  (see
footnote~\ref{footnote:IRSA}).   

\acknowledgments

We acknowledge useful conversations, comments and suggestions,
especially from Matt Ashby, Ivo Labb\'e, Gordon Richards, Nick Ross,
Eli Rykoff, Louis Strigari, and John Timlin.  We thank Benedikt Diemer
for providing his code to generate halo density profiles\ed{, and Justin
Howell for comments and help incorporating the SHELA data products into IRSA.  We also
thank the anonymous referee for a thorough reading and constructive
report.}  CP thanks the Space Telescope Science Institute for its
hospitality and for providing a scientifically conducive atmosphere
during the completion of this work.  This work is based on
observations made with the Spitzer Space Telescope, which is operated
by the Jet Propulsion Laboratory, California Institute of Technology
under a contract with NASA.   This work is supported by the National
Science Foundation through grant AST-1413317.  We acknowledge generous
support from the Texas A\&M University and the George P.\ and Cynthia
Woods Institute for Fundamental Physics and Astronomy.   GB is supported by
CONICYT/FONDECYT, Programa de Iniciacion, Folio 11150220. RC, CG, and
GZ acknowledge support from the Institute for Gravitation and the
Cosmos.  The Institute for Gravitation and the Cosmos is supported by
the Eberly College of Science and the Office of the Senior Vice
President for Research at the Pennsylvania State University.

This work made use of data from SDSS.  Funding for the SDSS and
SDSS-II has been provided by the Alfred P. Sloan Foundation, the
Participating Institutions, the National Science Foundation, the
U.S. Department of Energy, the National Aeronautics and Space
Administration, the Japanese Monbukagakusho, the Max Planck Society,
and the Higher Education Funding Council for England. The SDSS Web
Site is \myhref{http://www.sdss.org}. 

The SDSS is managed by the Astrophysical Research Consortium for the
Participating Institutions. The Participating Institutions are the
American Museum of Natural History, Astrophysical Institute Potsdam,
University of Basel, University of Cambridge, Case Western Reserve
University, University of Chicago, Drexel University, Fermilab, the
Institute for Advanced Study, the Japan Participation Group, Johns
Hopkins University, the Joint Institute for Nuclear Astrophysics, the
Kavli Institute for Particle Astrophysics and Cosmology, the Korean
Scientist Group, the Chinese Academy of Sciences (LAMOST), Los Alamos
National Laboratory, the Max-Planck-Institute for Astronomy (MPIA),
the Max-Planck-Institute for Astrophysics (MPA), New Mexico State
University, Ohio State University, University of Pittsburgh,
University of Portsmouth, Princeton University, the United States
Naval Observatory, and the University of Washington.

This research has made use of the NASA/ IPAC Infrared Science Archive,
which is operated by the Jet Propulsion Laboratory, California
Institute of Technology, under contract with the National Aeronautics
and Space Administration.  

This research has made use of the VizieR catalogue access tool, CDS,
Strasbourg, France.

This publication makes use of data products from the Two Micron All
Sky Survey, which is a joint project of the University of
Massachusetts and the Infrared Processing and Analysis
Center/California Institute of Technology, funded by the National
Aeronautics and Space Administration and the National Science
Foundation. 

This publication makes use of data products from the Wide-field
    Infrared Survey Explorer, which is a joint project of the University of
    California, Los Angeles, and the Jet Propulsion Laboratory/California
    Institute of Technology, funded by the National Aeronautics and Space
    Administration.

\medskip

\bibliography{alpharefs}{}
\bibliographystyle{apj}


\setcounter{table}{6}
\clearpage
\begin{turnpage}

\begin{deluxetable}{ccccccccccccccccc}
\scriptsize
\tablecaption{Preamble for all SHELA IRAC Catalogs\label{table:preamblecatalog}}

\tablehead{\colhead{ID} & \colhead{X} & \colhead{Y} &
  \colhead{RA(J2000)} & \colhead{DEC(J2000)} & \colhead{Isophotal
    Area} & \colhead{$a$} & \colhead{$e$} & \colhead{$\theta$} &
  \colhead{W(3.6)} & \colhead{W(4.5)} & \colhead{W(3.6)$_1$} &
  \colhead{W(4.5)$_1$} & \colhead{W(3.6)$_2$} & \colhead{W(4.5)$_2$} &
  \colhead{W(3.6)$_3$} & \colhead{W(4.5)$_3$}\\ \colhead{ } &
  \colhead{(pixel)} & \colhead{(pixel)} & \colhead{(deg)} &
  \colhead{(deg)} & \colhead{(arcsec$^2$)} & \colhead{(arcsec)} &
  \colhead{ } & \colhead{(deg)} & \colhead{ } & \colhead{ } &
  \colhead{ } & \colhead{ } & \colhead{ } & \colhead{ } & \colhead{ }
  & \colhead{ } \\ 
\colhead{(1)} & \colhead{(2)} & \colhead{(3)} & \colhead{(4)} &
\colhead{(5)} & \colhead{(6)} & \colhead{(7)} & \colhead{(8)} &
\colhead{(9)} & \colhead{(10)} & \colhead{(11)} & \colhead{(12)} &
\colhead{(13)} & \colhead{(14)} & \colhead{(15)} & \colhead{(16)} &
\colhead{(17)}  }
\startdata
100020 & 45386.8 & 1082.6 & 17.701277 & -1.132968 & 125.4 & 1.9 & 0.04 & 35.3 & 2.92 & 6.03 & 0.00 & 1.02 & 2.92 & 3.00 & 0.00 & 2.00 \\
100021 & 18859.7 & 1115.1 & 23.590947 & -1.125582 & 7.0 & 0.8 & 0.25 & -13.1 & 8.01 & 10.76 & 3.00 & 3.01 & 2.02 & 4.77 & 2.99 & 2.99 \\
100022 & 19340.1 & 1112.4 & 23.484482 & -1.126288 & 12.2 & 1.1 & 0.26 & -0.7 & 6.32 & 6.03 & 3.28 & 2.99 & 0.00 & 0.00 & 3.04 & 3.05 \\
100023 & 19528.0 & 1112.8 & 23.442827 & -1.126232 & 12.2 & 1.1 & 0.46 & -29.2 & 6.03 & 7.06 & 3.04 & 3.03 & 0.00 & 0.00 & 2.99 & 4.03 \\
100024 & 56605.8 & 1102.2 & 15.221331 & -1.125113 & 26.9 & 1.2 & 0.11 & -53.5 & 4.08 & 9.01 & 0.00 & 2.98 & 4.08 & 3.03 & 0.00 & 3.01 \\
100025 & 22941.5 & 1112.9 & 22.685727 & -1.126878 & 7.0 & 0.8 & 0.31 & -41.1 & 11.12 & 9.05 & 5.03 & 2.99 & 3.04 & 3.01 & 3.05 & 3.05 \\
100026 & 21391.8 & 1115.2 & 23.029557 & -1.126084 & 3.8 & 0.6 & 0.25 & -47.4 & 3.04 & 6.79 & 0.00 & 2.77 & 3.04 & 0.00 & 0.00 & 4.01 \\
100027 & 45127.6 & 1104.6 & 17.758739 & -1.128140 & 25.6 & 1.3 & 0.14 & -32.0 & 7.18 & 7.15 & 3.00 & 4.11 & 4.17 & 3.03 & 0.00 & 0.00 \\
100028 & 28402.5 & 1118.5 & 21.473082 & -1.126254 & 5.8 & 0.8 & 0.56 & -18.8 & 2.99 & 6.15 & 0.00 & 3.14 & 2.99 & 0.00 & 0.00 & 3.01 \\
100029 & 35643.7 & 1111.3 & 19.864062 & -1.127908 & 12.2 & 1.0 & 0.19 & -32.0 & 0.00 & 8.06 & 0.00 & 5.05 & 0.00 & 0.00 & 0.00 & 3.01 \\
100030 & 22402.3 & 1098.1 & 22.805387 & -1.130069 & 30.1 & 1.9 & 0.42 & -65.9 & 8.00 & 9.09 & 3.02 & 3.11 & 2.99 & 2.99 & 1.99 & 2.99 \\
100031 & 26946.7 & 1113.5 & 21.796466 & -1.127243 & 6.4 & 0.8 & 0.30 & -89.2 & 3.00 & 3.01 & 3.00 & 3.01 & 0.00 & 0.00 & 0.00 & 0.00 \\
100032 & 8436.4 & 1095.7 & 25.894587 & -1.126576 & 62.7 & 2.1 & 0.23 & -7.7 & 9.07 & 6.03 & 6.03 & 2.98 & 0.00 & 0.00 & 3.04 & 3.05 \\
100033 & 28542.3 & 1118.6 & 21.442033 & -1.126240 & 3.2 & 0.6 & 0.39 & -16.6 & 2.99 & 6.02 & 0.00 & 3.03 & 2.99 & 0.00 & 0.00 & 2.99 \\
100034 & 27031.2 & 1113.5 & 21.777696 & -1.127268 & 5.1 & 0.9 & 0.55 & 67.7 & 3.00 & 3.38 & 3.00 & 3.38 & 0.00 & 0.00 & 0.00 & 0.00 \\
100035 & 29155.7 & 1118.3 & 21.305746 & -1.126346 & 1.9 & 0.5 & 0.40 & 44.8 & 14.88 & 6.01 & 3.02 & 3.00 & 8.84 & 0.00 & 3.02 & 3.00 \\
100036 & 41240.4 & 1091.2 & 18.621071 & -1.131813 & 84.5 & 1.8 & 0.24 & 34.9 & 9.07 & 10.76 & 3.03 & 2.41 & 3.04 & 3.02 & 3.00 & 5.34 \\
100037 & 33861.8 & 1108.3 & 20.259993 & -1.128661 & 16.6 & 1.1 & 0.16 & -23.1 & 9.04 & 8.14 & 3.03 & 2.11 & 2.99 & 2.99 & 3.02 & 3.04 \\
100038 & 40805.0 & 1105.7 & 18.717715 & -1.128666 & 21.8 & 1.1 & 0.14 & 59.5 & 7.94 & 4.25 & 3.03 & 2.03 & 2.88 & 0.00 & 2.02 & 2.22 \\
100039 & 50844.3 & 1103.6 & 16.492991 & -1.126869 & 21.1 & 1.5 & 0.43 & -82.5 & 5.17 & 6.04 & 2.15 & 3.00 & 0.00 & 0.00 & 3.02 & 3.04 \\
100040 & 34697.3 & 1108.7 & 20.074340 & -1.128531 & 14.7 & 1.1 & 0.17 & 79.3 & 0.00 & 4.85 & 0.00 & 3.01 & 0.00 & 0.00 & 0.00 & 1.85 \\
100041 & 35355.1 & 1108.5 & 19.928174 & -1.128544 & 19.2 & 1.1 & 0.14 & -62.1 & 0.00 & 6.08 & 0.00 & 3.06 & 0.00 & 0.00 & 0.00 & 3.02
\enddata
\tablecomments{The full table is published in its entirety at
  \myhref{http://irsa.ipac.caltech.edu/data/SPITZER/SHELA/catalogs}.  This is a portion of the full table to provide form and guidance.  1.\ Unique object ID number, 2.\ central X pixel coordinate, 3.\ central Y pixel coordinate, 4.\ object right ascension (J2000) in decimal degrees, 5.\ object declination (J2000) in decimal degrees, 6.\ Isophotal area in the detection (combined 2 band, 3 epoch) image, 7.\ semimajor axis in the detection image, 8.\ ellipticity measured in the detection image, defined as $e = 1 - b/a$, where $b$ and $a$ are the semiminor and semimajor axes, respectively, 9.\ position angle measured in the detection image (degrees E from N), 10-17.\ values of the weight maps in the images at the location of the object.  The weight map is proportional to the exposure time map.  10--11.\ values in combined 3.6 and 4.5~\micron\ image weight maps, respectively.  12.--17.\ values in the 3.6 and 4.5~\micron\ weight maps for the individual epochs, respectively.}
\end{deluxetable}

\clearpage

\begin{deluxetable}{lcccccccccccccccccc}
\scriptsize
\tablewidth{0pt}
\tablecolumns{19}
\tablecaption{Photometry for Combined, 3 Epoch SHELA IRAC Catalogs\label{table:fullcatalog} }
\tablehead{\colhead{ID} & \colhead{Flags} & \colhead{Flags}
  & \colhead{f$^{(3.6)}_{\nu,\mathrm{ISO}}$} &
  \colhead{$\sigma^\mathrm{(3.6)}_{\mathrm{ISO}}$} &
  \colhead{f$^{(3.6)}_{\nu,\mathrm{AUTO}}$} &
  \colhead{$\sigma^\mathrm{(3.6)}_{\mathrm{AUTO}}$} &
  \colhead{f$^{(3.6)}_{\nu,\mathrm{4\arcsec}}$} &
  \colhead{$\sigma^\mathrm{(3.6)}_{\mathrm{4\arcsec}}$} &
  \colhead{f$^{(3.6)}_{\nu,\mathrm{6\arcsec}}$} &
  \colhead{$\sigma^\mathrm{(3.6)}_{\mathrm{6\arcsec}}$} &
  \colhead{f$^{(4.5)}_{\nu,\mathrm{ISO}}$} &
  \colhead{$\sigma^\mathrm{(4.5)}_{\mathrm{ISO}}$} &
  \colhead{f$^{(4.5)}_{\nu,\mathrm{AUTO}}$} &
  \colhead{$\sigma^\mathrm{(4.5)}_{\mathrm{AUTO}}$} &
  \colhead{f$^{(4.5)}_{\nu,\mathrm{4\arcsec}}$} &
  \colhead{$\sigma^\mathrm{(4.5)}_{\mathrm{4\arcsec}}$} &
  \colhead{f$^{(4.5)}_{\nu,\mathrm{6\arcsec}}$} &
  \colhead{$\sigma^\mathrm{(4.5)}_{\mathrm{6\arcsec}}$}\\ \colhead{ }
  & \colhead{(3.6$\mu$m)} & \colhead{(4.5$\mu$m)} &
  \colhead{($\mu$Jy)} & \colhead{($\mu$Jy)} & \colhead{($\mu$Jy)} &
  \colhead{($\mu$Jy)} & \colhead{($\mu$Jy)} & \colhead{($\mu$Jy)} &
  \colhead{($\mu$Jy)} & \colhead{($\mu$Jy)} & \colhead{($\mu$Jy)} &
  \colhead{($\mu$Jy)} & \colhead{ } & \colhead{($\mu$Jy)} &
  \colhead{($\mu$Jy)} & \colhead{($\mu$Jy)} & \colhead{($\mu$Jy)} &
  \colhead{($\mu$Jy)} \\
\colhead{(1)} & \colhead{(2)} & \colhead{(3)} & \colhead{(4)} &
\colhead{(5)} & \colhead{(6)} & \colhead{(7)} & \colhead{(8)} &
\colhead{(9)} & \colhead{(10)} & \colhead{(11)} & \colhead{(12)} &
\colhead{(13)} & \colhead{(14)} & \colhead{(15)} & \colhead{(16)} &
\colhead{(17)} & \colhead{(18)} & \colhead{(19)} } 
\startdata
100020 & 2 & 2 & 454 & 8.34 & 452 & 8.73 & 334 & 2.81 & 408 & 3.61 & 525 & 6.27 & 525 & 6.52 & 392 & 2.84 & 470 & 3.37 \\
100021 & 0 & 0 & 2.86 & 0.585 & 10.2 & 2.92 & 4.72 & 0.853 & 5.76 & 1.46 & 2.98 & 0.52 & 3.81 & 2.5 & 5.06 & 0.755 & 4.37 & 1.26 \\
100022 & 2 & 2 & 5.7 & 0.939 & 6.89 & 3.51 & 7.33 & 0.973 & 8.93 & 1.65 & 5.09 & 0.952 & 8.42 & 3.59 & 6.96 & 0.989 & 7.03 & 1.67 \\
100023 & 1 & 1 & 3.72 & 0.939 & 2.87 & 1.45 & 2.93 & 0.951 & 2.42 & 1.65 & 5.71 & 0.894 & 7.46 & 1.37 & 6.78 & 0.923 & 7.35 & 1.56 \\
100024 & 0 & 0 & 43.5 & 2.12 & 45 & 3.11 & 48.5 & 1.47 & 49.3 & 2.21 & 26.7 & 1.47 & 26.8 & 2.12 & 31.5 & 1.07 & 30.8 & 1.54 \\
100025 & 0 & 0 & 3.24 & 0.518 & 6.68 & 2.1 & 5.23 & 0.747 & 6.22 & 1.25 & 2.92 & 0.557 & 4.03 & 2.32 & 4.25 & 0.803 & 3.16 & 1.36 \\
100026 & 0 & 0 & 2.05 & 0.612 & 3.47 & 1.16 & 4.99 & 1.33 & 4.78 & 2.32 & 2.47 & 0.443 & 3.54 & 0.798 & 4.7 & 0.918 & 3.96 & 1.57 \\
100027 & 0 & 0 & 19 & 1.52 & 20 & 2.5 & 19.7 & 1.04 & 20.8 & 1.62 & 23.7 & 1.55 & 25.5 & 2.52 & 26.6 & 1.1 & 27 & 1.66 \\
100028 & 0 & 0 & 1.68 & 0.788 & 1.9 & 1.93 & 3.78 & 1.34 & 3.73 & 2.34 & 3.36 & 0.591 & 4.2 & 1.37 & 4.68 & 0.959 & 4.35 & 1.64 \\
100029 & 1 & 0 & 0 & 0 & 0 & 0 & 0 & 0 & 0 & 0 & 9.3 & 0.882 & 10.5 & 1.61 & 12.4 & 0.93 & 12.2 & 1.49 \\
100030 & 3 & 3 & 17.9 & 1.59 & 19.5 & 3.16 & 13.7 & 0.946 & 18.1 & 1.53 & 13.3 & 1.48 & 12.8 & 2.95 & 9.71 & 0.862 & 12.9 & 1.42 \\
100031 & 0 & 0 & 4.34 & 0.871 & 7.98 & 2.18 & 8.21 & 1.36 & 8.4 & 2.35 & 3.87 & 0.865 & 6.03 & 2.17 & 6.95 & 1.35 & 6.44 & 2.34 \\
100032 & 3 & 3 & 70.5 & 2.7 & 73.3 & 4.63 & 56.7 & 1.27 & 62.9 & 1.71 & 79.8 & 3.24 & 80.6 & 5.63 & 65.3 & 1.43 & 71.4 & 2 \\
100033 & 0 & 0 & 1.92 & 0.55 & 1.41 & 1.18 & 1.7 & 1.32 & -2.2 & 2.32 & 1.7 & 0.405 & 4.35 & 0.871 & 5.35 & 0.975 & 4.69 & 1.66 \\
100034 & 1 & 1 & 2.63 & 0.742 & 26.4 & 6.25 & 6.69 & 1.35 & 12.3 & 2.37 & 2.16 & 0.697 & 14.8 & 5.87 & 5.03 & 1.27 & 6.49 & 2.21 \\
100035 & 0 & 0 & 0.65 & 0.198 & 1.29 & 0.88 & 2.3 & 0.622 & 1.57 & 1.06 & 0.932 & 0.29 & 1.7 & 1.37 & 2.19 & 0.946 & 0.553 & 1.64 \\
100036 & 2 & 2 & 152 & 3.55 & 151 & 3.89 & 142 & 1.78 & 149 & 2.13 & 136 & 3.28 & 136 & 3.59 & 127 & 1.68 & 133 & 1.99 \\
100037 & 3 & 3 & 7.04 & 0.984 & 7.71 & 3.35 & 8.53 & 0.851 & 8.55 & 1.39 & 7.45 & 1.03 & 10.9 & 3.54 & 9 & 0.892 & 11.1 & 1.48 \\
100038 & 0 & 0 & 21.2 & 1.34 & 25.3 & 2.97 & 24.4 & 1.05 & 25.1 & 1.58 & 14.3 & 1.69 & 14.3 & 3.99 & 16.4 & 1.23 & 16.5 & 2.03 \\
100039 & 2 & 2 & 13.4 & 1.52 & 14.9 & 4.49 & 12.9 & 1.11 & 15.3 & 1.84 & 12.9 & 1.41 & 17.3 & 4.16 & 11.8 & 1.03 & 16.9 & 1.73 \\
100040 & 1 & 0 & 0 & 0 & 0 & 0 & 0 & 0 & 0 & 0 & 14.8 & 1.26 & 15.8 & 2.51 & 18.3 & 1.18 & 18.4 & 1.91 \\
100041 & 1 & 0 & 0 & 0 & 0 & 0 & 0 & 0 & 0 & 0 & 23.9 & 1.4 & 26.4 & 0.697 & 28.5 & 1.17 & 28.8 & 1.78
\enddata
\tablecomments{The full table is published in its entirety at
  \myhref{http://irsa.ipac.caltech.edu/data/SPITZER/SHELA/catalogs}. This is a portion of the full table to provide form and guidance. 1.\ Object ID number in Table~\ref{table:preamblecatalog}, 2.\ SExtractor flags in the 3.6~\micron\ image, 3.\ SExtractor flags in the 4.5~\micron\ image, 4.\ isophotal flux in the 3.6~\micron\ image, 5.\ error on isophotal flux, 6.\ total (Kron) flux in 3.6~\micron\ image, 7.\ error on total flux, 8.\ 3.6~\micron\ flux measured in 4\arcsec-diameter aperture, corrected to total, 9.\ error on 4\arcsec-diameter flux, 10.\ 3.6~\micron\ flux measured in 6\arcsec-diameter aperture, corrected to total, 11.\ error on 6\arcsec-diameter flux, 12.\ isophotal flux in the 4.5~\micron\ image, 13.\ error on isophotal flux, 14.\ total (Kron) flux in 4.5~\micron\ image, 15.\ error on total flux, 16.\ 4.5~\micron\ flux measured in 4\arcsec-diameter aperture, corrected to total, 17.\ error on 4\arcsec-diameter flux, 18.\ 4.5~\micron\ flux measured in 6\arcsec-diameter aperture, corrected to total, 19.\ error on 6\arcsec-diameter flux.}
\end{deluxetable}


\clearpage

\begin{deluxetable}{ccccccccccccccccccc}
\tablewidth{0pt}
\tablecolumns{19}
\scriptsize
\tablecaption{Photometry for SHELA Epoch 1 IRAC Catalogs\label{table:ep1catalog}}
\tablehead{\colhead{ID} & \colhead{Flags} & \colhead{Flags}
  & \colhead{f$^{(3.6)}_{\nu,\mathrm{ISO}}$} &
  \colhead{$\sigma^\mathrm{(3.6)}_{\mathrm{ISO}}$} &
  \colhead{f$^{(3.6)}_{\nu,\mathrm{AUTO}}$} &
  \colhead{$\sigma^\mathrm{(3.6)}_{\mathrm{AUTO}}$} &
  \colhead{f$^{(3.6)}_{\nu,\mathrm{4\arcsec}}$} &
  \colhead{$\sigma^\mathrm{(3.6)}_{\mathrm{4\arcsec}}$} &
  \colhead{f$^{(3.6)}_{\nu,\mathrm{6\arcsec}}$} &
  \colhead{$\sigma^\mathrm{(3.6)}_{\mathrm{6\arcsec}}$} &
  \colhead{f$^{(4.5)}_{\nu,\mathrm{ISO}}$} &
  \colhead{$\sigma^\mathrm{(4.5)}_{\mathrm{ISO}}$} &
  \colhead{f$^{(4.5)}_{\nu,\mathrm{AUTO}}$} &
  \colhead{$\sigma^\mathrm{(4.5)}_{\mathrm{AUTO}}$} &
  \colhead{f$^{(4.5)}_{\nu,\mathrm{4\arcsec}}$} &
  \colhead{$\sigma^\mathrm{(4.5)}_{\mathrm{4\arcsec}}$} &
  \colhead{f$^{(4.5)}_{\nu,\mathrm{6\arcsec}}$} &
  \colhead{$\sigma^\mathrm{(4.5)}_{\mathrm{6\arcsec}}$}\\ \colhead{ }
  & \colhead{(3.6$\mu$m)} & \colhead{(4.5$\mu$m)} &
  \colhead{($\mu$Jy)} & \colhead{($\mu$Jy)} & \colhead{($\mu$Jy)} &
  \colhead{($\mu$Jy)} & \colhead{($\mu$Jy)} & \colhead{($\mu$Jy)} &
  \colhead{($\mu$Jy)} & \colhead{($\mu$Jy)} & \colhead{($\mu$Jy)} &
  \colhead{($\mu$Jy)} & \colhead{ } & \colhead{($\mu$Jy)} &
  \colhead{($\mu$Jy)} & \colhead{($\mu$Jy)} & \colhead{($\mu$Jy)} &
  \colhead{($\mu$Jy)} \\
\colhead{(1)} & \colhead{(2)} & \colhead{(3)} & \colhead{(4)} &
\colhead{(5)} & \colhead{(6)} & \colhead{(7)} & \colhead{(8)} &
\colhead{(9)} & \colhead{(10)} & \colhead{(11)} & \colhead{(12)} &
\colhead{(13)} & \colhead{(14)} & \colhead{(15)} & \colhead{(16)} &
\colhead{(17)} & \colhead{(18)} & \colhead{(19)} } 
\startdata
100020 & 3 & 2 & 0 & 0 & 0 & 0 & 0 & 0 & 0 & 0 & 506 & 12.2 & 514 & 12.7 & 385 & 3.46 & 463 & 4.86 \\
100021 & 0 & 0 & 3.78 & 0.904 & 9.91 & 4.45 & 6.02 & 1.33 & 5.85 & 2.29 & 2.46 & 0.889 & -2.32 & 4.42 & 3.5 & 1.31 & 0.925 & 2.27 \\
100022 & 2 & 2 & 8.11 & 1.27 & 9.44 & 4.54 & 10.6 & 1.31 & 13.4 & 2.23 & 4.43 & 1.29 & 10.2 & 4.76 & 5.82 & 1.33 & 5.77 & 2.3 \\
100023 & 1 & 1 & 3.32 & 1.28 & 1.02 & 1.98 & 2.42 & 1.3 & -0.875 & 2.25 & 5.19 & 1.29 & 6.8 & 2.01 & 5.91 & 1.32 & 5.83 & 2.28 \\
100024 & 1 & 0 & 0 & 0 & 0 & 0 & 0 & 0 & 0 & 0 & 27.2 & 2.31 & 28.1 & 3.42 & 30.7 & 1.5 & 31.7 & 2.4 \\
100025 & 0 & 0 & 2.4 & 0.7 & 3.13 & 2.95 & 3.14 & 1.02 & 3.39 & 1.77 & 3.04 & 0.898 & 5.76 & 3.82 & 5.29 & 1.33 & 4.44 & 2.29 \\
100026 & 1 & 0 & 0 & 0 & 0 & 0 & 0 & 0 & 0 & 0 & 2.17 & 0.626 & 3.71 & 1.19 & 5.17 & 1.38 & 3.62 & 2.37 \\
100027 & 0 & 0 & 22.4 & 2.21 & 26.4 & 3.65 & 22.5 & 1.44 & 25.4 & 2.37 & 24.2 & 1.93 & 26.3 & 3.14 & 26.6 & 1.3 & 27.5 & 2.06 \\
100028 & 1 & 0 & 0 & 0 & 0 & 0 & 0 & 0 & 0 & 0 & 3.47 & 0.779 & 4.3 & 1.87 & 4.69 & 1.29 & 4.92 & 2.24 \\
100029 & 1 & 0 & 0 & 0 & 0 & 0 & 0 & 0 & 0 & 0 & 9.98 & 1.06 & 13 & 1.96 & 13.4 & 1.11 & 13.9 & 1.82 \\
100030 & 3 & 3 & 18.6 & 2.43 & 16.8 & 4.74 & 14.5 & 1.38 & 20.6 & 2.34 & 11.5 & 2.37 & 12.1 & 4.66 & 8.36 & 1.32 & 12.1 & 2.28 \\
100031 & 0 & 0 & 4.28 & 0.857 & 7.92 & 2.14 & 8.11 & 1.35 & 8.37 & 2.3 & 3.75 & 0.849 & 5.89 & 2.13 & 6.7 & 1.33 & 6.27 & 2.29 \\
100032 & 3 & 3 & 71.1 & 3.06 & 71.4 & 5.03 & 56.6 & 1.37 & 63.4 & 1.93 & 84.9 & 4.23 & 84.5 & 7.07 & 68.9 & 1.71 & 76 & 2.56 \\
100033 & 1 & 0 & 0 & 0 & 0 & 0 & 0 & 0 & 0 & 0 & 0.851 & 0.516 & 2.37 & 1.16 & 2.18 & 1.3 & 0.543 & 2.26 \\
100034 & 1 & 1 & 2.67 & 0.729 & 26 & 5.72 & 6.61 & 1.34 & 12.1 & 2.32 & 2.14 & 0.684 & 15.1 & 5.37 & 5.06 & 1.25 & 6.55 & 2.17 \\
100035 & 0 & 0 & 0.233 & 0.36 & -1.22 & 1.89 & 0.9 & 1.29 & -0.846 & 2.26 & 1.01 & 0.38 & 2.14 & 1.9 & 2.5 & 1.31 & 1.23 & 2.27 \\
100036 & 2 & 2 & 148 & 5.28 & 149 & 5.8 & 141 & 2.06 & 148 & 2.8 & 147 & 5.86 & 146 & 6.45 & 128 & 2.1 & 137 & 2.99 \\
100037 & 3 & 3 & 5.01 & 1.59 & 3.26 & 5.31 & 5.99 & 1.33 & 6.37 & 2.28 & 9.81 & 1.92 & 17.8 & 6.38 & 11.1 & 1.6 & 17.4 & 2.77 \\
100038 & 0 & 0 & 18 & 1.96 & 19.1 & 4.44 & 20.9 & 1.42 & 21.1 & 2.34 & 13.7 & 2.35 & 14 & 5.41 & 17.2 & 1.67 & 15.4 & 2.81 \\
100039 & 2 & 2 & 15 & 2.25 & 19.5 & 6.39 & 14 & 1.6 & 15.6 & 2.73 & 12.6 & 1.91 & 22.5 & 5.42 & 10.2 & 1.36 & 15.1 & 2.33 \\
100040 & 1 & 0 & 0 & 0 & 0 & 0 & 0 & 0 & 0 & 0 & 16.3 & 1.54 & 17.9 & 3.07 & 20.4 & 1.43 & 20.3 & 2.35 \\
100041 & 1 & 0 & 0 & 0 & 0 & 0 & 0 & 0 & 0 & 0 & 22 & 1.83 & 24.2 & 0.668 & 26.4 & 1.45 & 27 & 2.35
\enddata
\tablecomments{The full table is published in its entirety at
  \myhref{http://irsa.ipac.caltech.edu/data/SPITZER/SHELA/catalogs}. This is a portion of the full table to provide form and guidance. 1.\ Object ID number in Table~\ref{table:preamblecatalog}, 2.\ SExtractor flags in the 3.6~\micron\ image, 3.\ SExtractor flags in the 4.5~\micron\ image, 4.\ isophotal flux in the 3.6~\micron\ image, 5.\ error on isophotal flux, 6.\ total (Kron) flux in 3.6~\micron\ image, 7.\ error on total flux, 8.\ 3.6~\micron\ flux measured in 4\arcsec-diameter aperture, corrected to total, 9.\ error on 4\arcsec-diameter flux, 10.\ 3.6~\micron\ flux measured in 6\arcsec-diameter aperture, corrected to total, 11.\ error on 6\arcsec-diameter flux, 12.\ isophotal flux in the 4.5~\micron\ image, 13.\ error on isophotal flux, 14.\ total (Kron) flux in 4.5~\micron\ image, 15.\ error on total flux, 16.\ 4.5~\micron\ flux measured in 4\arcsec-diameter aperture, corrected to total, 17.\ error on 4\arcsec-diameter flux, 18.\ 4.5~\micron\ flux measured in 6\arcsec-diameter aperture, corrected to total, 19.\ error on 6\arcsec-diameter flux.}
\end{deluxetable}


\clearpage
\begin{deluxetable}{ccccccccccccccccccc}
\tablewidth{0pt}
\tablecolumns{19}
\scriptsize
\tablecaption{Photometry for SHELA Epoch 2 IRAC Catalogs\label{table:ep2catalog}}
\tablehead{\colhead{ID} & \colhead{Flags} & \colhead{Flags}
  & \colhead{f$^{(3.6)}_{\nu,\mathrm{ISO}}$} &
  \colhead{$\sigma^\mathrm{(3.6)}_{\mathrm{ISO}}$} &
  \colhead{f$^{(3.6)}_{\nu,\mathrm{AUTO}}$} &
  \colhead{$\sigma^\mathrm{(3.6)}_{\mathrm{AUTO}}$} &
  \colhead{f$^{(3.6)}_{\nu,\mathrm{4\arcsec}}$} &
  \colhead{$\sigma^\mathrm{(3.6)}_{\mathrm{4\arcsec}}$} &
  \colhead{f$^{(3.6)}_{\nu,\mathrm{6\arcsec}}$} &
  \colhead{$\sigma^\mathrm{(3.6)}_{\mathrm{6\arcsec}}$} &
  \colhead{f$^{(4.5)}_{\nu,\mathrm{ISO}}$} &
  \colhead{$\sigma^\mathrm{(4.5)}_{\mathrm{ISO}}$} &
  \colhead{f$^{(4.5)}_{\nu,\mathrm{AUTO}}$} &
  \colhead{$\sigma^\mathrm{(4.5)}_{\mathrm{AUTO}}$} &
  \colhead{f$^{(4.5)}_{\nu,\mathrm{4\arcsec}}$} &
  \colhead{$\sigma^\mathrm{(4.5)}_{\mathrm{4\arcsec}}$} &
  \colhead{f$^{(4.5)}_{\nu,\mathrm{6\arcsec}}$} &
  \colhead{$\sigma^\mathrm{(4.5)}_{\mathrm{6\arcsec}}$}\\ \colhead{ }
  & \colhead{(3.6$\mu$m)} & \colhead{(4.5$\mu$m)} &
  \colhead{($\mu$Jy)} & \colhead{($\mu$Jy)} & \colhead{($\mu$Jy)} &
  \colhead{($\mu$Jy)} & \colhead{($\mu$Jy)} & \colhead{($\mu$Jy)} &
  \colhead{($\mu$Jy)} & \colhead{($\mu$Jy)} & \colhead{($\mu$Jy)} &
  \colhead{($\mu$Jy)} & \colhead{ } & \colhead{($\mu$Jy)} &
  \colhead{($\mu$Jy)} & \colhead{($\mu$Jy)} & \colhead{($\mu$Jy)} &
  \colhead{($\mu$Jy)}\\ 
\colhead{(1)} & \colhead{(2)} & \colhead{(3)} & \colhead{(4)} &
\colhead{(5)} & \colhead{(6)} & \colhead{(7)} & \colhead{(8)} &
\colhead{(9)} & \colhead{(10)} & \colhead{(11)} & \colhead{(12)} &
\colhead{(13)} & \colhead{(14)} & \colhead{(15)} & \colhead{(16)} &
\colhead{(17)} & \colhead{(18)} & \colhead{(19)}  }
\startdata
100020 & 2 & 2 & 453 & 7.49 & 451 & 7.74 & 334 & 2.82 & 408 & 3.64 & 523 & 7.48 & 521 & 7.73 & 391 & 2.99 & 469 & 3.76 \\
100021 & 0 & 0 & 1.05 & 1.07 & 11.7 & 5.56 & 3.25 & 1.63 & 6.56 & 2.89 & 4.23 & 0.747 & 7.81 & 3.62 & 7.25 & 1.11 & 6.32 & 1.9 \\
100022 & 3 & 3 & 0 & 0 & 0 & 0 & 0 & 0 & 0 & 0 & 0 & 0 & 0 & 0 & 0 & 0 & 0 & 0 \\
100023 & 1 & 1 & 0 & 0 & 0 & 0 & 0 & 0 & 0 & 0 & 0 & 0 & 0 & 0 & 0 & 0 & 0 & 0 \\
100024 & 0 & 0 & 43.5 & 2.14 & 45.1 & 3.1 & 48.4 & 1.47 & 49.3 & 2.23 & 28.3 & 2.37 & 28.7 & 3.51 & 32.2 & 1.52 & 32.3 & 2.47 \\
100025 & 0 & 0 & 4.27 & 0.913 & 1.29 & 3.91 & 6.68 & 1.36 & 4.18 & 2.36 & 4.44 & 0.919 & 9 & 3.95 & 6.81 & 1.37 & 6.67 & 2.38 \\
100026 & 0 & 1 & 2.09 & 0.597 & 3.5 & 1.16 & 4.95 & 1.35 & 4.36 & 2.36 & 0 & 0 & 0 & 0 & 0 & 0 & 0 & 0 \\
100027 & 0 & 0 & 16.8 & 1.94 & 15.8 & 3.19 & 17.9 & 1.26 & 17.8 & 2.08 & 22 & 2.27 & 22.8 & 3.74 & 25.6 & 1.48 & 24.8 & 2.44 \\
100028 & 0 & 1 & 1.7 & 0.779 & 2.02 & 1.96 & 3.8 & 1.35 & 3.82 & 2.37 & 0 & 0 & 0 & 0 & 0 & 0 & 0 & 0 \\
100029 & 1 & 1 & 0 & 0 & 0 & 0 & 0 & 0 & 0 & 0 & 0 & 0 & 0 & 0 & 0 & 0 & 0 & 0 \\
100030 & 3 & 3 & 17.3 & 2.53 & 19.1 & 4.88 & 13.8 & 1.42 & 16.1 & 2.42 & 18.5 & 2.53 & 20 & 4.88 & 13.2 & 1.41 & 18.4 & 2.43 \\
100031 & 1 & 1 & 0 & 0 & 0 & 0 & 0 & 0 & 0 & 0 & 0 & 0 & 0 & 0 & 0 & 0 & 0 & 0 \\
100032 & 3 & 3 & 0 & 0 & 0 & 0 & 0 & 0 & 0 & 0 & 0 & 0 & 0 & 0 & 0 & 0 & 0 & 0 \\
100033 & 0 & 1 & 1.9 & 0.533 & 1.42 & 1.19 & 1.72 & 1.33 & -2.22 & 2.36 & 0 & 0 & 0 & 0 & 0 & 0 & 0 & 0 \\
100034 & 1 & 1 & 0 & 0 & 0 & 0 & 0 & 0 & 0 & 0 & 0 & 0 & 0 & 0 & 0 & 0 & 0 & 0 \\
100035 & 0 & 1 & 0.744 & 0.233 & -0.221 & 1.14 & 1.39 & 0.786 & -0.559 & 1.37 & 0 & 0 & 0 & 0 & 0 & 0 & 0 & 0 \\
100036 & 2 & 2 & 146 & 5.36 & 145 & 5.85 & 139 & 2.07 & 146 & 2.86 & 125 & 5.34 & 124 & 5.83 & 126 & 2.01 & 129 & 2.81 \\
100037 & 3 & 3 & 7.01 & 1.66 & 7.93 & 5.43 & 8.07 & 1.38 & 7.97 & 2.39 & 5.79 & 1.65 & 8.11 & 5.43 & 7.35 & 1.37 & 8.29 & 2.39 \\
100038 & 0 & 1 & 22.7 & 2.1 & 30.6 & 4.7 & 26.6 & 1.52 & 27.2 & 2.5 & 0 & 0 & 0 & 0 & 0 & 0 & 0 & 0 \\
100039 & 3 & 3 & 0 & 0 & 0 & 0 & 0 & 0 & 0 & 0 & 0 & 0 & 0 & 0 & 0 & 0 & 0 & 0 \\
100040 & 1 & 1 & 0 & 0 & 0 & 0 & 0 & 0 & 0 & 0 & 0 & 0 & 0 & 0 & 0 & 0 & 0 & 0 \\
100041 & 1 & 1 & 0 & 0 & 0 & 0 & 0 & 0 & 0 & 0 & 0 & 0 & 0 & 0 & 0 & 0
& 0 & 0
\enddata
\tablecomments{The full table is published in its entirety at
  \myhref{http://irsa.ipac.caltech.edu/data/SPITZER/SHELA/catalogs}. This is a portion of the full table to provide form and guidance. 1.\ Object ID number in Table~\ref{table:preamblecatalog}, 2.\ SExtractor flags in the 3.6~\micron\ image, 3.\ SExtractor flags in the 4.5~\micron\ image, 4.\ isophotal flux in the 3.6~\micron\ image, 5.\ error on isophotal flux, 6.\ total (Kron) flux in 3.6~\micron\ image, 7.\ error on total flux, 8.\ 3.6~\micron\ flux measured in 4\arcsec-diameter aperture, corrected to total, 9.\ error on 4\arcsec-diameter flux, 10.\ 3.6~\micron\ flux measured in 6\arcsec-diameter aperture, corrected to total, 11.\ error on 6\arcsec-diameter flux, 12.\ isophotal flux in the 4.5~\micron\ image, 13.\ error on isophotal flux, 14.\ total (Kron) flux in 4.5~\micron\ image, 15.\ error on total flux, 16.\ 4.5~\micron\ flux measured in 4\arcsec-diameter aperture, corrected to total, 17.\ error on 4\arcsec-diameter flux, 18.\ 4.5~\micron\ flux measured in 6\arcsec-diameter aperture, corrected to total, 19.\ error on 6\arcsec-diameter flux.}
\end{deluxetable}


\clearpage

\begin{deluxetable}{ccccccccccccccccccc}
\tablewidth{0pt}
\tablecolumns{19}
\scriptsize
\tablecaption{Photometry from SHELA Epoch 3 IRAC Catalogs\label{table:ep3catalog}}
\tablehead{\colhead{ID} & \colhead{Flags} & \colhead{Flags}
  & \colhead{f$^{(3.6)}_{\nu,\mathrm{ISO}}$} &
  \colhead{$\sigma^\mathrm{(3.6)}_{\mathrm{ISO}}$} &
  \colhead{f$^{(3.6)}_{\nu,\mathrm{AUTO}}$} &
  \colhead{$\sigma^\mathrm{(3.6)}_{\mathrm{AUTO}}$} &
  \colhead{f$^{(3.6)}_{\nu,\mathrm{4\arcsec}}$} &
  \colhead{$\sigma^\mathrm{(3.6)}_{\mathrm{4\arcsec}}$} &
  \colhead{f$^{(3.6)}_{\nu,\mathrm{6\arcsec}}$} &
  \colhead{$\sigma^\mathrm{(3.6)}_{\mathrm{6\arcsec}}$} &
  \colhead{f$^{(4.5)}_{\nu,\mathrm{ISO}}$} &
  \colhead{$\sigma^\mathrm{(4.5)}_{\mathrm{ISO}}$} &
  \colhead{f$^{(4.5)}_{\nu,\mathrm{AUTO}}$} &
  \colhead{$\sigma^\mathrm{(4.5)}_{\mathrm{AUTO}}$} &
  \colhead{f$^{(4.5)}_{\nu,\mathrm{4\arcsec}}$} &
  \colhead{$\sigma^\mathrm{(4.5)}_{\mathrm{4\arcsec}}$} &
  \colhead{f$^{(4.5)}_{\nu,\mathrm{6\arcsec}}$} &
  \colhead{$\sigma^\mathrm{(4.5)}_{\mathrm{6\arcsec}}$}\\ \colhead{ }
  & \colhead{(3.6$\mu$m)} & \colhead{(4.5$\mu$m)} &
  \colhead{($\mu$Jy)} & \colhead{($\mu$Jy)} & \colhead{($\mu$Jy)} &
  \colhead{($\mu$Jy)} & \colhead{($\mu$Jy)} & \colhead{($\mu$Jy)} &
  \colhead{($\mu$Jy)} & \colhead{($\mu$Jy)} & \colhead{($\mu$Jy)} &
  \colhead{($\mu$Jy)} & \colhead{ } & \colhead{($\mu$Jy)} &
  \colhead{($\mu$Jy)} & \colhead{($\mu$Jy)} & \colhead{($\mu$Jy)} &
  \colhead{($\mu$Jy)} \\ 
\colhead{(1)} & \colhead{(2)} & \colhead{(3)} & \colhead{(4)} &
\colhead{(5)} & \colhead{(6)} & \colhead{(7)} & \colhead{(8)} &
\colhead{(9)} & \colhead{(10)} & \colhead{(11)} & \colhead{(12)} &
\colhead{(13)} & \colhead{(14)} & \colhead{(15)} & \colhead{(16)} &
\colhead{(17)} & \colhead{(18)} & \colhead{(19)} } 
\startdata
100020 & 3 & 2 & 0 & 0 & 0 & 0 & 0 & 0 & 0 & 0 & 529 & 8.89 & 528 & 9.23 & 395 & 3.14 & 474 & 4.09 \\
100021 & 0 & 0 & 2.94 & 0.903 & 7.83 & 4.48 & 3.94 & 1.34 & 4.06 & 2.33 & 1.06 & 0.884 & -0.863 & 4.47 & 2.14 & 1.32 & 2.93 & 2.32 \\
100022 & 2 & 2 & 2.87 & 1.29 & 3.54 & 4.73 & 3.65 & 1.32 & 3.73 & 2.31 & 5.37 & 1.3 & 6.02 & 4.73 & 7.6 & 1.35 & 7.79 & 2.32 \\
100023 & 1 & 1 & 4 & 1.31 & 4.04 & 2.04 & 3.03 & 1.33 & 4.67 & 2.33 & 5.94 & 1.15 & 7.72 & 1.78 & 7.25 & 1.18 & 8.22 & 2.03 \\
100024 & 1 & 0 & 0 & 0 & 0 & 0 & 0 & 0 & 0 & 0 & 24.4 & 2.32 & 23.2 & 3.43 & 31.4 & 1.51 & 28 & 2.41 \\
100025 & 0 & 0 & 3.38 & 0.898 & 15.7 & 3.85 & 6.51 & 1.34 & 11.6 & 2.33 & 1.01 & 0.873 & -3.61 & 3.81 & 0.423 & 1.3 & -2.04 & 2.29 \\
100026 & 1 & 0 & 0 & 0 & 0 & 0 & 0 & 0 & 0 & 0 & 2.67 & 0.54 & 3.52 & 1.01 & 4.62 & 1.17 & 4.44 & 2.02 \\
100027 & 1 & 1 & 0 & 0 & 0 & 0 & 0 & 0 & 0 & 0 & 0 & 0 & 0 & 0 & 0 & 0 & 0 & 0 \\
100028 & 1 & 0 & 0 & 0 & 0 & 0 & 0 & 0 & 0 & 0 & 3.16 & 0.792 & 3.74 & 1.93 & 4.47 & 1.33 & 3.16 & 2.31 \\
100029 & 1 & 0 & 0 & 0 & 0 & 0 & 0 & 0 & 0 & 0 & 7.98 & 1.33 & 6.38 & 2.53 & 10.7 & 1.38 & 9.06 & 2.34 \\
100030 & 3 & 3 & 16.4 & 3.01 & 21.2 & 5.86 & 12.3 & 1.67 & 16.4 & 2.89 & 9.7 & 2.45 & 4.07 & 4.76 & 7.64 & 1.36 & 7.77 & 2.34 \\
100031 & 1 & 1 & 0 & 0 & 0 & 0 & 0 & 0 & 0 & 0 & 0 & 0 & 0 & 0 & 0 & 0 & 0 & 0 \\
100032 & 3 & 3 & 68.3 & 4.2 & 74.9 & 6.91 & 56.2 & 1.65 & 60.7 & 2.52 & 73.5 & 4.2 & 75 & 6.91 & 61.5 & 1.68 & 66.2 & 2.54 \\
100033 & 1 & 0 & 0 & 0 & 0 & 0 & 0 & 0 & 0 & 0 & 2.55 & 0.547 & 6.13 & 1.21 & 8.2 & 1.36 & 8.53 & 2.35 \\
100034 & 1 & 1 & 0 & 0 & 0 & 0 & 0 & 0 & 0 & 0 & 0 & 0 & 0 & 0 & 0 & 0 & 0 & 0 \\
100035 & 0 & 0 & 0.787 & 0.37 & 6.71 & 1.95 & 5.44 & 1.34 & 8.56 & 2.34 & 0.86 & 0.373 & 1 & 1.93 & 1.78 & 1.32 & -0.328 & 2.31 \\
100036 & 2 & 2 & 159 & 5.34 & 158 & 5.84 & 145 & 2.09 & 151 & 2.85 & 135 & 4.11 & 135 & 4.47 & 127 & 1.82 & 133 & 2.33 \\
100037 & 3 & 3 & 8.98 & 1.64 & 11.1 & 5.33 & 11.4 & 1.38 & 11.2 & 2.34 & 7.24 & 1.62 & 7.34 & 5.31 & 8.96 & 1.36 & 8.64 & 2.33 \\
100038 & 0 & 0 & 23.4 & 2.43 & 26.3 & 5.47 & 26.1 & 1.74 & 27.7 & 2.9 & 14.6 & 2.29 & 14 & 5.2 & 15.4 & 1.61 & 17.1 & 2.74 \\
100039 & 2 & 2 & 12.2 & 1.93 & 10.4 & 5.39 & 12.1 & 1.38 & 15.1 & 2.36 & 12.6 & 1.93 & 10.2 & 5.37 & 12.9 & 1.39 & 17.8 & 2.36 \\
100040 & 1 & 0 & 0 & 0 & 0 & 0 & 0 & 0 & 0 & 0 & 11.9 & 1.92 & 11.7 & 3.93 & 14.4 & 1.74 & 14.4 & 2.99 \\
100041 & 1 & 0 & 0 & 0 & 0 & 0 & 0 & 0 & 0 & 0 & 25.7 & 1.88 & 27.7 &
0.714 & 30.5 & 1.5 & 30.2 & 2.42 
\enddata
\tablecomments{The full table is published in its entirety at
  \myhref{http://irsa.ipac.caltech.edu/data/SPITZER/SHELA/catalogs}. This is a portion of the full table to provide form and guidance. 1.\ Object ID number in Table~\ref{table:preamblecatalog}, 2.\ SExtractor flags in the 3.6~\micron\ image, 3.\ SExtractor flags in the 4.5~\micron\ image, 4.\ isophotal flux in the 3.6~\micron\ image, 5.\ error on isophotal flux, 6.\ total (Kron) flux in 3.6~\micron\ image, 7.\ error on total flux, 8.\ 3.6~\micron\ flux measured in 4\arcsec-diameter aperture, corrected to total, 9.\ error on 4\arcsec-diameter flux, 10.\ 3.6~\micron\ flux measured in 6\arcsec-diameter aperture, corrected to total, 11.\ error on 6\arcsec-diameter flux, 12.\ isophotal flux in the 4.5~\micron\ image, 13.\ error on isophotal flux, 14.\ total (Kron) flux in 4.5~\micron\ image, 15.\ error on total flux, 16.\ 4.5~\micron\ flux measured in 4\arcsec-diameter aperture, corrected to total, 17.\ error on 4\arcsec-diameter flux, 18.\ 4.5~\micron\ flux measured in 6\arcsec-diameter aperture, corrected to total, 19.\ error on 6\arcsec-diameter flux.}
\end{deluxetable}

%

\clearpage
\setcounter{table}{12}

\begin{deluxetable}{lccccccccccccccc}
\tablewidth{0pt}
\tablecolumns{16}
\scriptsize
\tablecaption{SDSS Stripe 82 Coadd Photometry for sources matched to SHELA\label{table:sdsscatalog}}
\tablehead{\colhead{ID} & \colhead{SDSS ID} & \colhead{SDSS RA} &
  \colhead{SDSS DEC} & \colhead{TYPE} & \colhead{SDSS FLAGS} &
  \colhead{$\sdssu$} & \colhead{$\sigma_{\sdssu}$} & \colhead{$\sdssg$} &
  \colhead{$\sigma_{\sdssg}$} & \colhead{$\sdssr$} & \colhead{$\sigma_{\sdssr}$} &
  \colhead{$\sdssi$} & \colhead{$\sigma_{\sdssi}$} & \colhead{$\sdssz$} &
  \colhead{$\sigma_{\sdssz}$}\\ 
 \colhead{ } & \colhead{ } & \colhead{(deg)} &
  \colhead{(deg)} & \colhead{ } & \colhead{ } & \colhead{(mag)} &
  \colhead{(mag)} & \colhead{(mag)} & \colhead{(mag)} &
  \colhead{(mag)} & \colhead{(mag)} & \colhead{(mag)} &
  \colhead{(mag)} & \colhead{(mag)} & \colhead{(mag)} \\
\colhead{(1)} & \colhead{(2)} & \colhead{(3)} & \colhead{(4)} &
\colhead{(5)} & \colhead{(6)} & \colhead{(7)} & \colhead{(8)} &
\colhead{(9)} & \colhead{(10)} & \colhead{(11)} & \colhead{(12)} &
\colhead{(13)} & \colhead{(14)} & \colhead{(15)} & \colhead{(16)} }
\startdata
100020 & 8647474690342256787 & 17.701219 & -1.132932 & 3 & 103347650576 & 19.377 & 0.016 & 18.439 & 0.004 & 17.740 & 0.003 & 17.434 & 0.003 & 17.175 & 0.006 \\
100024 & 8647474690341142937 & 15.221294 & -1.125124 & 6 & 34628174080 & 24.438 & 0.551 & 21.697 & 0.025 & 20.347 & 0.009 & 19.601 & 0.007 & 19.192 & 0.014 \\
100029 & 8647474690343175313 & 19.864137 & -1.127867 & 3 & 2450547277824 & 22.385 & 0.151 & 22.097 & 0.066 & 21.192 & 0.029 & 20.948 & 0.037 & 20.984 & 0.123 \\
100036 & 8647474690342650333 & 18.621063 & -1.131875 & 3 & 103347650560 & 22.802 & 0.243 & 21.048 & 0.027 & 19.731 & 0.009 & 19.213 & 0.010 & 18.837 & 0.019 \\
100038 & 8647474690342651302 & 18.717668 & -1.128721 & 3 & 70439879574360 & 25.901 & 3.217 & 24.574 & 0.547 & 23.981 & 0.294 & 23.269 & 0.231 & 22.215 & 0.290 \\
100041 & 8647474690343175190 & 19.928157 & -1.128586 & 3 & 68987912448 & 27.804 & 3.907 & 23.321 & 0.116 & 21.927 & 0.031 & 21.062 & 0.023 & 20.654 & 0.051 \\
100044 & 8647474690342258061 & 17.794618 & -1.127763 & 3 & 68987912448 & 23.361 & 0.328 & 22.871 & 0.114 & 21.835 & 0.042 & 21.323 & 0.039 & 21.003 & 0.095 \\
100048 & 8647474690342782172 & 18.936888 & -1.128068 & 3 & 281543964622848 & 24.181 & 0.547 & 23.570 & 0.170 & 23.485 & 0.150 & 23.571 & 0.253 & 24.049 & 1.245 \\
100049 & 8647474690342717098 & 18.867537 & -1.128079 & 3 & 68987912448 & 26.888 & 3.737 & 23.853 & 0.183 & 23.418 & 0.112 & 22.700 & 0.091 & 22.371 & 0.220 \\
100052 & 8647474690341797917 & 16.668147 & -1.135992 & 3 & 34628173840 & 16.636 & 0.003 & 15.029 & 0.003 & 14.560 & 0.003 & 14.510 & 0.003 & 14.477 & 0.003
\enddata
\tablecomments{The full table is published in its entirety at
  \myhref{http://irsa.ipac.caltech.edu/data/SPITZER/SHELA/catalogs}. This is a portion of the full table to provide form and guidance. 1.\ Object ID number in Table~\ref{table:preamblecatalog}, 2.\ SDSS ID number, 3.\ SDSS right ascension (J2000) in decimal degrees, 4.\ SDSS declination (J2000) in decimal degrees, 5.\ SDSS Type (common values are Type=3 for galaxy and Type=6 for star), 6.\ SDSS Flags value, 7.\ SDSS $u$ magnitude, 8.\ error on $u$ magnitude, 9.\ SDSS $g$ magnitude, 10.\ error on $g$ magnitude, 11.\ SDSS $r$ magnitude, 12.\ error on $r$ magnitude, 13.\ SDSS $i$ magnitude, 14.\ error on $i$ magnitude, 15.\ SDSS $z$ magnitude, 16.\ error on $z$ magnitude}
\end{deluxetable}

\end{turnpage}

\clearpage
\global\pdfpageattr\expandafter{\the\pdfpageattr/Rotate 90}

\end{document}